\documentclass[12pt]{article}
\usepackage{graphicx}

\def\ignore#1{{}}

\let\oldtheequation=\theequation
\def\doteqs#1{\setcounter{equation}{0}            
\def\theequation{{#1}.\oldtheequation}}
\newcounter{sxn}
\def\sx#1{\addtocounter{sxn}{1} \vskip 1.cm  \goodbreak
\noindent{\large\bf\leftline{\thesxn.~~#1}} \nobreak \vskip -.5cm}
\def\sxn#1{\sx{#1} \doteqs{\thesxn}}

\newcounter{axn}

\date{}

\newdimen\mybaselineskip
\newcommand{\beeq}{\begin{equation}}
\newcommand{\eneq}{\end{equation}}
\newcommand{\beqn}{\begin{eqnarray}}
\newcommand{\eeqn}{\end{eqnarray}}

\def\la{\raise.16ex\hbox{$\langle$}\lower.16ex\hbox{}  }
\def\ra{\, \raise.16ex\hbox{$\rangle$}\lower.16ex\hbox{} }

\def\psibar{ \psi \kern-.65em\raise.6em\hbox{$-$} \lower.6em\hbox{} }
\def\psibarb{ \psi \kern-.65em\raise.6em\hbox{$-$}  }

\begin{document}

\thispagestyle{empty}

\baselineskip=12pt

%{\small \noindent \mydate  \hfill UW}

%{\small \noindent Ramin\hfill   }

\vspace*{3.cm}

\begin{center}  
{\LARGE \bf  A Detailed Analytic Study of the Asymptotic Quasinormal Modes of Schwarzschild-Anti De Sitter Black Holes}
\end{center}

\baselineskip=14pt

\vspace{2cm}
\begin{center}
{\bf  Ramin G. Daghigh$\sharp$ and Michael D. Green$\dagger$}
\end{center}

\vspace{0.25 cm}
\centerline{\small \it $\sharp$ Natural Sciences Department, Metropolitan State University, Saint Paul, Minnesota, USA 55106}
\vskip 0 cm
\centerline{} 

\centerline{\small \it $\dagger$ Mathematics Department, Metropolitan State University, Saint Paul, Minnesota, USA 55106}
\vskip 0 cm
\centerline{} 

\vspace{1cm}
\begin{abstract}
We analyze analytically the asymptotic regions of the quasinormal mode frequency spectra with infinitely large overtone numbers for $D$-dimensional Schwarzschild black holes in anti de Sitter spacetimes.  In this limit, we confirm the analytic results obtained previously in the literature using different methods.  In addition, we show that in certain spacetime dimensions these techniques imply the existence of other regions of the asymptotic quasinormal mode frequency spectrum which have not previously appeared in the literature.  For large black holes, some of these modes have a damping rate of $1.2T_H$, where $T_H$ is the Hawking temperature.  This is less than the damping rate of the lowest overtone quasinormal mode calculated by other authors.  It is not completely clear whether these modes actually exist or are an artifact of an unknown flaw in the analytic techniques being used.  We discuss the possibility of the existence of these modes and explore some of the consequences.  We also examine the possible connection between the asymptotic quasinormal modes of Schwarzschild-anti de Sitter black holes and the quantum level spacing of their horizon area spectrum. 

\vspace{0.5cm}

\noindent PACS numbers: 04.20.-q, 04.30.-w, 04.50.Gh, 04.60.-m, 04.70.-s,
04.70.Bw, 04.70.Dy

\baselineskip=20pt plus 1pt minus 1pt
\end{abstract}

%%%%%%%%%%% 1 %%%%%%%%%%
\newpage

\sxn{Introduction}

\vskip 0.3cm

Black hole quasinormal modes (QNMs) are the natural vibrational modes of perturbations in the spacetime exterior to an event horizon. The corresponding frequency spectrum is discrete and complex. The imaginary part of the frequency signals the presence of damping, a necessary consequence of boundary conditions that require energy to be carried away from the system.

The study of QNMs was originally motivated by the question of stability of black holes against the emission of these modes.  Regge and Wheeler \cite{Regge} were the first who tested the stability of a black hole against the emission of QNMs in asymptotically flat spacetimes.  QNMs are also relevant in astrophysical observations.  For a review of this topic see the references in \cite{gwaves}.  The QNMs which are important for observational purposes are the slowly damped ones since they describe the frequency spectrum of the gravitational radiation that is expected to emerge from black hole formation during late times.  QNM frequencies can also be considered as the poles in black hole greybody factors which are important in the study of Hawking radiation.  The analytical monodromy techniques which were developed by Motl and Neitzke in \cite{Motl1, Motl2} to calculate the highly damped QNM frequencies of black holes are used by Neitzke \cite{Neitzke} and Harmark {\it et al} \cite{HarmarkNS} to compute the asymptotic greybody factors for static, spherically symmetric black holes.  The QNMs of black holes may also shed some light on the quantum nature of gravity via Bohr's correspondence principal.  One such correspondence is suggested by Hod \cite{Hod} who proposes that there may exist a connection between the highly damped QNMs of Schwarzschild black holes and the semi-classical level spacing in the black hole quantum area spectrum suggested by Bekenstein and Mukhanov \cite{Bekenstein-M}.  This proposal was motivated in part by the apparent universality of the real part
of the frequency in the high damping limit, as well as by its special numerical value for
Schwarzschild black holes, where
\beeq
8\pi M \omega \mathop{\longrightarrow}_{|\omega_I|\to\infty}\ln(3)+2\pi i\left(n+{1\over 2}\right)+O(n^{-{1\over 2}}) ~.
\label{HD-limit}
\eneq
Here $M$ is the mass of the black hole, $\omega_I$ is the imaginary part of the frequency, which is the damping rate, and $n \gg 1$ is the overtone number.  (Throughout this paper we use Planck units, so that $c = \hbar = G = k_B =1$).  The result (\ref{HD-limit}) was obtained numerically by Nollert \cite{Nollert} and confirmed analytically by Motl and Neitzke \cite{Motl1, Motl2} and using a different analytic technique was reconfirmed by Andersson and Howls \cite{Andersson}.  The coefficient $\ln(3)$ in (\ref{HD-limit}) allows an elegant statistical interpretation \cite{Hod, Bekenstein-1} of the resulting Bekenstein-Hawking entropy spectrum with minimum equidistant spacing of
\beeq
\Delta S=\ln (3) ~.
\eneq
Recently, Corichi {\it et al} \cite{Corichi} performed a detailed numerical analysis of the number of microstates compatible with
a given black hole horizon area within Loop Quantum Gravity (LQG) formalism and showed that there is indeed
a deep relation between the concept of entropy in LQG and in Bekenstein's heuristic picture supplemented
by Hod's proposal when oscillatory behavior in
the entropy-area relation is properly interpreted.  The resulting minimum spacing of the quantized entropy spectrum for a large black hole is
\beeq
\Delta S \approx 2 \gamma_0 \ln (3) ~,
\eneq
where $\gamma_0 \approx 0.274\dots$ is the Barbero-Immirzi parameter.  A new version of Hod's proposal was put forward recently by Maggiore \cite{Maggiore} who considers black hole perturbations in terms of a collection of damped harmonic oscillators and argues that the more relevant physical frequency in connecting QNMs to the quantum structure of the horizon area is the proper frequency $\omega_0$ of the undamped oscillators rather than the frequency of the damped oscillation, which coincides with the real part of the QNM frequency ($\omega_R$).  Since 
\beeq
\omega_0=\sqrt{\omega_R^2+\omega_I^2} ~,
\eneq
in the large damping limit we have $\omega_0\approx \omega_I$.  Maggiore's interpretation results in a different spacing of the quantized entropy spectrum where 
\beeq
\Delta S = 2 \pi ~.
\eneq
The main advantage of Maggiore's interpretation over Hod's interpretation is in the extent of its universality.  Unlike $\ln(3)$ which does not appear for certain field perturbations coupled to the background of a Schwarzschild black hole \cite{Motl2} and for certain classes of single horizon black holes in $2$-$D$ dilaton gravity theories \cite{Ramin1}, Maggiore's entropy spacing $2\pi$ appears for all types of perturbations and all single horizon black holes.  %Recently, it was also shown that Maggiore's entropy spacing of $2\pi$ appears even in Kerr black holes \cite{Kerr}.

In this paper we focus on the asymptotic regions of the QNM frequency spectrum with infinitely large overtone numbers for $D$-dimensional Schwarzschild black holes in anti de Sitter (AdS) spacetimes.  Black holes in AdS space have attracted a great deal of attention.  It was shown by Hawking and Page \cite{Page} that, unlike black holes in flat spacetimes, large black holes in AdS spacetimes have positive specific heat and can be in stable equilibrium with thermal radiation at a fixed temperature.  Also, according to the conjecture proposed by Maldacena \cite{Maldacena}, a large static black hole in AdS space corresponds to a thermal state in conformal field theory (CFT).  The decay of a test field in the black hole spacetime describes the return of a perturbed thermal state to its thermal equilibrium in the CFT.  The dynamical time scale for the return to thermal equilibrium in the CFT is simply equal to the decay rate of the perturbation of a large black hole in AdS spacetime.

The low overtone QNMs of Schwarzschild-AdS black holes for scalar perturbations was first addressed numerically by Chan and Mann \cite{low-tone-1}.  Horowitz and Hubeny \cite{Horowitz} computed these modes for scalar perturbations in four, five, and seven spacetime dimensions which are of interest in the context of AdS/CFT correspondence.  This work was completed by Konoplya \cite{Konoplya0} who analyzed the scalar QNMs for small Schwarzschild-AdS black holes in detail.  Cardoso and Lemos \cite{Cardoso-L} extended these calculations to include electromagnetic and gravitational perturbations.  Berti and Kokkotas confirmed all the above results in \cite{Berti-K} and extended their numerical calculations to Reissner-Nordstr$\ddot{\rm o}$m black holes in AdS spacetimes.  Finally, the low overtone QNMs of Dirac spinors were addressed by Giammatteo and Jing in \cite{low-tone-2}

The high overtone QNMs of Schwarzschild-AdS black holes were calculated numerically for scalar, electromagnetic, and gravitational perturbations in four spacetime dimensions by Cardoso, Konoplya, and Lemos in \cite{Cardoso-K-L}.  Motivated by AdS/CFT correspondence, many authors have calculated the asymptotic QNMs in five spacetime dimensions (some examples are given in \cite{high-tone-5d}).  Analytic calculations of the infinitely high overtone (asymptotic) QNMs of four dimensional Schwarzschild-AdS black holes were done by Cardoso {\it et al} in \cite{Cardoso-N-S} using a method based on the monodromy technique developed by Motl and Neitzke \cite{Motl1, Motl2}.  Natario and Schiappa \cite{Natario-S} generalized the analytic results to include Schwarzschild-AdS black holes in dimensions greater than four and Ghosh {\it et al} \cite{Ghosh} generalized the analytic results further by studying the infinitely high overtone QNMs of asymptotically non-flat black holes in a much more generic way which includes Schwarzschild-AdS black holes as a subset.   

In the absence of a black hole, most fields in an AdS space oscillate with discrete purely real frequencies (normal modes) because the negative cosmological constant provides an effective confining box in such a space.  Natario and Schiappa \cite{Natario-S} showed that as the size of a black hole in AdS space approaches zero the asymptotic QNM frequencies approach the pure AdS normal modes.  In this limit, the real part of the QNM frequency becomes much larger than the imaginary part.  We explore the asymptotic regions of the QNM frequency spectrum of Schwarzschild-AdS black holes using the analytic technique developed by Andersson and Howls \cite{Andersson}.  In addition to confirming the previous results, we show that, as far as the analytic calculations are concerned, there exist other regions of the asymptotic QNM frequency spectrum.  For example, in even spacetime dimensions we find asymptotic QNM frequencies that resemble the pure AdS normal mode frequencies.  We call these modes the highly real QNMs in which the real part of the frequency approaches infinity while the damping rate approaches a finite value.  These analytic techniques indicate that this region of the spectrum is not specific to small black holes but appears for Schwarzschild-AdS black holes of any size.  Assuming these modes actually exist, we show that they have a damping rate less than the one found by Horowitz and Hubeny \cite{Horowitz}, and therefore will be the dominant modes in the context of AdS/CFT correspondence.  In other words, in four spacetime dimensions, the highly real QNMs are the ones which determine the timescale for the approach to thermal equilibrium in field theory via AdS$_4$/CFT$_3$ correspondence.  In higher spacetime dimensions $D\ge 7$, other asymptotic regions of the QNM spectrum appear which will be discussed in this work.  

The paper is organized as follows.  In section 2, we describe the general formalism.  In section 3, we calculate the asymptotic QNM frequencies of Schwarzschild-AdS black holes in $D\ge 4$ spacetime dimensions using analytic phase-integral methods based on the WKB approximation.  In section 4, we explore the highly real QNMs in even spacetime dimensions.  In section 5, we explore the highly damped QNMs which are defined as the modes where the damping rate goes to infinity while the real part of the frequency approaches a finite value.  In section 6, we explore other possible asymptotic QNMs.  In section 7, we end the paper with conclusions and discussions.

\sxn{General Formalism and the Asymptotic Quasinormal Modes}
\vskip 0.3cm

Ishibashi and Kodama have shown in \cite{Ishibashi1}, \cite{Ishibashi2}, and \cite{Ishibashi3} that various classes of non-rotating black hole metric perturbations in a spacetime with dimension $D\ge 4$ are governed generically by a Schr$\ddot{\mbox o}$dinger wave-like equation of the form
\beeq
{d^2\psi \over dz^2}+\left[ \omega^2-V(r) \right]\psi =0 ~,
\label{Schrodinger}
\eneq
where $V(r)$ is the QNM potential obtained by Ishibashi and Kodama.  In this paper, we assume the perturbations depend on time as $e^{-i\omega t}$.  Consequently, in order to have damping, the imaginary part of $\omega$ must be negative.  
The Tortoise coordinate $z$ is defined by 
\beeq
dz ={dr \over f(r) }~,
\label{tortoise}
\eneq
where $f(r)$ is related to the spacetime geometry, and is given by 
\beeq
f(r)=1-{2\mu \over r^{D-3}}-\lambda r^2~.
\label{function f}
\eneq
The ADM mass, $M$, of the black hole is related to the parameter $\mu$ by
\beeq
M = {(D-2)A_{D-2} \over 8 \pi G_D}\mu~,
\eneq
where $G_D$ is the Newton gravitational constant in spacetime dimension $D$ and $A_n$ is the area of a unit $n$-sphere,
\beeq
A_n={2\pi^{n+1 \over 2} \over \Gamma\left({n+1 \over 2}\right) }~.
\eneq
The value of the cosmological constant, $\Lambda$, is given by 
\beeq
\Lambda = { (D-1)(D-2)\over 2 }\lambda~.
\eneq

The effective potential $V(r)$, was found explicitly by Ishibashi and Kodama \cite{Ishibashi1,Ishibashi2,Ishibashi3} for scalar (reducing to polar at $D=4$), vector (reducing to axial at $D=4$), and tensor (non-existing at $D=4$) perturbations.  The effective potential for tensor perturbations has been shown \cite{Gibbons, Konoplya} to be equivalent to that of the decay of a test scalar field in a black hole background. The effective potential in AdS space is zero at the event horizon ($z\rightarrow -\infty$) and diverges at spatial infinity ($z\rightarrow \eta$, where $\eta$ is a constant discussed in section 3).  There are many ways of choosing the boundary conditions.  In this paper, we choose our boundary conditions so that the asymptotic behavior of the solutions is
\beeq
\psi(z) \approx \left\{ \begin{array}{ll}
                   e^{-i\omega z}  & \mbox{as $z \rightarrow -\infty$ }~,\\
                   0  & \mbox{as $z\rightarrow \eta$ ~,}
                   \end{array}
           \right.        
\label{asymptotic}
\eneq 
which represents an in-going wave at the event horizon and no waves at infinity. 

Since the tortoise coordinate is multi-valued, it is more convenient to work in the complex $r$-plane.  After rescaling the wavefunction $\psi=\Psi/\sqrt{f}$ we obtain
\beeq
\frac{d^2\Psi}{dr^2}+R(r)\Psi=0 ~,
\label{Schrodinger-r}
\eneq 
where
\beeq
R(r)= {\omega^2\over f^2(r)}-U(r)~,
\label{Rr}
\eneq
with
\beeq
U(r)={V(r)\over f^2}+{1\over 2}\frac{f''}{f}-\frac{1}{4}\left(\frac{f'}{f}\right)^2 ~.
\label{Ur}
\eneq
Here prime denotes differentiation with respect to $r$.

%\sxn{Large Damping Limit and Universality}

In the asymptotic limit, where
$|\omega| \to \infty$,
it has been pointed out in \cite{Ramin1, Daghigh-RN} that when the cosmological constant is zero the potential $U(r)$ is negligible everywhere on the complex plane except when $r\rightarrow 0$.  In this region
\beeq
U(r)\sim \frac{J^2-1}{4r^2}~,
\label{Rr2}
\eneq
where $J$ is given by
\beeq
J= \left\{ \begin{array}{ll}
                   0~~~ \mbox{Tensor Perturbation ($D>4$)}~,\\
                   2(D-2)~~~ \mbox{Vector Perturbation}~,\\
                   0~~~ \mbox{Scalar Perturbation}~.\\
                   \end{array}
           \right.        
\label{J}
\eneq
In the presence of a cosmological constant, the potential $U(r)$ also becomes important relative to the term $\omega^2/f^2$ in the region where $r \rightarrow \infty$.  In this region 
\beeq
U(r)\sim {V(r) \over f^2} \sim \frac{J_{\infty}^2-1}{4r^2}~,
\label{Rr3}
\eneq
where $J_{\infty}$ is given by
\beeq
J_{\infty}= \left\{ \begin{array}{ll}
                   D-1~~~ \mbox{Tensor Perturbation}~,\\
                   
                   D-3~~~ \mbox{Vector Perturbation}~,\\
                 
                   D-5~~~ \mbox{Scalar Perturbation}~.\\
                   \end{array}
           \right.        
\label{Jinfinite}
\eneq

\sxn{The WKB Condition on Asymptotic Quasinormal Modes}
\vskip 0.3cm

For a differential equation of the form (\ref{Schrodinger-r}), the WKB approximation is valid as long as
\beeq
\left |R^{-{3\over 2}}{dR\over dr}\right| \ll 1~.
\eneq
For the function $R(r)$ given in Eq.\ (\ref{Rr}), we find
\beeq
\left |R^{-{3\over 2}}{dR\over dr}\right| \propto {1\over |\omega|} ~, 
\eneq
which means that as long as $|\omega|\rightarrow \infty$ the WKB approximation is valid.
  
In order to determine the asymptotic QNM frequency of Schwarzschild-AdS black holes, we use the two WKB solutions to Eq.\ (\ref{Schrodinger-r}) which are given by
\beeq
\left\{ \begin{array}{ll}
                   f_1^{(t)}(r)=\frac{1}{\sqrt {Q(r)}}e^{+i\int_{t}^rQ(r')dr'}~,\\
                   \\
                   f_2^{(t)}(r)={1 \over \sqrt{Q(r)}}e^{-i\int_{t}^rQ(r')dr'}~,
                   \end{array}
           \right.        
\label{WKB}
\eneq
where
\beeq
Q(r)=\sqrt{R(r)-\frac{1}{4r^2}}~.
\label{Q^2-general}
\eneq
Note that we have shifted the function $R(r)$ by $1/(4r^2)$ in order to guarantee the correct behavior of the WKB solutions at the origin \cite{Andersson}.  In the presence of a cosmological constant, it turns out that the same shift of $1/(4r^2)$ is also necessary to guarantee the correct behavior of the WKB solutions at infinity.  The proof for this shift in the region near $r=\infty$ is almost identical to the proof shown in \cite{Andersson} for the region near $r = 0$ and therefore is not presented in this paper again.  Using Eqs.\ (\ref{Rr}), (\ref{Rr2}), and  (\ref{Rr3}), when $|\omega| \rightarrow \infty$ we can write
\beeq
Q(r)\sim \sqrt{\frac{\omega^2}{f^2}-\frac{J^2}{4r^2}}~,
\label{Q-general-0}
\eneq
if we exclude the region close to $r=\infty$, and
\beeq
Q(r)\sim \sqrt{\frac{\omega^2}{f^2}-\frac{J_{\infty}^2}{4r^2}}~,
\label{Q-general-infty}
\eneq
if we exclude the region close to $r=0$.  In Eq.\ (\ref{WKB}), $t$ is a simple zero of the function $Q^2$.  It is important to point out that the function $Q$ is multi-valued because of the square-root.  To make $Q$ single-valued, we have to introduce branch cuts from the simple zeros of $Q^2$.  We choose the phase of $Q$ such that the in-going wave solution at the event horizon is proportional to $f_2$.

In order to extract the WKB condition on the asymptotic QNM frequencies, first, we need to determine the zeros and poles of the function $Q$ and consequently the behavior of the Stokes and anti-Stokes lines in the complex $r$-plane.  Stokes lines are the lines on which 
\beeq
\mbox{Re}\int_t^r Q(r')dr'=0~,
\label{Stokes}
\eneq   
and anti-Stokes lines are the lines on which 
\beeq
\mbox{Im}\int_t^r Q(r')dr'=0~.
\label{anti-Stokes}
\eneq  
The location of the poles of the function $Q$ can be found by solving $|\lambda|r^{D-1}+r^{D-3}-2\mu=0$.  There is no simple analytic expression for this equation and the roots need to be found numerically.  For even spacetime dimensions, we get an odd number of roots of the form
\beeq
R_n=R_H,~R_1,~ \bar{R}_1,~ \dots,~ R_{D-2 \over 2},~ \bar{R}_{D-2 \over 2},
\eneq
where $R_H$ is the location of the event horizon on the real axis.  Note that the summation of these roots has to be zero since the coefficient of $r^{D-2}$ is zero. For odd spacetime dimensions, we get an even number of roots of the form
\beeq
R_n=R_H,~-R_H,~R_1,~-R_1,~ \dots,~ R_{D-3 \over 2},~-R_{D-3 \over 2},
\eneq
where again the roots add up to zero for the same reason mentioned above.

The location of the zeros of the function $Q$ depends on the phase of the QNM frequency $\omega$.  In the other black hole spacetimes, the asymptotic limit of the QNM frequency was taken to be the highly damped limit where $\omega$ was almost purely imaginary.  It was shown numerically by Cardoso and Lemos \cite{Cardoso-L} and analytically by Cardoso {\it et al} \cite{Cardoso-N-S} and Natario {\it et al} \cite{Natario-S} that in the case of black holes in AdS spacetimes the asymptotic limit of the QNM frequency does not coincide with the highly damped limit.  To explain this, we need to take a closer look at the phase integral $\int_t^r Q(r')dr'$.  Unfortunately, this integral does not have an analytic solution everywhere on the complex $r$-plane, but we can divide the complex plane into three different regions where we can solve this integral analytically.  These regions are:

\noindent Region I: $r \le \tilde{r}$ where $\tilde{r} << R_n$:
\begin{eqnarray}
\int_{t}^r Qdr &=& \int_{t}^r {\omega \over f} \sqrt{1-{J^2f^2\over 4\omega^2 r^2}}dr \approx \int_{t}^r {-\omega r^{D-3} \over 2\mu }\sqrt{1-{J^2\mu^2\over \omega^2r^{2(D-2)}}}dr \nonumber \\
&=& {J \over 2(D-2)}\left[-y\left(\sqrt{1-{1\over y^2}}-{1\over 2 i \sqrt{y^2}}\ln{\sqrt{y^2-1}-i\over \sqrt{y^2-1}+i} \right) \right]_{y_t=\pm1}^{y_r={\omega r^{D-2} \over J \mu}}\nonumber \\
&=&{J \over 2(D-2)}\left[-y_r\left(\sqrt{1-{1\over y_r^2}}-{1\over 2 i \sqrt{y_r^2}}\ln{\sqrt{y_r^2-1}-i\over \sqrt{y_r^2-1}+i} \right) - \left(\pm {\pi \over 2}\right)\right]
~,
\label{region-I}
\end{eqnarray}
where we have used the change of variable $y={\omega r^{D-2}\over J \mu}$ to solve the integral above.

\noindent Region II: $\tilde{r}< r \le \bar{r}$ where $\bar{r} \gg R_n$:
\begin{eqnarray}
\int_{\tilde{r}}^r Qdr \approx \int_{\tilde{r}}^r {\omega \over f}dr &=& \omega \sum_{n=0}^{D-2} {1\over 2k_n} \left[ \ln\left(1-{r\over R_n}\right) -\ln\left(1- {\tilde{r}\over R_n}\right)\right]
\nonumber\\
&\approx&  \omega \left[\sum_{n=0}^{D-2} {1\over 2k_n} \ln\left(1- {r \over R_n}\right)\right] +  {\omega\tilde{r}^{D-2} \over 2(D-1)\mu}~,
\label{region-II}
\end{eqnarray}
where $R_0=R_H$ and $k_n=k(R_n)={f'(R_n) \over 2}~$.

\noindent
Region III: $r > \bar{r}$:
\begin{eqnarray}
\int_{t_\infty}^r Qdr &=& \int_{t_\infty}^r {\omega \over f} \sqrt{1-{J_\infty^2f^2\over 4\omega^2 r^2}}dr \approx \int_{t_\infty}^r {\omega  \over |\lambda|r^2}\sqrt{1-{J_\infty^2|\lambda|^2 r^2 \over 4\omega^2}}dr \nonumber \\
&=& {J_\infty \over 2}\left[-y\left(\sqrt{1-{1\over y^2}}-{1\over 2 i \sqrt{y^2}}\ln{\sqrt{y^2-1}-i\over \sqrt{y^2-1}+i} \right) \right]_{y_{t_\infty}=\pm1}^{y_r={2\omega \over J_\infty |\lambda| r}}\nonumber \\
&=&{J_\infty \over 2}\left[-y_r\left(\sqrt{1-{1\over y_r^2}}-{1\over 2 i \sqrt{y_r^2}}\ln{\sqrt{y_r^2-1}-i\over \sqrt{y_r^2-1}+i} \right) - \left(\pm {\pi \over 2}\right)\right]
~,
\label{region-III}
\end{eqnarray}
where we have used the change of variable $y={2\omega \over J_{\infty} |\lambda|r}$ to solve the integral above.
In region II, we notice that when $r \gg R_n$ the phase integral can be approximated to be:
\beeq
\int_{\tilde{r}}^r Qdr  \approx  \omega \eta~,
\label{large-r}
\eneq
where
\beeq
\eta = \sum_{n=0}^{D-2} {1\over 2k_n} \ln\left(-{1 \over R_n}\right)~
\label{eta}
\eneq
is a constant which can be determined numerically.  

Note that in approximation (\ref{large-r}) the last term on the right hand side of Eq.\ (\ref{region-II}) is ignored.  This is possible due to the fact that when $|\omega| \rightarrow \infty$ the approximation in region II is valid as long as $\tilde{r}$ and $\bar{r}$ are finite but they can be taken to be extremely small.  Equation (\ref{large-r}) tells us that in order to have an anti-Stokes line extending to large values of $r$ in the complex plane we need to have $\mbox{Im}~(\omega \eta)=0$ according to (\ref{anti-Stokes}).  This means that $\arg(\omega)+\arg(\eta)=n\pi$, where $n=0,1,2,\dots$.  With no anti-Stokes lines going to infinity, we cannot solve this problem because we cannot impose the boundary condition at infinity on the QNM frequency.  Therefore, the constant $\eta$ puts a restriction on the values that $\omega$ can take.  For example, consider the case of Schwarzschild-AdS black holes with a horizon radius much larger than the AdS radius $R_{AdS}=1/\sqrt{|\lambda|}$. For these large black holes the poles can be found analytically using $|\lambda|r^{D-1}-2\mu=0$, with the roots at
\beeq
R_n=\left|\left({2 \mu \over |\lambda|}\right)^{1\over D-1}\right|\exp\left({2\pi i \over D-1}n\right)~~~~~n=0,1,\dots,D-2~.
\eneq
Using these roots, we can determine the constant $\eta$.  It is clear from Eq.\ (\ref{eta}) that $\eta$ is multivalued.  Therefore, in order to determine $\eta$ we need to introduce branch cuts in the complex $r$-plane.  Depending on how we choose these branch cuts, we get different results for $\eta$ which are
\beeq
\eta = |\eta| e^{i\pi \over D-1},~|\eta| e^{3i\pi \over D-1},~|\eta| e^{5i\pi \over D-1}, ~\dots, ~|\eta| e^{(2D-3)i\pi \over D-1} ~,
\label{complex-eta1}
\eneq
where
\beeq
|\eta|={1\over 4 T_H \sin\left({\pi \over D-1}\right)}~.
\eneq
Here $T_H=k_H/2\pi=f'(R_H)/4\pi$ is the Hawking temperature for large Schwarzschild-AdS black holes.  Therefore, in the case of large black holes the possible QNM frequencies that allow the presence of an anti-Stokes line stretching to infinity can take the form
\beeq
\omega = |\omega| e^{-{n i\pi \over D-1}}~,
\label{omega}
\eneq
where $n=1, 3, 5, \dots$ and in order to exclude the unphysical cases where $\omega$ has a negative real part or a positive imaginary part we need to impose the condition that $n \le {D-1 \over 2}$.  In addition to (\ref{omega}), in even spacetime dimensions we also have the option of $\omega=|\omega|$ which results in the highly real QNMs explained in the next section.  The authors of \cite{Natario-S} have only investigated the case where $\omega = |\omega| e^{-{i\pi \over D-1}}$.  In this section we will explore this case using the analytic technique developed in \cite{Andersson}.  In the following sections we will explore all the other possible cases.

\begin{figure}[tb]
\begin{center}
\includegraphics[height=10cm]{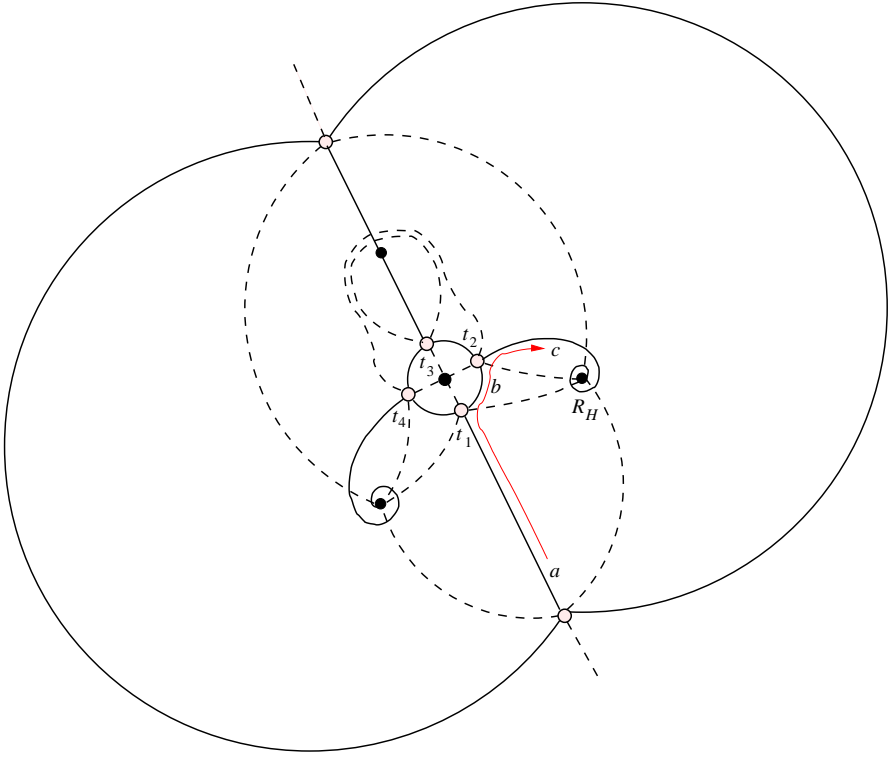}
\end{center}
\caption{Schematic presentation of the Stokes (dashed) and anti-Stokes (solid) lines in the complex $r$-plane for large Schwarzschild-AdS black holes in four spacetime dimensions.  The open circles are the zeros and the filled circles are the poles of the function $Q$.  $R_H$ is the event horizon.  The thin arrow which starts on the anti-Stokes line that stretches to infinity at point $a$ and ends on the anti-Stokes line that connects to the event horizon at point $c$ is the path we take to determine the WKB condition on QNM frequencies.}
\label{schem-d4-AdS}
\end{figure}

\begin{figure}[tb]
\begin{center}
\includegraphics[height=3.5cm]{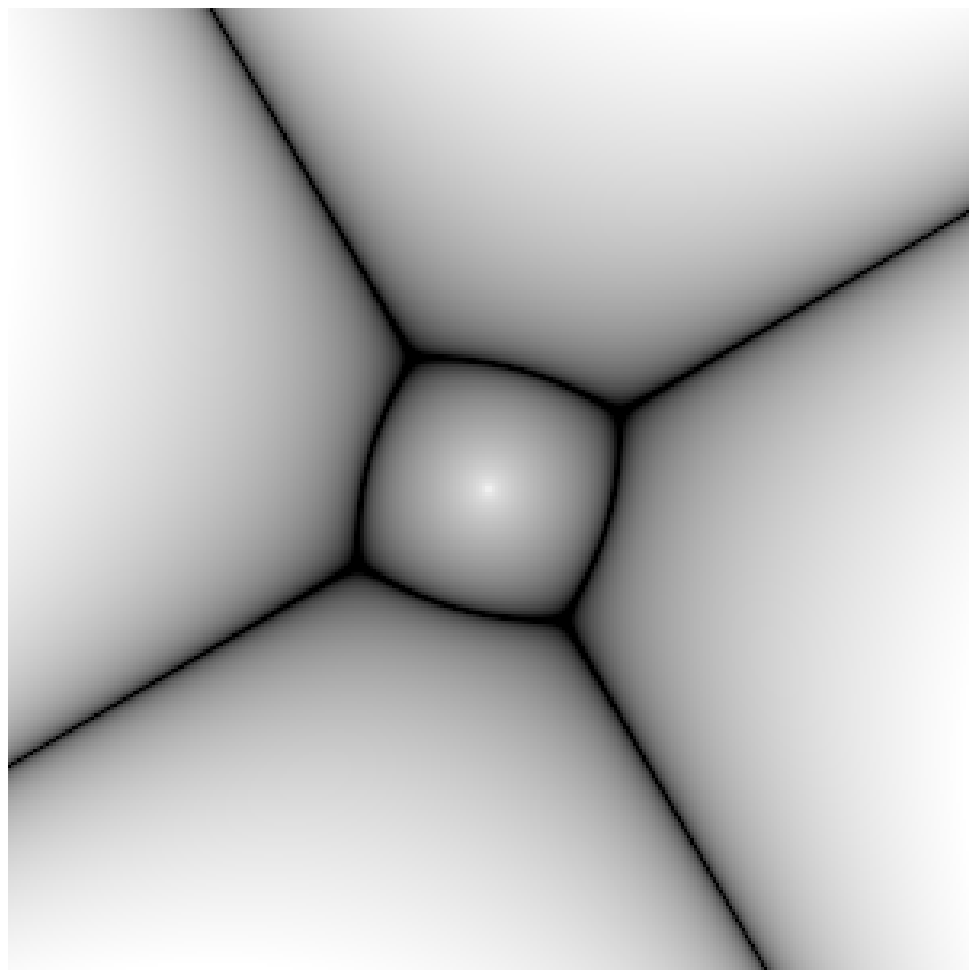}
\includegraphics[height=3.5cm]{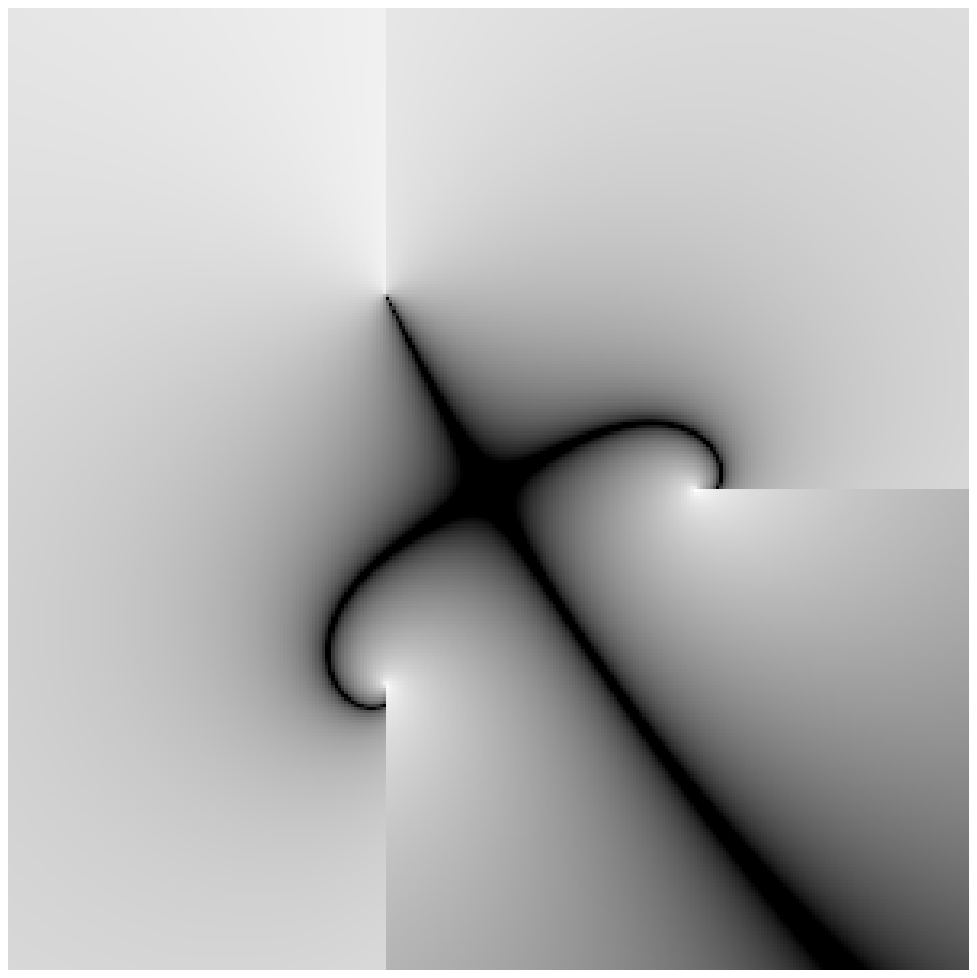}
\includegraphics[height=3.5cm]{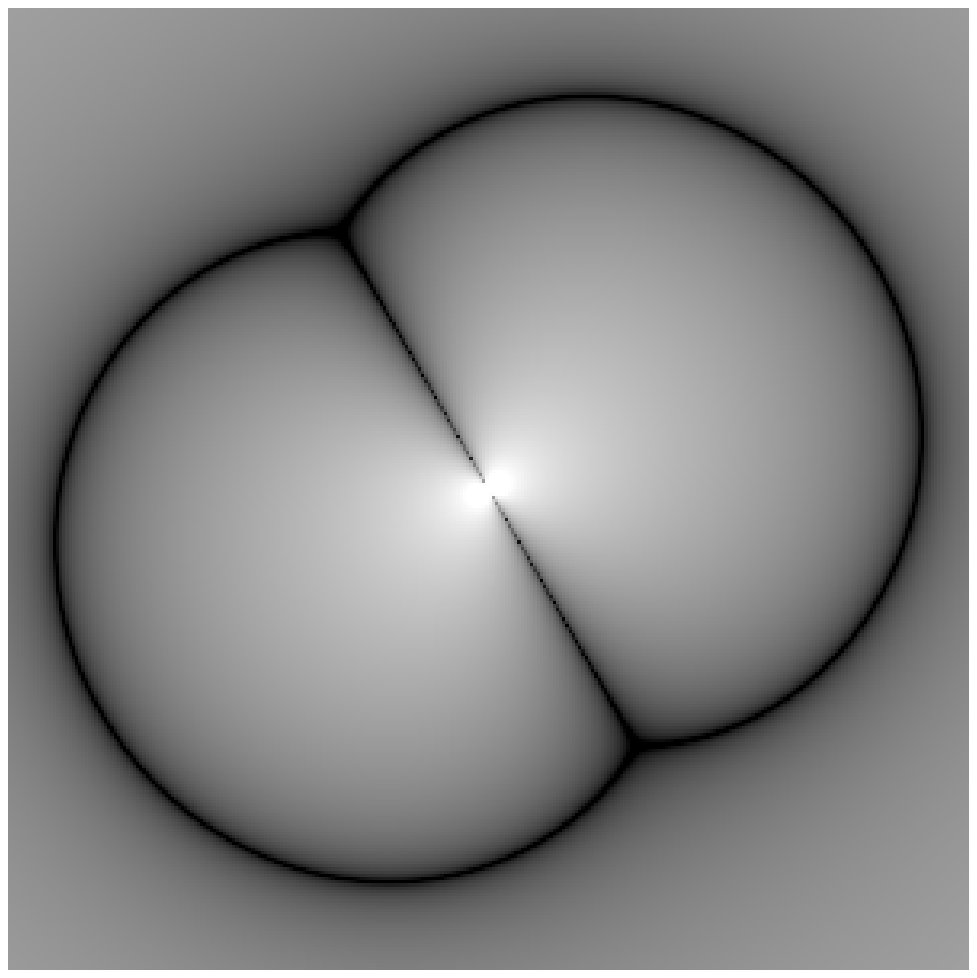}\\
\includegraphics[height=3.5cm]{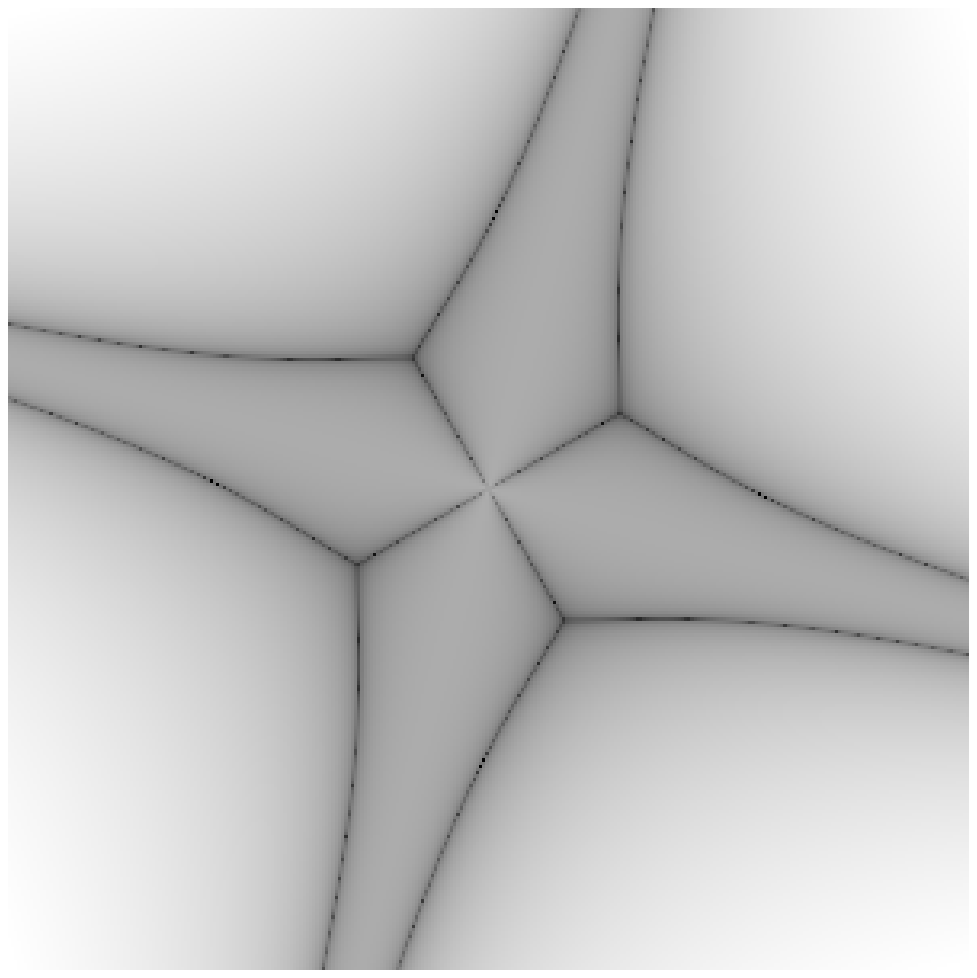}
\includegraphics[height=3.5cm]{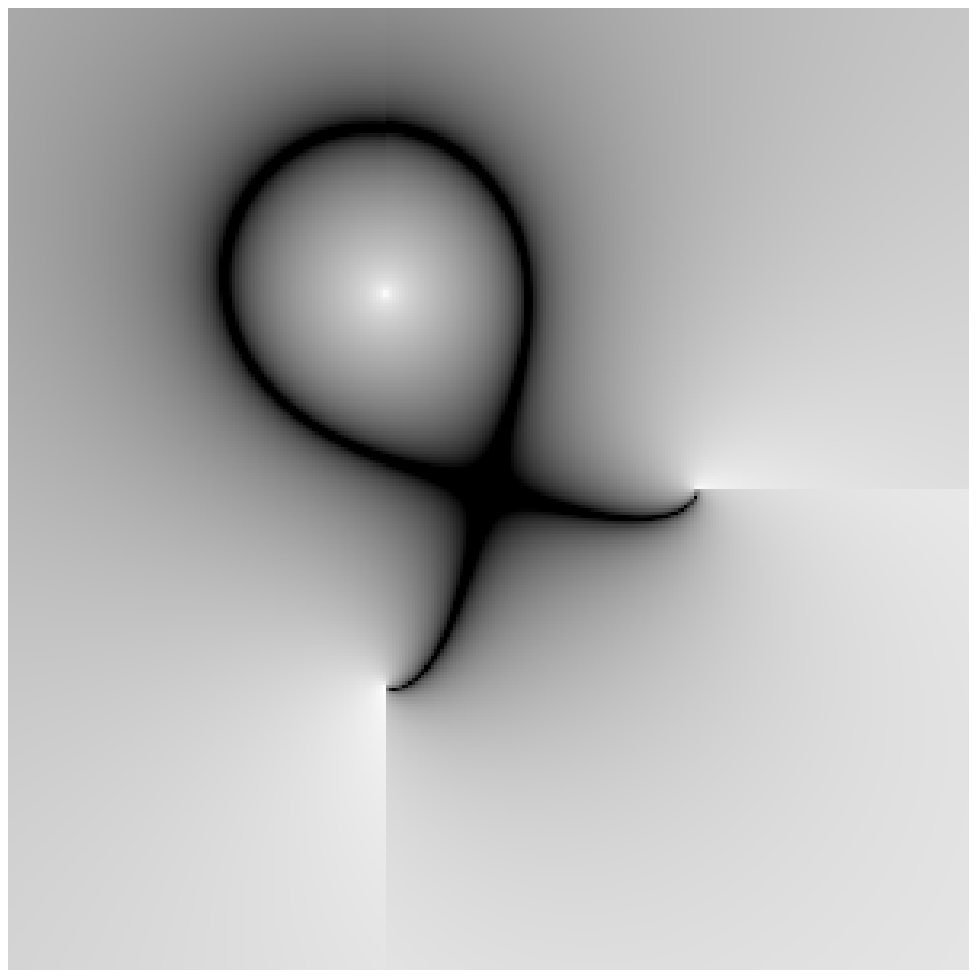}
\includegraphics[height=3.5cm]{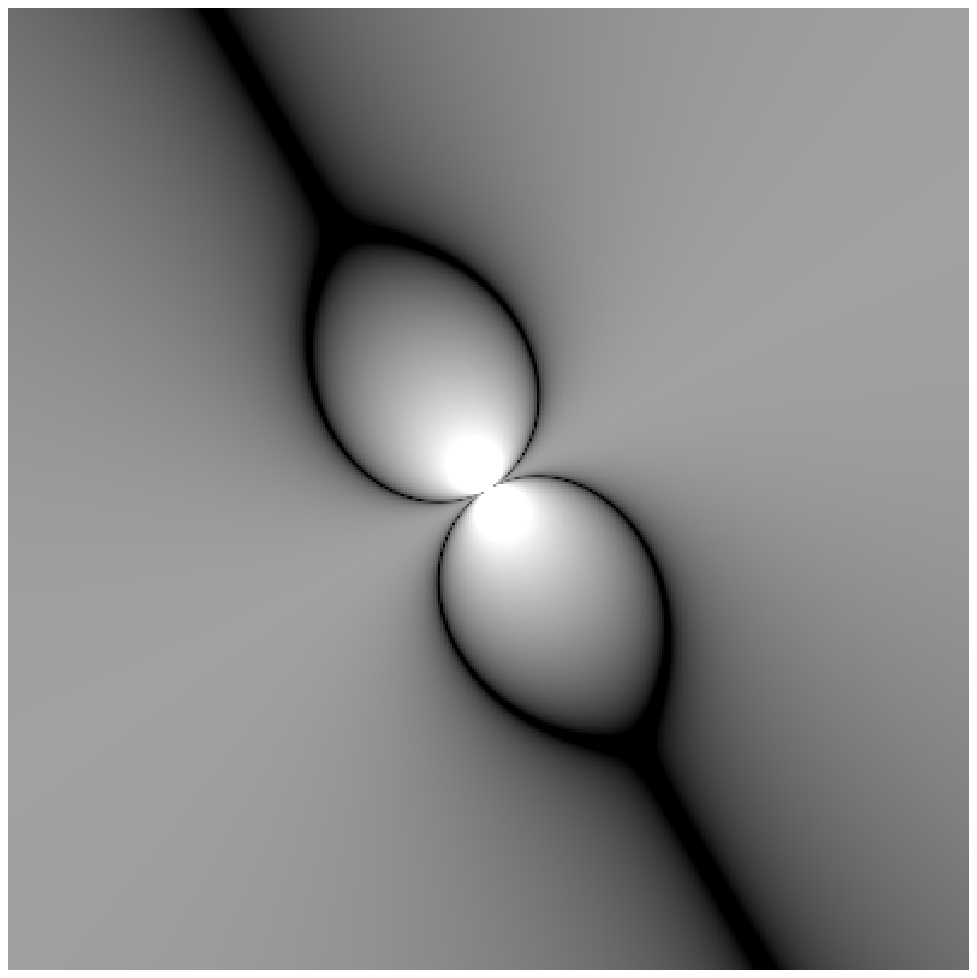}
\end{center}
\caption{Numerically generated pictures of the Stokes and anti-Stokes lines for large Schwarzschild-AdS black holes in $D=4$ when $\arg(\omega)=-\pi/3$.  The top three pictures from left to right show the anti-Stokes lines in the complex $r$-plane close to the origin, away from the origin, and close to infinity respectively.  The bottom three pictures show the corresponding Stokes lines.  For the complete structure of Stokes/anti-Stokes lines, see Fig. \ref{schem-d4-AdS}.}
\label{numerical1}
\end{figure}

\begin{figure}[tb]
\begin{center}
\includegraphics[height=3.5cm]{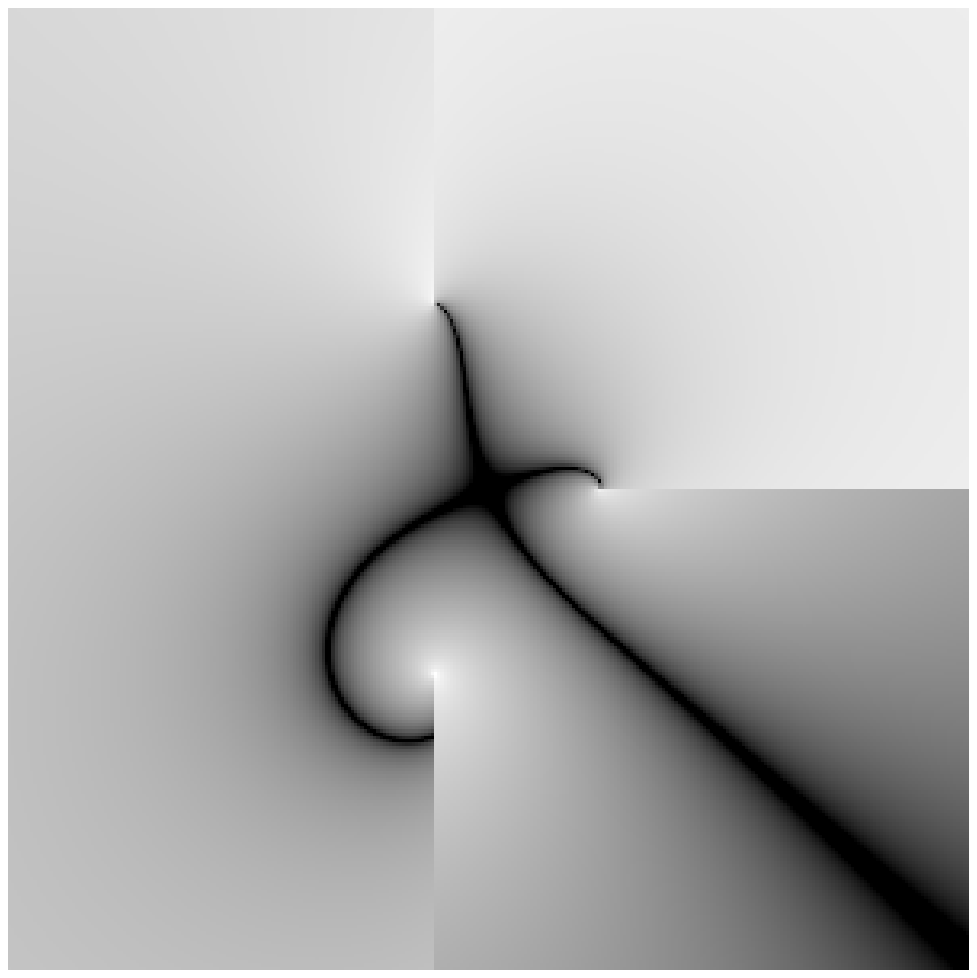}
\includegraphics[height=3.5cm]{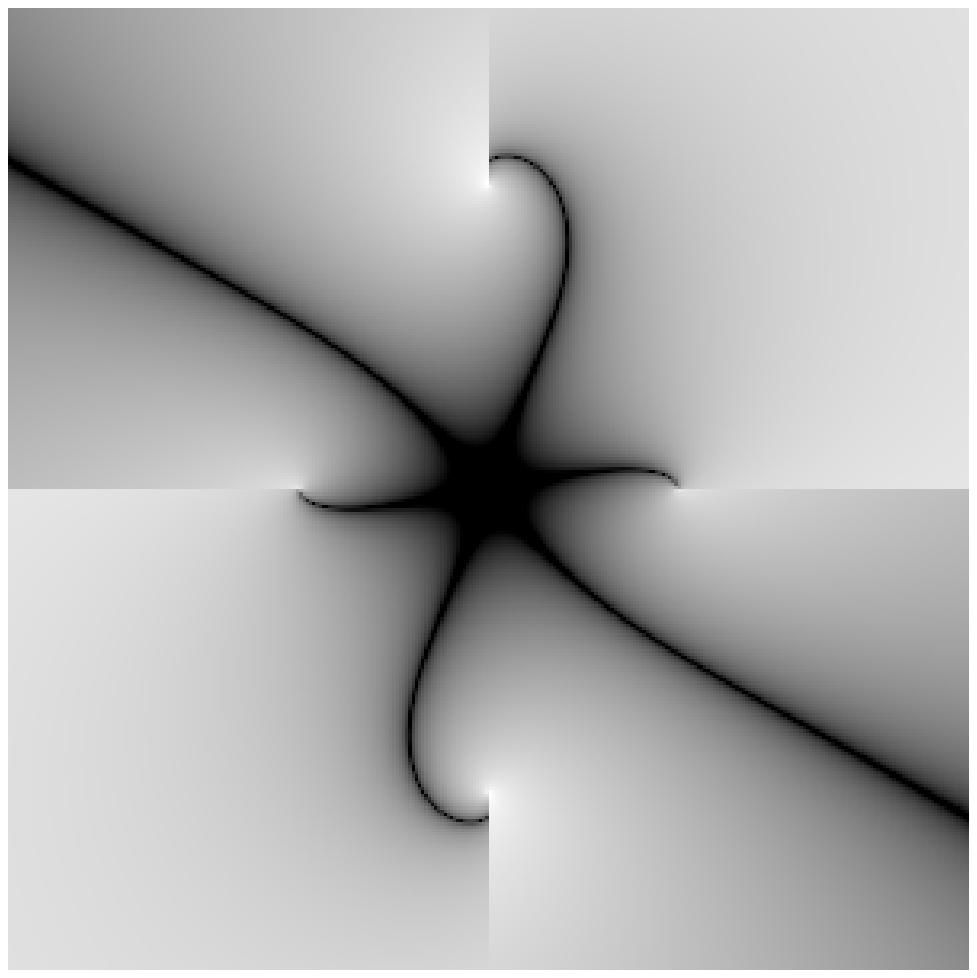}
\includegraphics[height=3.5cm]{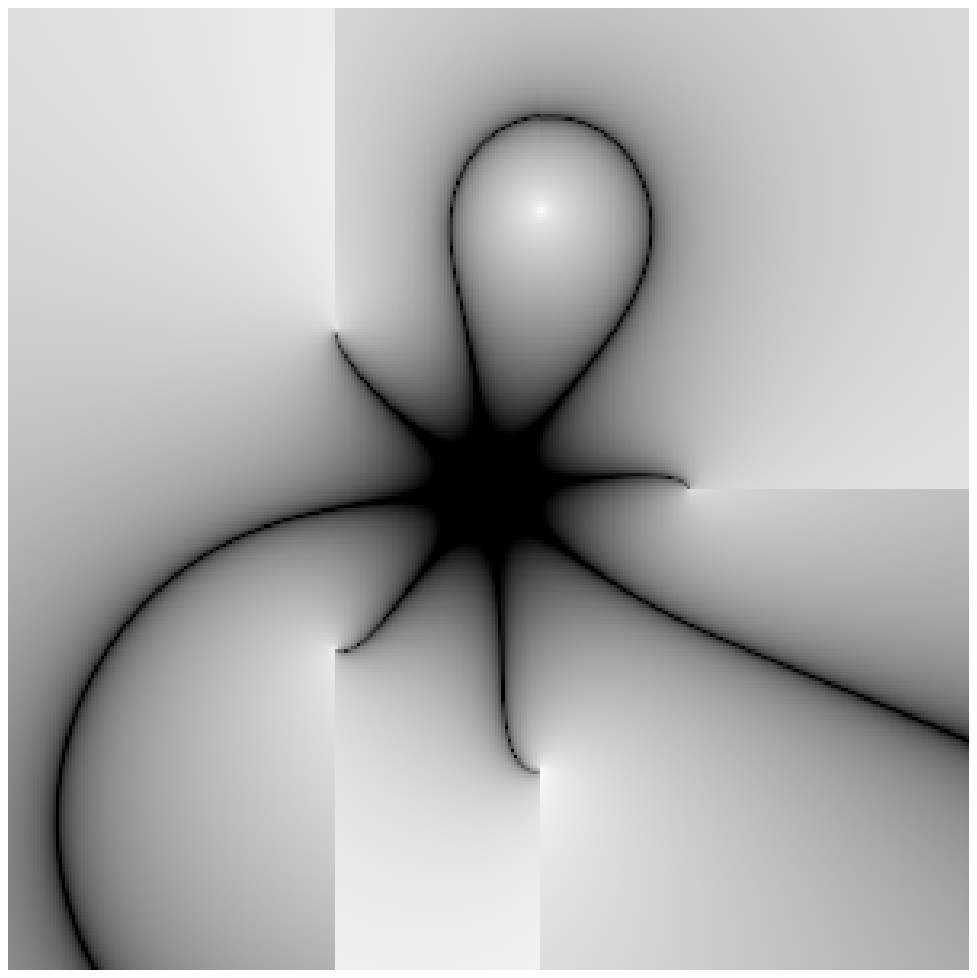}
\includegraphics[height=3.5cm]{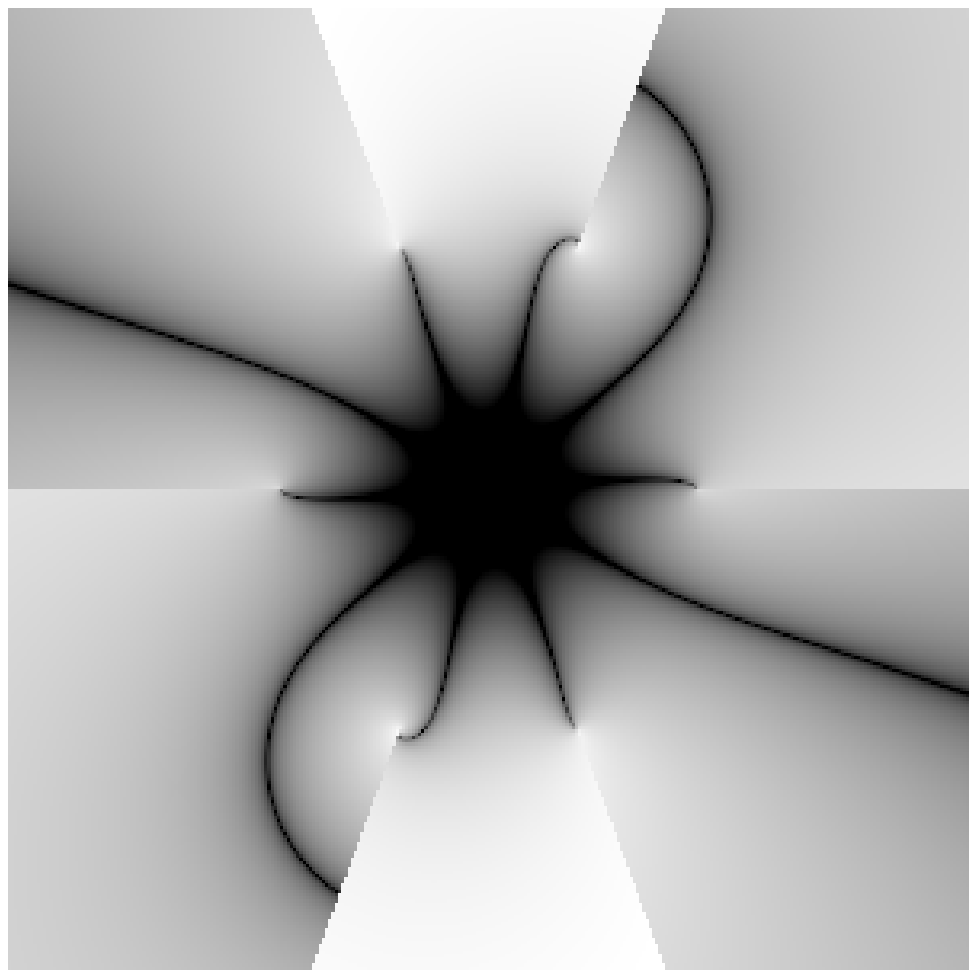}
\\
\includegraphics[height=3.5cm]{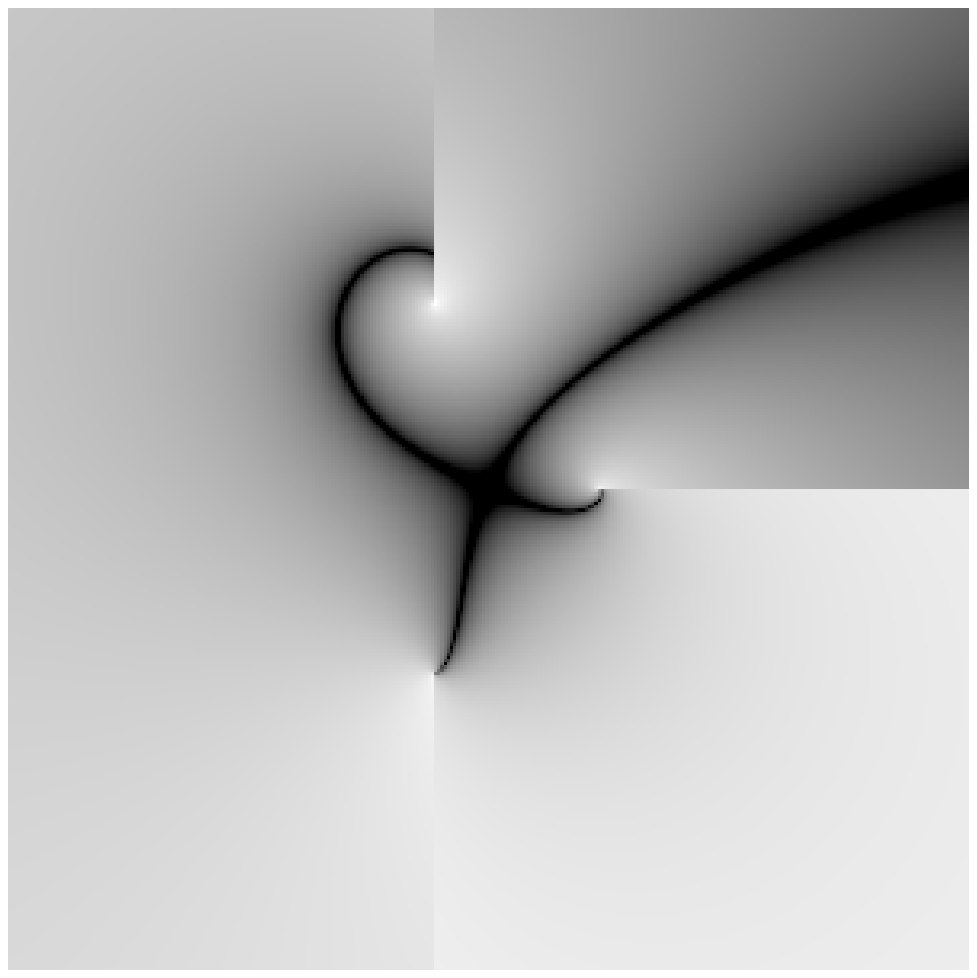}
\includegraphics[height=3.5cm]{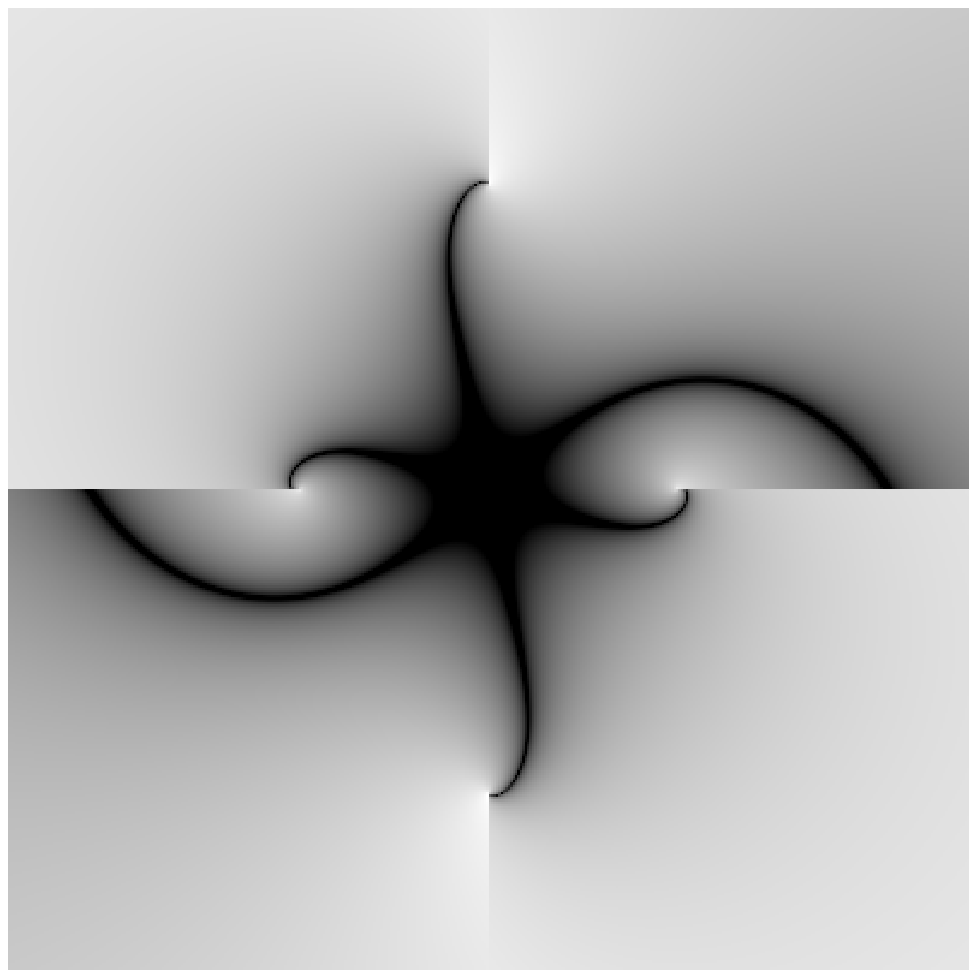}
\includegraphics[height=3.5cm]{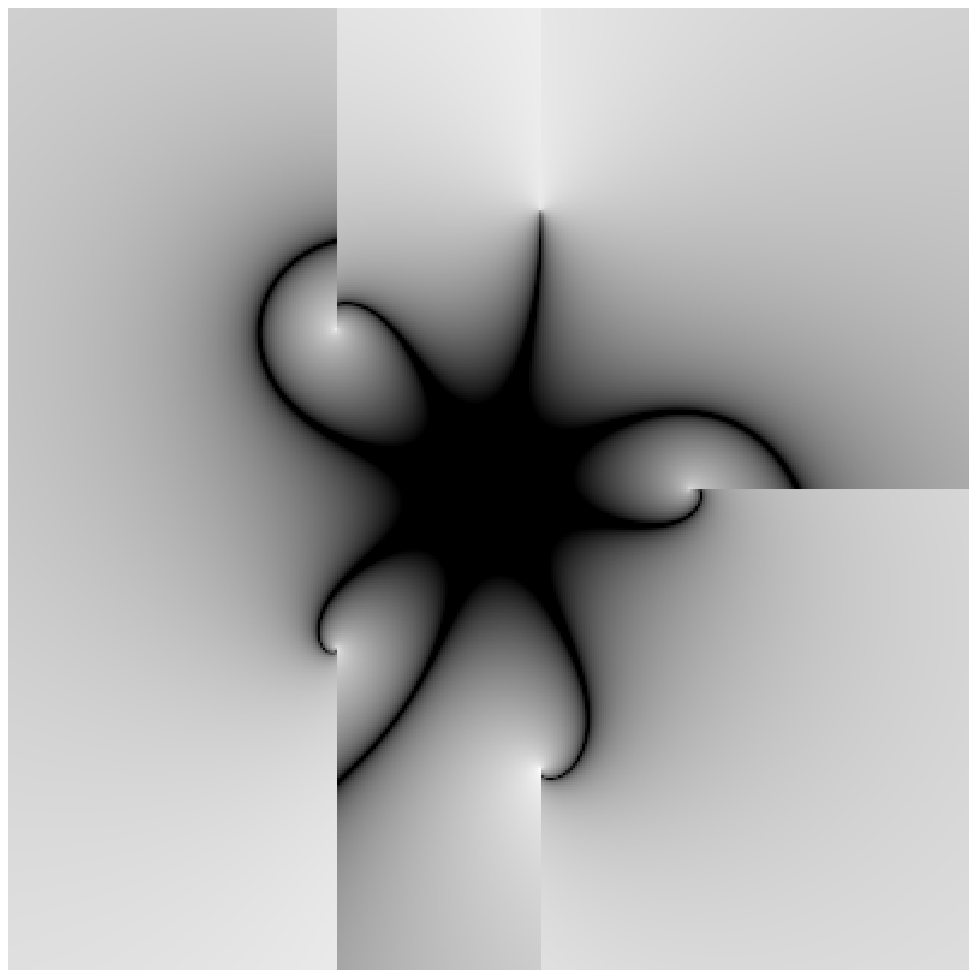}
\includegraphics[height=3.5cm]{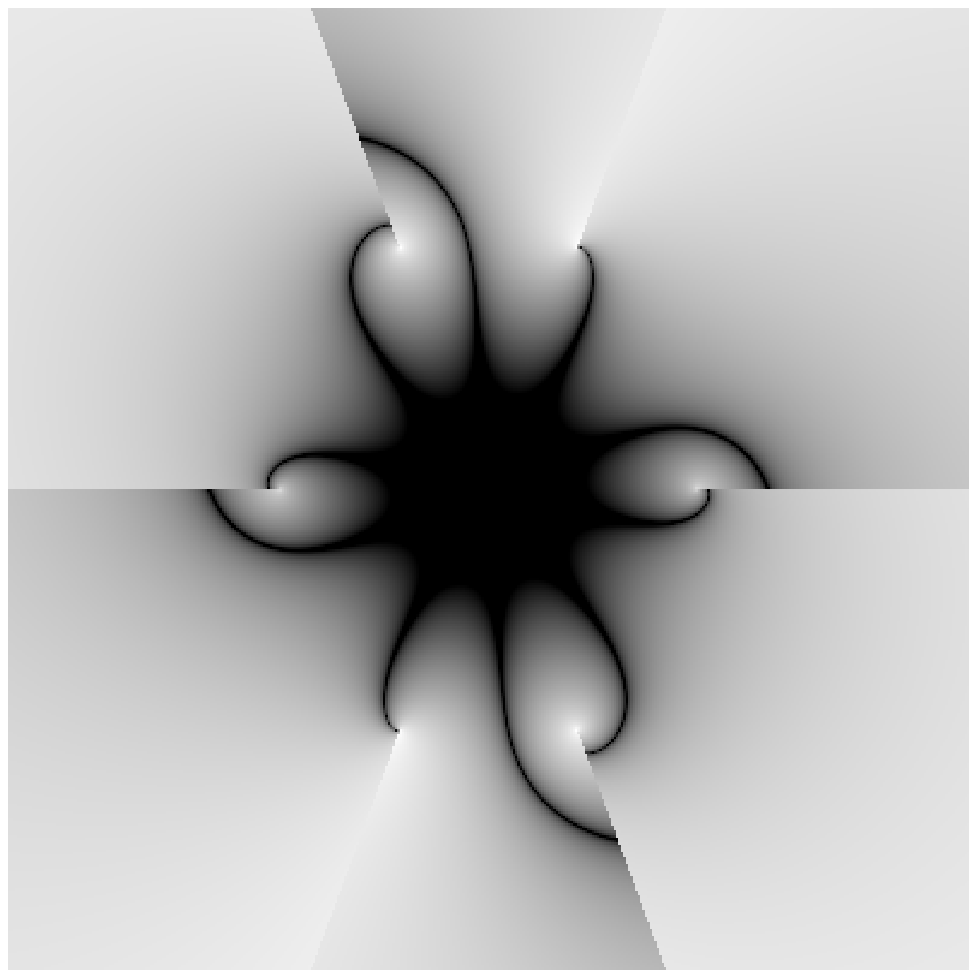}
\end{center}
\caption{Numerically generated pictures of the Stokes and anti-Stokes lines for intermediate Schwarzschild-AdS black holes with $-{\pi \over D-1}< \arg(\omega) <0$.  The top three pictures from left to right show the anti-Stokes lines in the complex $r$-plane in $D=4$, $D=5$, $D=6$, and $D=7$ respectively.  The bottom three pictures show the corresponding Stokes lines.}
\label{numerical2}
\end{figure}

In the general case, we can determine the argument of $\eta$, and consequently the argument of $\omega$, numerically.  Using this information, we can determine the location of the zeros of the function $Q$ using Eq.\ (\ref{Q-general-0}):
\beeq
t=\left({J\mu \over |\omega|}\right)^{1\over D-2} \exp\left({-i\arg(\omega)\over D-2}+{in\pi \over D-2}\right)~,~~\mbox{where $n=0,\dots, 2D-5$}~.
\eneq
We can also determine the location of the zeros of the function $Q$ close to infinity using Eq.\ (\ref{Q-general-infty}):
\beeq
t_\infty={2|\omega| \over J_\infty|\lambda|} \exp\left[i\arg(\omega)+i n\pi \right]~,~\mbox{where $n=0,1$}~.
\eneq
From each of these simple zeros of the function $Q^2$ emanates three Stokes and three anti-Stokes lines in a way that the angle between a Stokes line and an anti-Stokes line is $60$ degrees.  The orientation of these lines can be determined by studying the WKB phase $\int Qdr$ in the vicinity of the zeros of $Q$.  After determining the orientation of the lines emanating from the zeros, it is easy to guess the structure of the Stokes/anti-Stokes lines in the whole complex $r$-plane.  
We have plotted the rough schematic behavior of these lines in Fig. \ref{schem-d4-AdS} for large Schwarzschild-AdS black holes in four spacetime dimensions in the case where $\arg(\omega)=-\pi/3$.  This was done with the aid of Fig. \ref{numerical1}, where these lines are plotted numerically for large Schwarzschild-AdS black holes in spacetime dimension $D=4$ in regions I, II, and III using Eqs.\ (\ref{region-I}), (\ref{region-II}), and (\ref{region-III}) respectively.  In Fig. \ref{numerical2}, for intermediate Schwarzschild-AdS black holes, the behavior of Stokes/anti-Stokes lines are plotted numerically in region II using Eq.\ (\ref{region-II}) for spacetime dimensions $D=4$, $D=5$, $D=6$, and $D=7$.

To determine the WKB condition on QNMs, we will partly follow the method developed by Natario and Schiappa in \cite{Natario-S}.  Natario and Schiappa did their calculations using the tortoise coordinate.  In this paper, we will do all the calculations in the $r$-coordinate.   In the large $r$ limit, Eq.\ (\ref{Schrodinger-r}) can be approximated as 
\beeq
\frac{d^2\Psi}{dr^2}+\left({\omega^2 \over |\lambda|r^4}-{J_\infty^2-1 \over 4r^2}\right)\Psi=0 ~.
\label{Schrodinger-large-r}
\eneq 
This differential equation can be solved exactly and the solution is 
\beeq
\Psi(r)=A_+ \sqrt{2\pi r}J_{J_\infty \over 2}\left({\omega\over |\lambda|r}\right)+ A_- \sqrt{2\pi r}J_{-{J_\infty \over 2}}\left({\omega\over |\lambda|r}\right)~,
\label{Psi-large-r}
\eneq 
where $J_j$ represent Bessel functions of the first kind and $A_\pm$ are complex constants.
The boundary condition at infinity tells us that the wave function $\Psi$ should die off as $r\rightarrow \infty$.  This requires $A_-=0$.  Using the approximation
\beeq
J_j(x) \sim \sqrt{2\over \pi x}\cos(x-{j\pi \over 2}-{\pi \over 4})~~\mbox{when}~x\gg 1~,
\label{approx-bessel}
\eneq 
we can rewrite Eq.\ (\ref{Psi-large-r}) as
\beeq
\Psi(r)\sim \sqrt{|\lambda|\over \omega }r A_+\left(e^{-i\beta} e^{i{\omega\over |\lambda|r}}+  e^{i\beta} e^{-i{\omega\over |\lambda|r}}\right) ~,
\label{Psi-large-r1}
\eneq 
where 
\beeq
\beta = {\pi \over 4}(1+J_{\infty}) ~.
\label{beta}
\eneq 
For the rest of the calculation we will use the method explained by Andersson and Howls in \cite{Andersson}.  In order to apply this method, we need to write the wave function (\ref{Psi-large-r1}) in terms of the WKB solutions (\ref{WKB}).  To do that we need to evaluate $\int_t^r Q(r')dr'$ with $r \gg R_n$.  This integral can be solved using the results in Eqs.\ (\ref{region-I}) and (\ref{region-II}):
\beeq
\int_{t_1}^r Q(r')dr'= \int_{t_1}^{\tilde{r}} Q(r')dr'+ \int_{\tilde{r}}^r Q(r')dr'\sim {J\over2(D-2)}{\pi\over2}+ \omega \left(\eta-{1\over |\lambda|r}\right)~,
\label{phase-t1-r}
\eneq 
where $t_1$ is the zero indicated in Fig. \ref{schem-d4-AdS}.  We can now evaluate the WKB solutions to be
\beeq
\left\{ \begin{array}{ll}
                   f_1^{(t_1)}(r)\sim \frac{r}{\sqrt {\omega/|\lambda|}}e^{+i\left[\omega \left(\eta-{1\over |\lambda|r}\right)+{J\over2(D-2)}{\pi\over2}\right]}~,\\
                   \\
                   f_2^{(t_1)}(r)  \sim \frac{r}{\sqrt {\omega/|\lambda|}}e^{-i\left[\omega \left(\eta-{1\over |\lambda|r}\right)+{J\over2(D-2)}{\pi\over2}\right]}~,
                   \end{array}
           \right.        
\label{WKB-large-r}
\eneq
and rewrite Eq.\ (\ref{Psi-large-r1}) in terms of these solutions where
\beeq
\Psi(r)\sim  A_+ e^{i\beta} e^{-i\omega\eta} e^{-i{J\over2(D-2)}{\pi\over2}} f_1^{(t_1)}+  A_+e^{-i\beta} e^{i\omega\eta} e^{i{J\over2(D-2)}{\pi\over2}} f_2^{(t_1)}~.
\label{Psi-a}
\eneq 
The topology shown in Fig. \ref{schem-d4-AdS} is generic in the sense that in all spacetime dimensions there is always an unbounded anti-Stokes line which extends to infinity next to an anti-Stokes line that ends at the event horizon.  This generic behavior can be observed in Fig. \ref{numerical2} where we show the topology of Stokes/anti-Stokes lines in higher spacetime dimensions.  To derive the WKB condition, we follow the path shown in Fig. \ref{schem-d4-AdS} starting on the unbounded anti-Stokes line at point $a$ and ending on an anti-Stokes line which connects to the event horizon at point $c$.  Note that, since the topology shown in Fig. \ref{schem-d4-AdS} is generic, the following results hold for arbitrary spacetime dimensions greater than three.  The solution at point $a$ is given by Eq.\ (\ref{Psi-a}), which can be written as 
\beeq
\Psi_{a}=c_1 f_1^{(t_1)}+c_2 f_2^{(t_1)}~,
\label{Psi-a-short}
\eneq 
where 
\beeq
c_1 = A_+ e^{i\beta} e^{-i\omega\eta} e^{-i{J\over2(D-2)}{\pi\over2}}~,
\label{c1}
\eneq
and 
\beeq
c_2 = A_+e^{-i\beta} e^{i\omega\eta} e^{i{J\over2(D-2)}{\pi\over2}}~.
\label{c2}
\eneq 
The rules on how to move along anti-Stokes lines are explained in detail by Andersson and Howls \cite{Andersson}. In this method, the WKB solutions do not change character along anti-Stokes lines except when we extend the solutions to a neighboring anti-Stokes line in the vicinity of a zero of the function $Q$ where we have to cross a Stokes line .  Therefore, we can extend the solution at $a$ to the vicinity of $t_1$ with no change.  To extend the solution to point $b$ on a neighboring anti-Stokes line we must cross a Stokes line on which we choose the ingoing solution $f_2$ to be dominant since this Stokes line ends at the event horizon.  After accounting for the Stokes phenomenon we find:
\beeq
\Psi_b = (c_1+ic_2) f_1^{(t_1)} +c_2 f_2^{(t_1)}~.
\eneq
Changing the lower limit of the phase integral from $t_1$ to $t_2$ gives
\beeq
\Psi_b (c_1+ic_2)e^{i\gamma_{12}} f_1^{(t_2)} + c_2 e^{-i\gamma_{12}} f_2^{(t_2)}~,
\eneq
where
\beeq
\gamma_{12} =\int_{t_1}^{t_2} Q dr \approx {J\over 2(D-2)} \pi   ~
\label{gamma1}
\eneq
via Eq.\ (\ref{region-I}).  To extend the solution to point $c$ we need to cross another Stokes line which ends at the event horizon and consequently the solution $f_2$ is dominant on this line again.  After crossing this Stokes line, we get
\beeq
\Psi_c = \left[c_1e^{i\gamma_{12}}+ ic_2(e^{i\gamma_{12}}+e^{-i\gamma_{12}})\right] f_1^{(t_2)} + c_2 e^{-i\gamma_{12}} f_2^{(t_2)}~.
\eneq
Since the line labeled $c$ is connected to the event horizon, on this line we need to impose the boundary condition where we have purely in-going waves.  This means that the coefficient of $f_1$, which represents an out-going wave in our choice of phase for $Q$, must be zero.  By setting the coefficient of $f_1$ to zero, we arrive at the WKB condition on highly damped QNM frequencies
\beeq
e^{2i\omega \eta} = i e^{2i\beta} {{e^{i\gamma_{12}}} \over 1+e^{2i\gamma_{12}}}~.
\eneq
This result translates to
\beeq
\omega \eta = \beta + n\pi + {\pi\over 4}+{i\over2}\ln\left[2\cos\left({J\over 2(D-2)}\pi\right)\right]~~~~\mbox{as $n\rightarrow \infty$}~.
\eneq
After entering the values for $J$ and $\beta$ for tensor, vector, and scalar perturbations using Eqs.\ (\ref{J}), (\ref{Jinfinite}), and (\ref{beta}), it is easy to show that the asymptotic QNM frequency of all types of perturbations follow the condition
\beeq
\omega \eta = n\pi + {\pi\over 4}(D+1) +{i\over2}\ln2~.
\label{wkb-complex}
\eneq
This result matches exactly with the result obtained by Natario and Schiappa in section $3.3.1$ of \cite{Natario-S} (there is a sign difference because our perturbations depend on time as $e^{-i\omega t}$ while Natario and Schiappa use $e^{i\omega t}$).  For comparison of the analytical result (\ref{wkb-complex}) with the numerical calculations of the asymptotic QNMs of Schwarzschild-AdS black holes see \cite{Natario-S}.  Equation (\ref{wkb-complex}) is valid for all types of perturbations in all spacetime dimensions greater than three, except for scalar perturbations in spacetime dimensions four and five.  As explained by Natario and Schiappa \cite{Natario-S}, in the case of four spacetime dimensions $J_\infty=-1$, but we can set $J_\infty=1$ since all we need in the differential Eq.\ (\ref{Schrodinger-large-r}) is to have $J_\infty^2=1$.  Following the same calculations with $J_\infty=1$, we get
\beeq
\omega \eta = n\pi + {3\pi\over 4} +{i\over2}\ln2~,
\label{wkb-complex-d=4}
\eneq
for scalar perturbations in $D=4$.  In five spacetime dimensions, the situation is completely different.  The QNM potential 
$V(r) \approx -{|\lambda|r^2/4}$
diverges to $-\infty$ instead of $+\infty$ as $r\rightarrow \infty$.  In such a case the boundary condition at $r\sim \infty$ needs to be modified since clearly the wavefunction cannot go to zero in this region.  Further investigation is necessary on this case.

\begin{figure}[tb]
\begin{center}
\includegraphics[height=3.5cm]{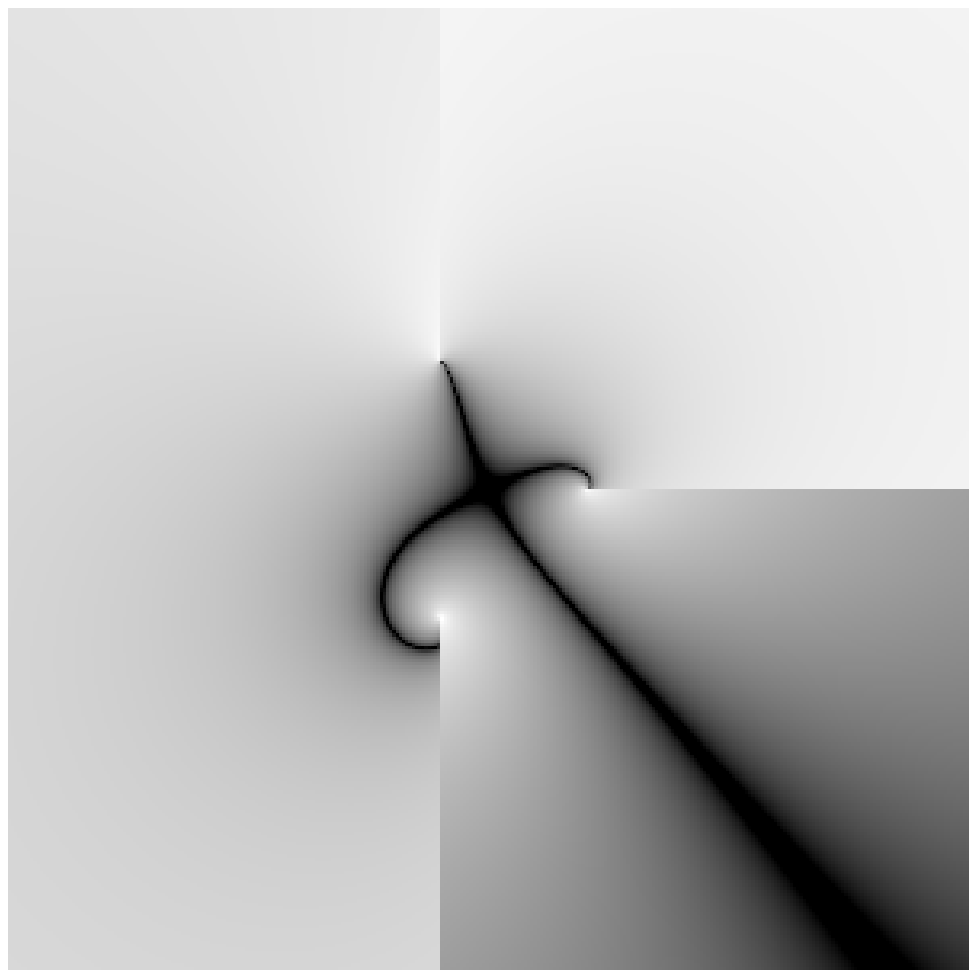}
\includegraphics[height=3.5cm]{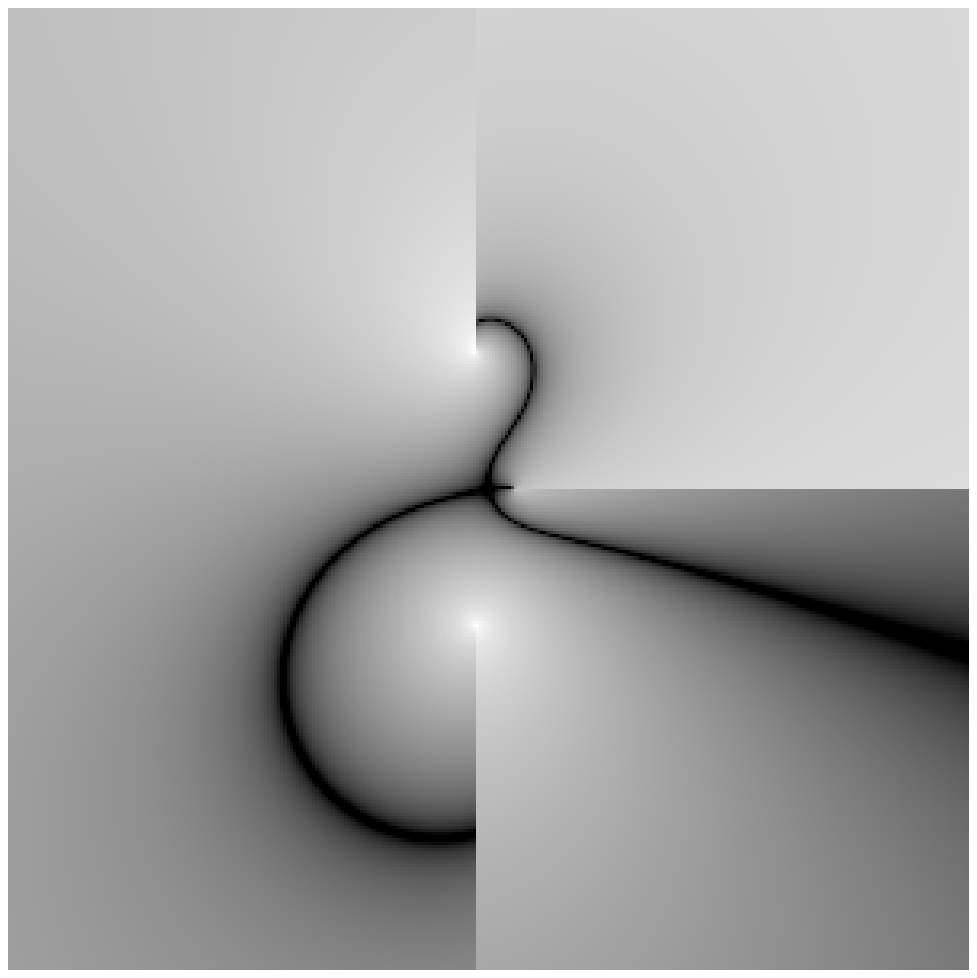}
\includegraphics[height=3.5cm]{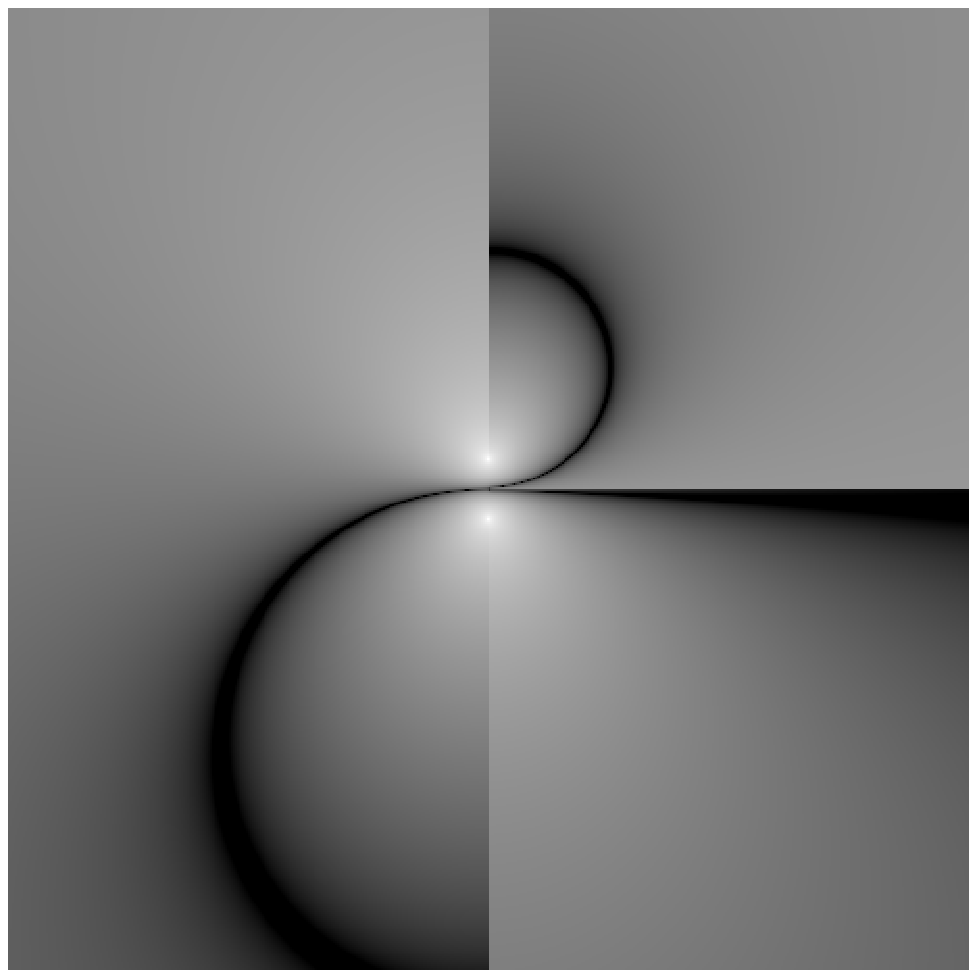}\\
\includegraphics[height=3.5cm]{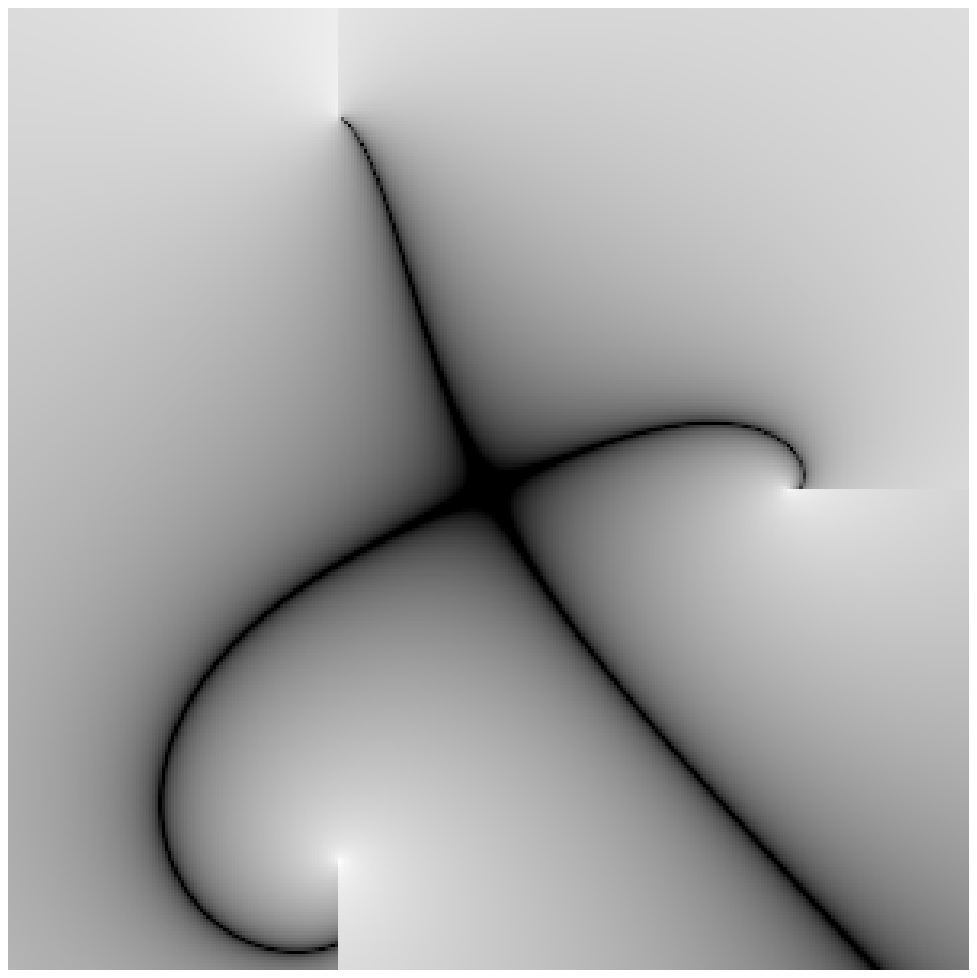}
\includegraphics[height=3.5cm]{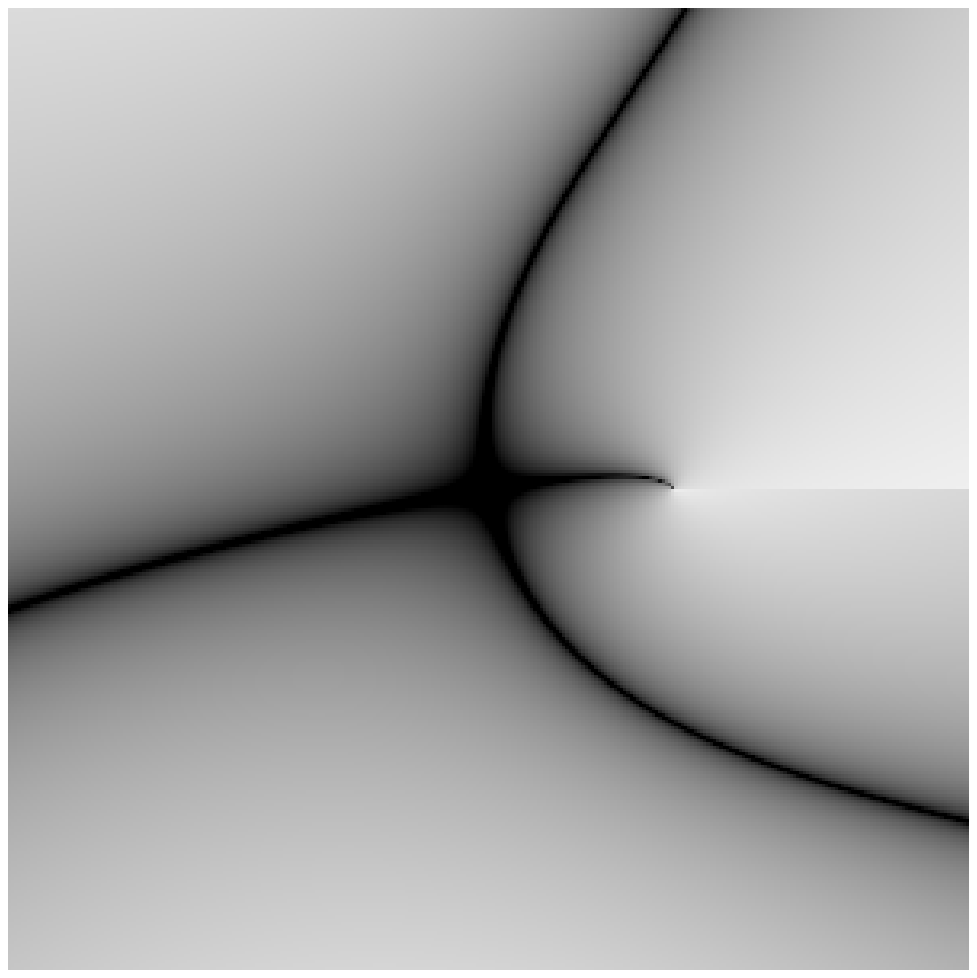}
\includegraphics[height=3.5cm]{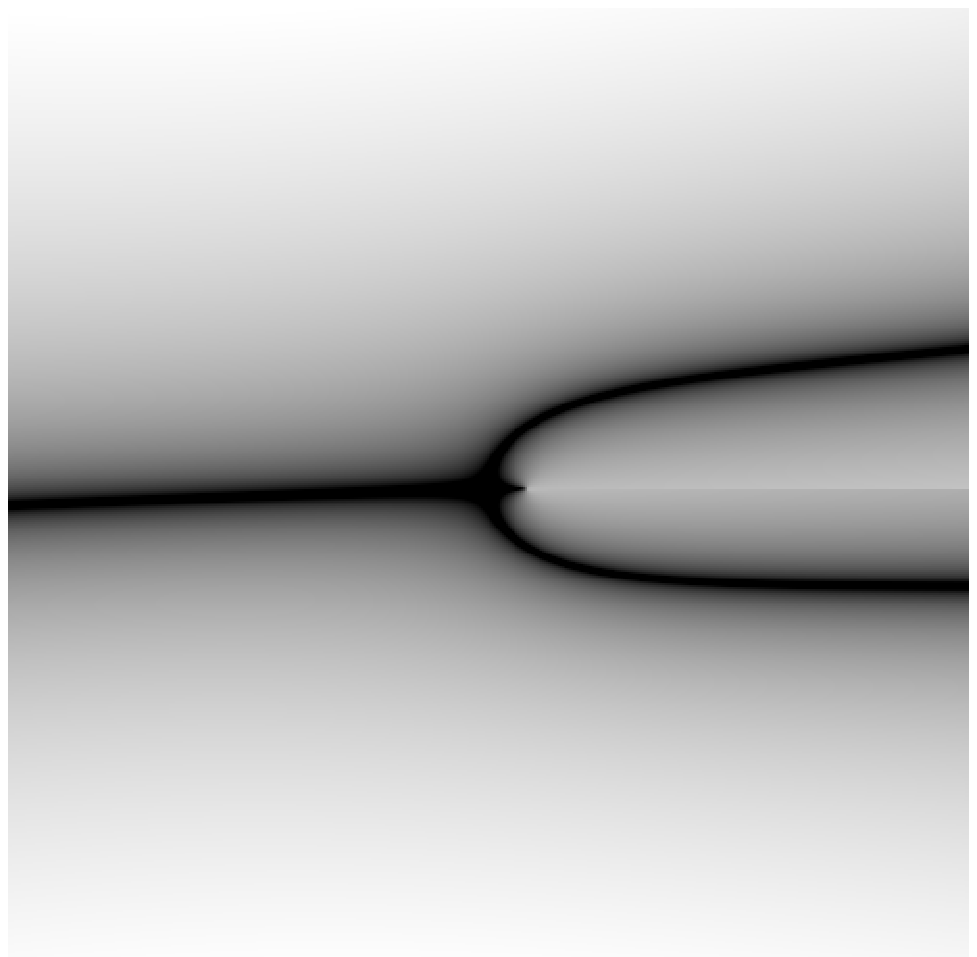}
\end{center}
\caption{Numerically generated pictures of the anti-Stokes lines in the complex $r$-plane for Schwarzschild-AdS black holes in $D=4$ when $\arg(\omega)=-\arg(\eta)$ approaches zero.  The top three pictures from left to right show how the topology of the anti-Stokes lines change as $\arg(\omega)\rightarrow 0$.  The bottom three pictures are the enlarged version of the top three pictures close to the origin.}
\label{numerical-smallBH}
\end{figure}

Before ending this section, we would like to explore the case of small black holes which have a horizon radius much less than the AdS length scale.  If we evaluate $\eta$ for small black holes with the same choice of branch that resulted in
\beeq
\eta={1\over 4 T_H \sin\left({\pi \over D-1}\right)}e^{{i\pi \over D-1}}~
\eneq
for large black holes, we get
\beeq
\eta\approx {\pi \over 2\sqrt{|\lambda|}}e^{i\theta}~,
\eneq
with $\theta \rightarrow 0$.  This means that for small black holes, the asymptotic QNM frequency is
\beeq
\omega \approx 2\sqrt{|\lambda|} \left\{ n+ {D+1 \over 4} + i {\ln2 \over 2\pi} \right\}~.
\label{complex-wkb-final}
\eneq
This result is also identical to the one obtained by Natario and Schiappa in section 3.3.1 of \cite{Natario-S} (again with a difference in sign because our perturbations depend on time as $e^{-i\omega t}$ instead of $e^{i\omega t}$).  Equation (\ref{complex-wkb-final}) shows that for small Schwarzschild-AdS black holes, the QNM frequency approaches the limit $2\sqrt{|\lambda |}n$ as n approaches infinity.  As one should expect, this result agrees with the normal modes in pure AdS space in the large n limit, see section $4.2$ of \cite{Natario-S}.  One last important issue to explore is that the result (\ref{complex-wkb-final}) is valid only if the topology of anti-Stokes lines stays in the same general form shown in Figs. \ref{schem-d4-AdS} and \ref{numerical2} as the size of the black hole approaches zero.   In Fig. \ref{numerical-smallBH} we show that the topology of anti-Stokes lines does not change form as the horizon radius becomes much smaller than the AdS length scale.  It is important to mention that the asymptotic QNM frequency (\ref{complex-wkb-final}) of small black holes leads to the correct result, which is the large $n$ normal mode frequency of a pure AdS space with no black holes \cite{Cardoso-K-L, Natario-S, Konoplya-0}.

\sxn{Highly Real Quasinormal Modes}
\vskip 0.3cm

\begin{figure}[tb]
\begin{center}
\includegraphics[height=9cm]{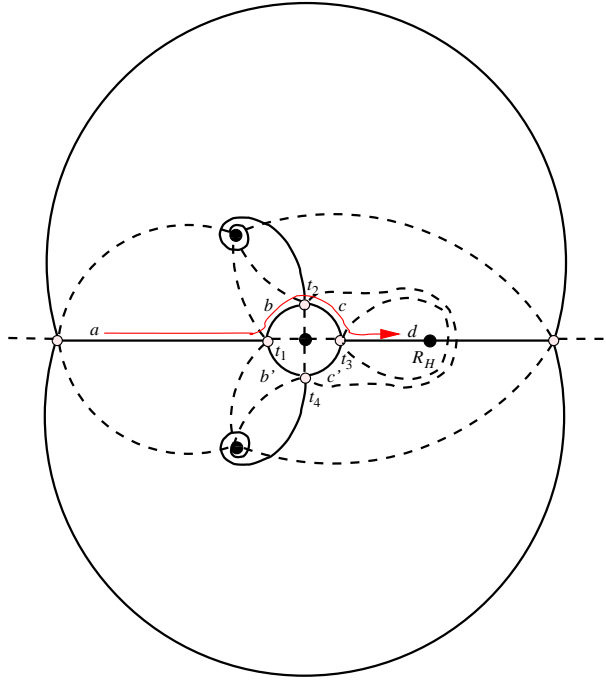}
\end{center}
\caption{Schematic presentation of the Stokes (dashed) and anti-Stokes (solid) lines in the complex $r$-plane for the highly real QNMs of large Schwarzschild-AdS black holes in four spacetime dimensions.  The open circles are the zeros and the filled circles are the poles of the function $Q$.  $R_H$ is the event horizon.  The thin arrow which starts on the anti-Stokes line that stretches to infinity at point $a$ and ends on the anti-Stokes line that connects to the event horizon at point $d$ is the path we take to determine the WKB condition on QNM frequencies.}
\label{schem-d4-AdS-real}
\end{figure}

\begin{figure}[tb]
\begin{center}
\includegraphics[height=3.5cm]{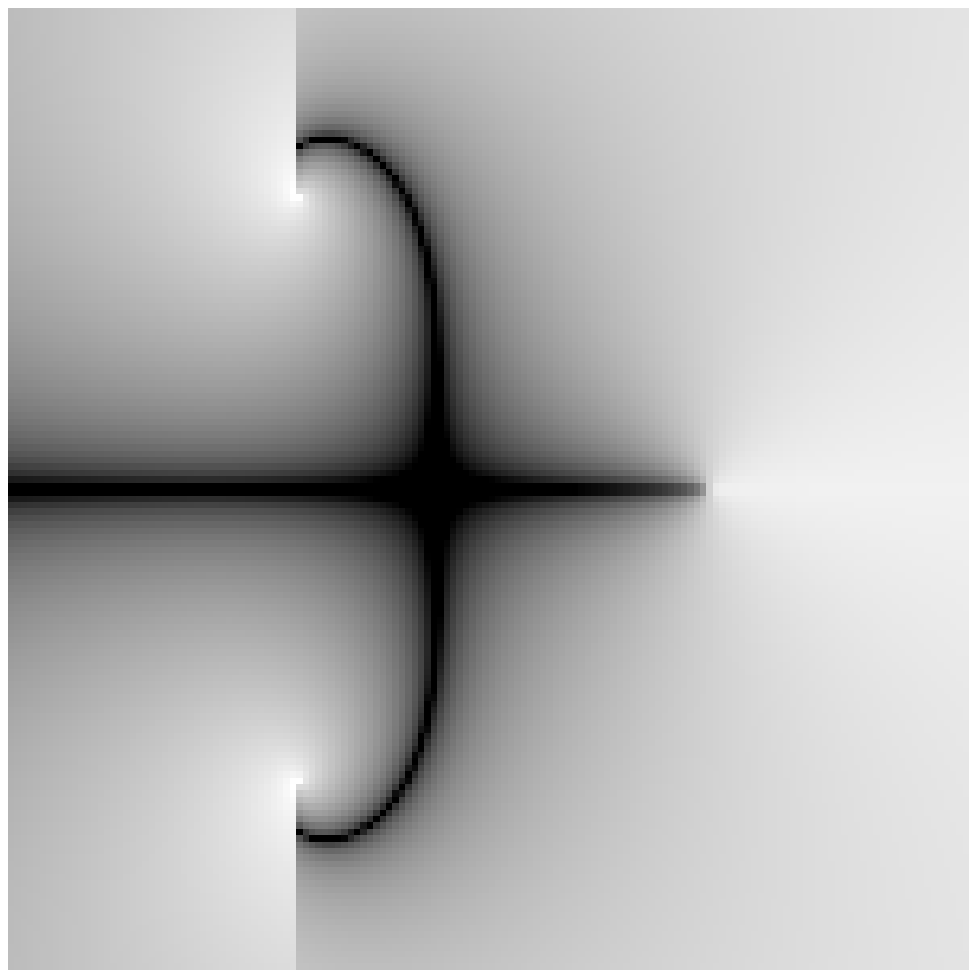}
\includegraphics[height=3.5cm]{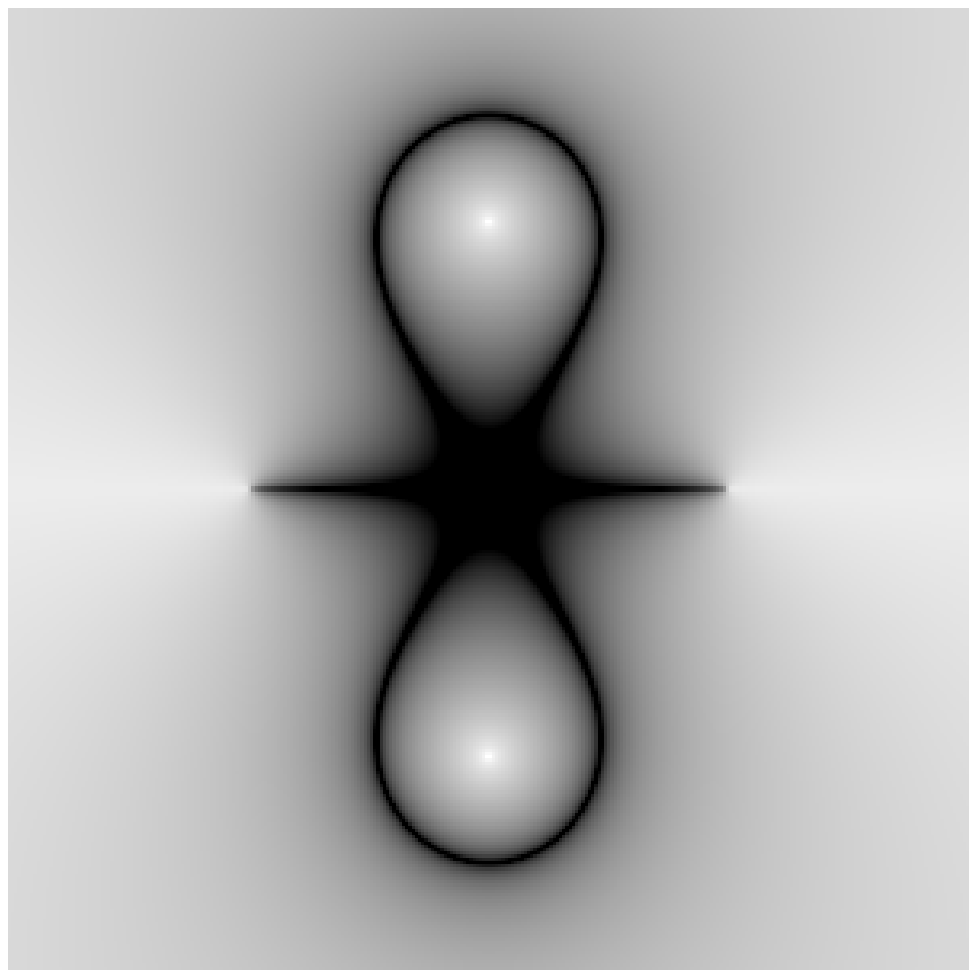}
\includegraphics[height=3.5cm]{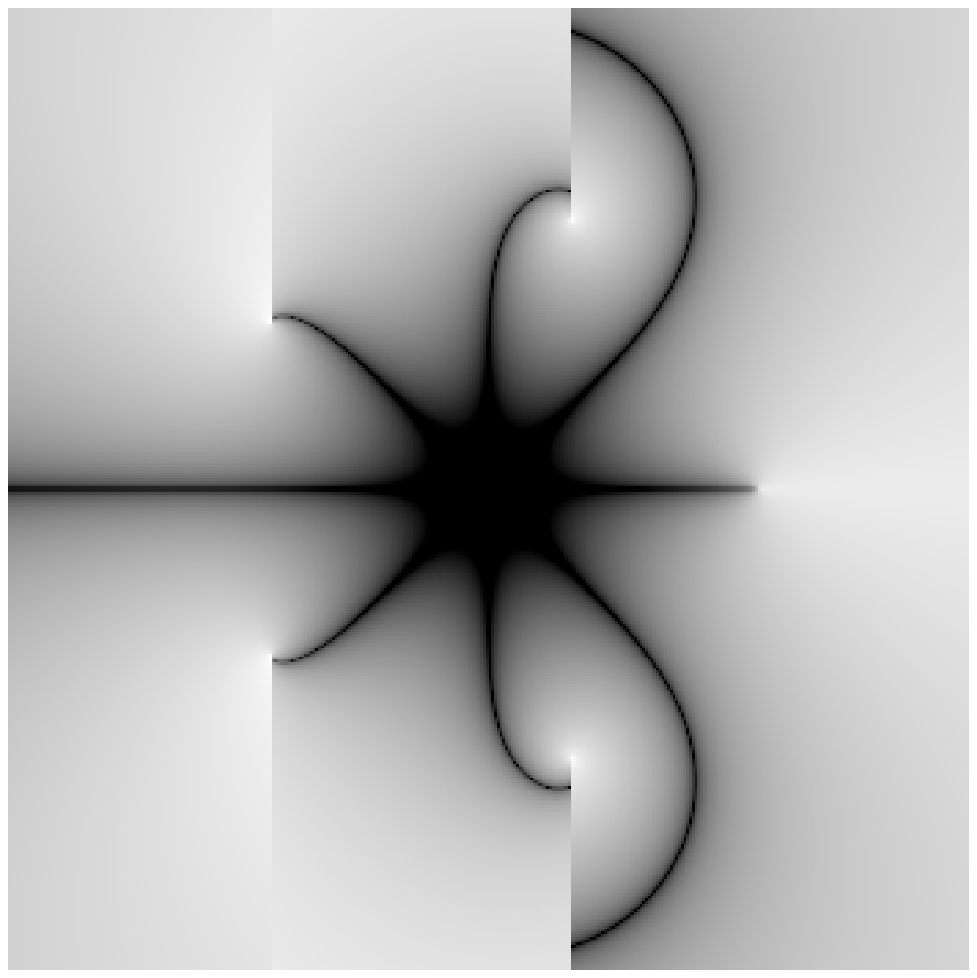}
\includegraphics[height=3.5cm]{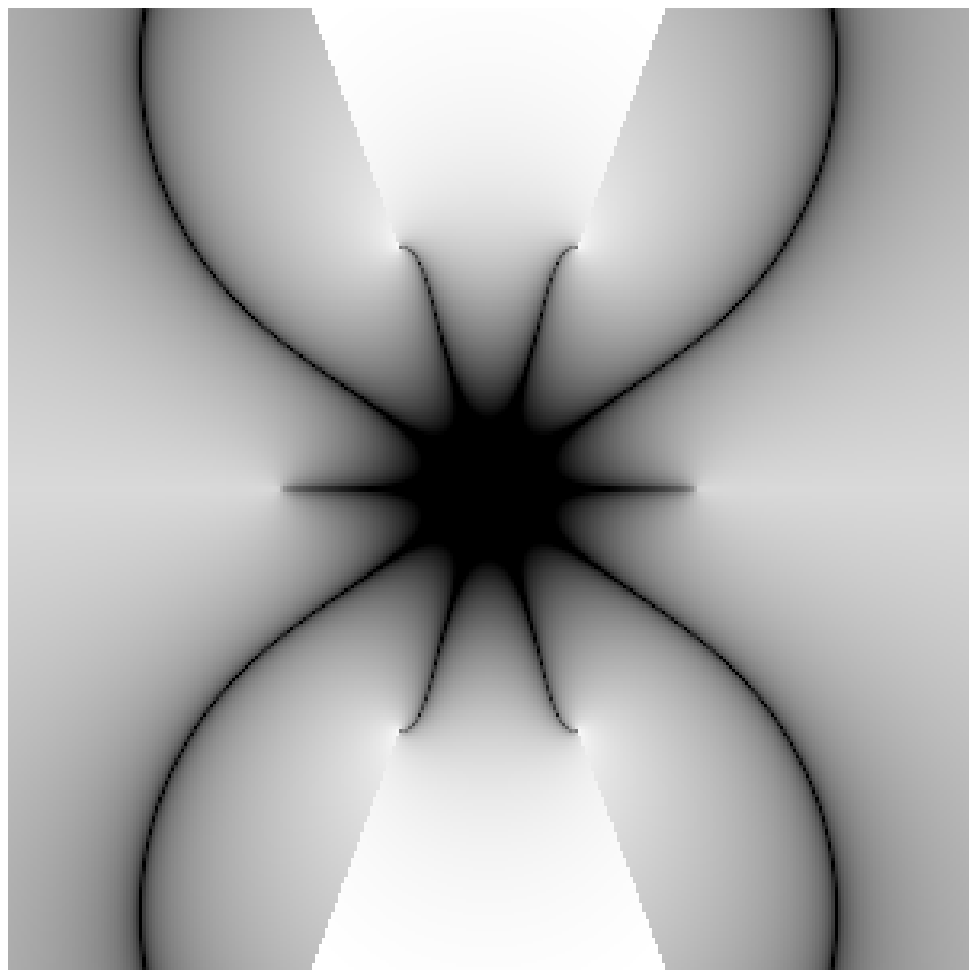}\\
\includegraphics[height=3.5cm]{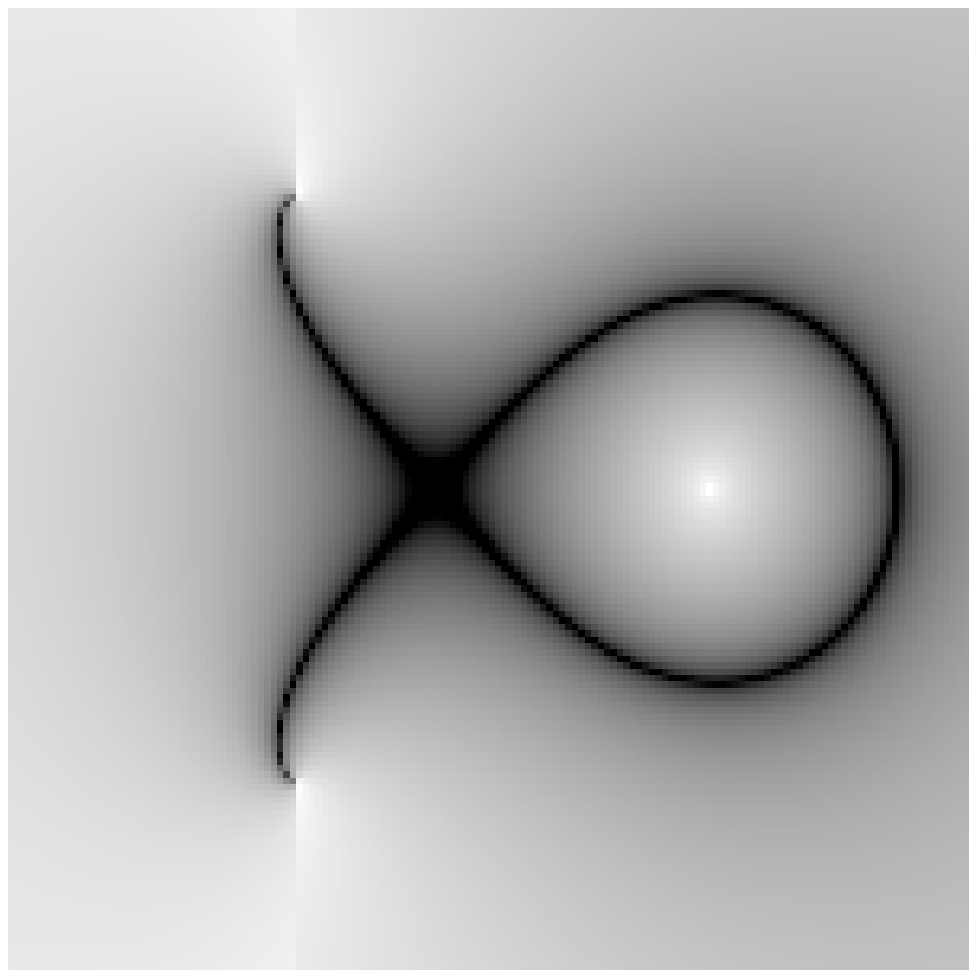}
\includegraphics[height=3.5cm]{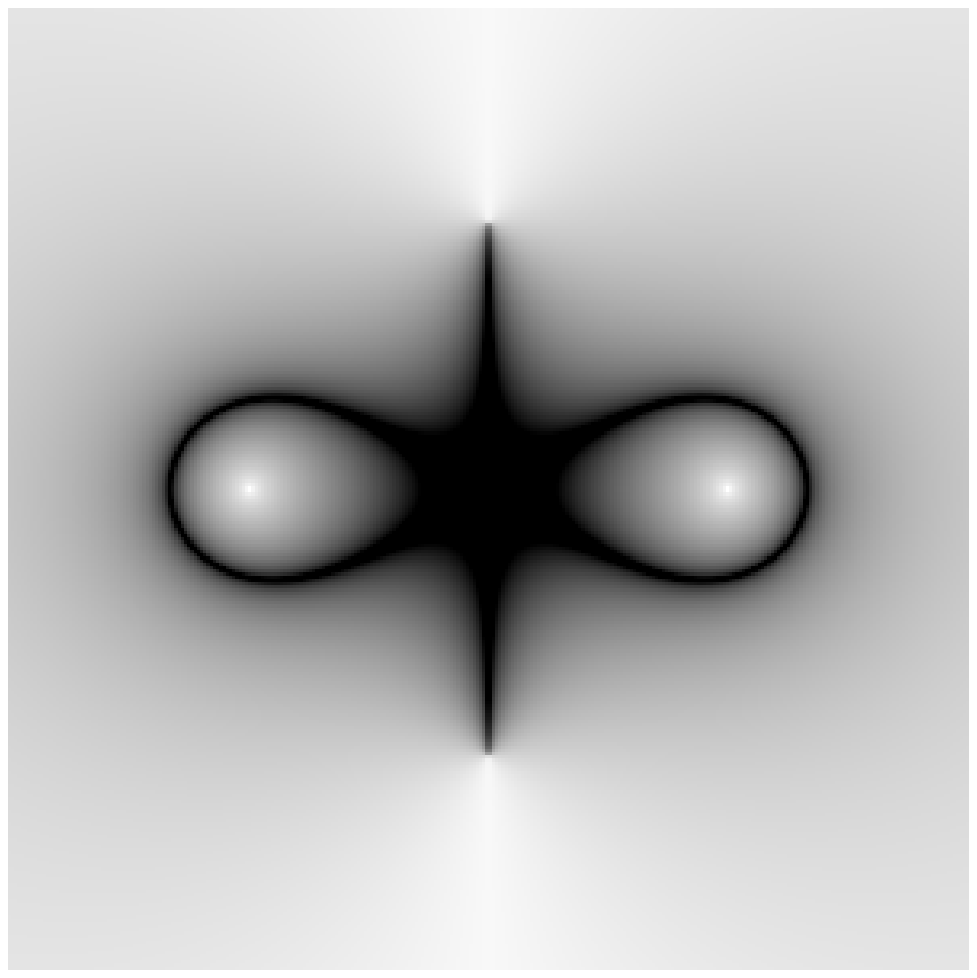}
\includegraphics[height=3.5cm]{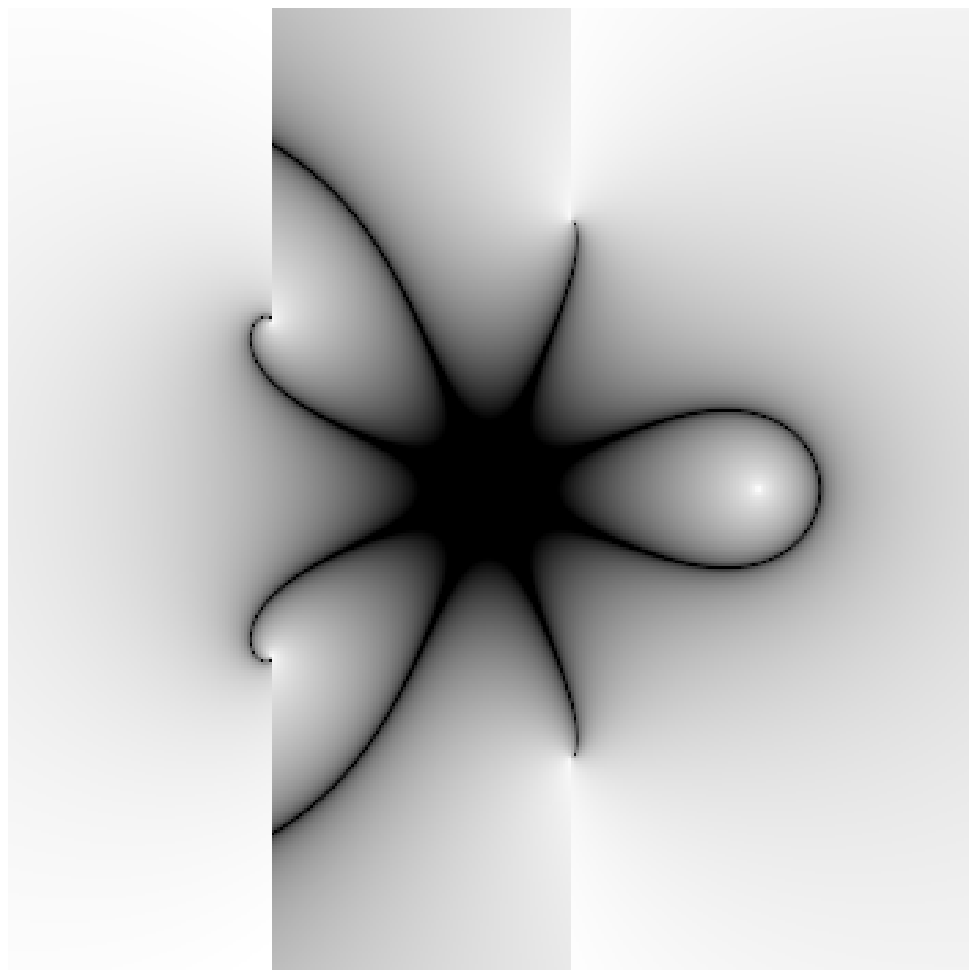}
\includegraphics[height=3.5cm]{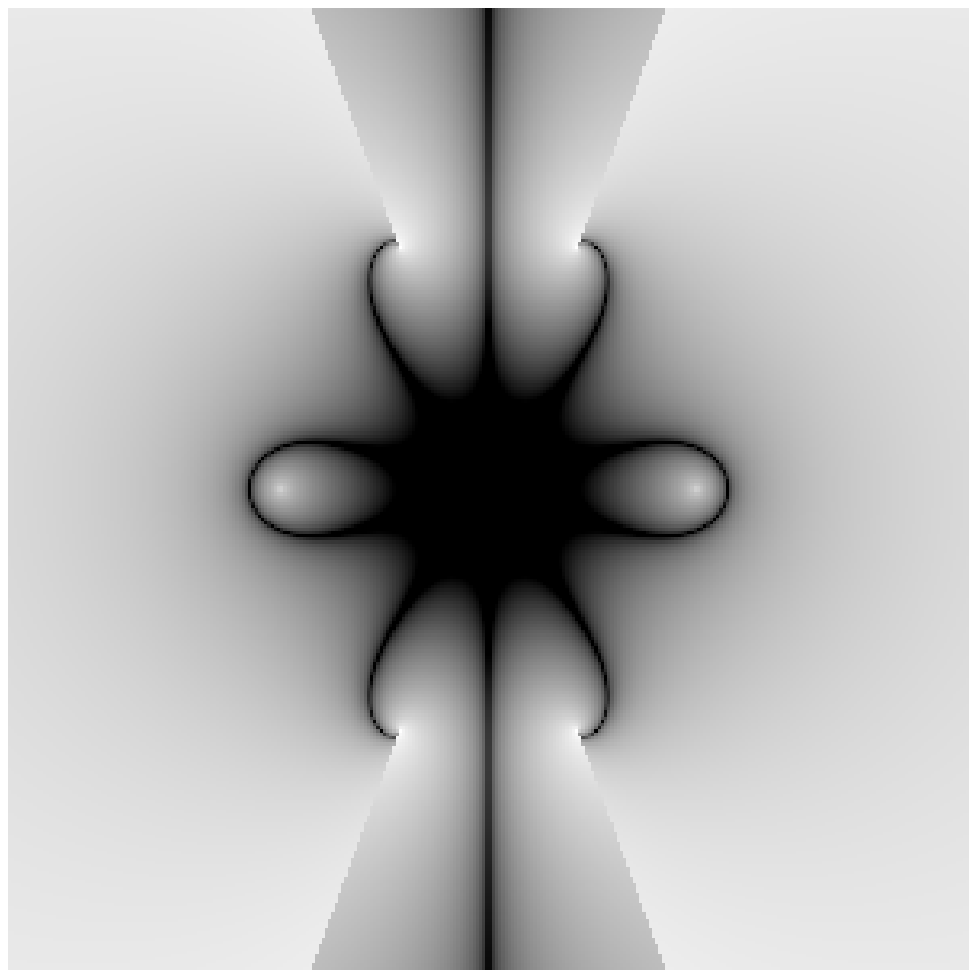}
\end{center}
\caption{Numerically generated pictures of the Stokes and anti-Stokes lines for the highly real QNMs of Schwarzschild-AdS black holes.  The top four pictures from left to right show the anti-Stokes lines in the complex $r$-plane in $D=4$, $D=5$, $D=6$, and $D=7$ respectively.  The bottom four pictures show the corresponding Stokes lines.  Note that in odd dimensions no anti-Stokes lines extend to infinity.}
\label{numerical3}
\end{figure}

We now derive the WKB condition for the case where $\eta$ is a negative real constant and consequently $\arg(\omega)\approx 0$.  In four spacetime dimensions the topology of Stokes/anti-Stokes lines is shown in Fig. \ref{schem-d4-AdS-real}.
Unfortunately this topology is not generic.  As one can see in Fig. \ref{numerical3}, in odd spacetime dimensions there is no anti-Stokes line extending to infinity which makes it impossible to derive a WKB condition for these dimensions and in higher even spacetime dimensions we have more Stoke/anti-Stokes lines in between the unbounded anti-Stokes line and the anti-Stokes line which ends at the horizon.  To get the WKB condition when $D = 4$, we follow the path shown in Fig. \ref{schem-d4-AdS-real} starting on the unbounded anti-Stokes line at point $a$ and ending on an anti-Stokes line which connects to the horizon at point $d$.  To find the WKB solution at point $a$, we can repeat the steps (\ref{Schrodinger-large-r}) through (\ref{c2}) with the exception that in Eq.\ (\ref{Psi-large-r})  ${\omega \over |\lambda|r}$ has a negative real value (instead of a positive real value) on the anti-Stokes line which extends to negative infinity on the real axis.  Therefore, we need to use the approximation
\beeq
J_j(x) \sim \sqrt{2\over \pi x}\cos(x+{j\pi \over 2}+{\pi \over 4})~~\mbox{when}~x\ll -1~,
\eneq 
instead of the one in Eq.\ (\ref{approx-bessel}).  This results in a minor modification where we need to replace $\beta$ with
\beeq
\bar{\beta}=-\beta=-{\pi \over 4}(1+J_{\infty}) ~
\label{beta-bar}
\eneq 
in Eq.\ (\ref{Psi-large-r1}) and also the rest of the calculations.  We also need to modify the phase integral in Eq.\ (\ref{phase-t1-r}) to
\beeq
\int_{t_1}^r Q(r')dr'\sim -{J\over2(D-2)}{\pi\over2}+ \omega \left(\eta-{1\over |\lambda|r}\right)~,
\eneq 
since the zero at $t_1$ is now at a different location.  These changes will result in a slight modification of the constants $c_1$ and $c_2$ in Eqs.\ (\ref{c1}) and (\ref{c2}) where we get
\beeq
c_1 = A_+ e^{i\bar{\beta}} e^{-i\omega\eta} e^{i{J\over2(D-2)}{\pi\over2}}~,
\label{c1-real}
\eneq
and 
\beeq
c_2 = A_+e^{-i\bar{\beta}} e^{i\omega\eta} e^{-i{J\over2(D-2)}{\pi\over2}}~.
\label{c2-real}
\eneq 
Interestingly, it turns out that in the topology shown in Fig. \ref{schem-d4-AdS-real} we have the freedom to choose which function is dominant on different pairs of the Stokes lines.  We can either take $f_1$ dominant on the pair of Stokes lines which end at a pole and $f_2$ dominant on the pair of Stokes lines which encircle the event horizon or vice versa.  The choice of $f_1$ or $f_2$ being dominant on both pairs of the Stokes lines is restricted due to the fact that we need to introduce the branch cuts emanating from the zeros away from the path that we take to derive the WKB condition.  Introducing the branch cut at $t_2$ away from the path forces us to have different functions dominant on the two Stokes lines which cross our path.  Also, we are not allowed to take different functions to be dominant on the same pair of Stokes lines because in the limit $|\omega| \rightarrow \infty$ the Stokes lines in each pair join each other as soon as they move away from the origin.  If we move along the path shown in Fig. \ref{schem-d4-AdS-real} with the choice of $f_1$ dominant on the pair of Stokes lines ending at the pole and $f_2$ dominant on the pair encircling the event horizon and using the fact that 
\beeq
\gamma_{12}=-\gamma_{23}=-{J\over 2(D-2)}\pi =-\gamma~,
\label{gamma}
\eneq
we get the WKB condition:
\beeq
e^{2i\omega \eta} = i e^{2i\bar{\beta}} {1+e^{-2i\gamma}+ e^{2i\gamma} \over e^{-i\gamma}+e^{i\gamma} }~.
\label{WKB-real-01}
\eneq
We are also allowed to take a path which is the mirror reflection of the path shown in Fig. \ref{schem-d4-AdS-real} in the lower half of the complex plane.  In other words, instead of moving through points $b$ and $c$ we can go through points $b'$ and $c'$ shown in Fig. \ref{schem-d4-AdS-real} and the final WKB condition is
\beeq
e^{2i\omega \eta} = -i e^{2i\bar{\beta}} {1+e^{-2i\gamma}+ e^{2i\gamma} \over e^{-i\gamma}+e^{i\gamma} }~.
\label{WKB-real-02}
\eneq
The results (\ref{WKB-real-01}) and (\ref{WKB-real-02}) translate to
\beeq
\omega \eta = \bar{\beta} - n\pi  \pm {\pi\over 4}-{i\over2}\ln\left[2\cos\left({J\over 4}\pi\right)-{1 \over 2\cos\left({J\over 4}\pi\right)}\right]~~~~\mbox{as $n\rightarrow \infty$}
\eneq
in four spacetime dimensions.  After entering the values for $J$ and $\bar{\beta}$ in $D=4$ for tensor, vector, and scalar perturbations using Eqs.\ (\ref{J}), (\ref{Jinfinite}), and (\ref{beta}), it is easy to show that tensor and vector perturbations satisfy the condition
\beeq
\omega \eta = -n\pi \pm {\pi\over 4} -{i\over2}\ln{3\over 2}
\label{3/2case}
\eneq
and scalar perturbations satisfy
\beeq
\omega \eta = -n\pi -{\pi\over2}\pm {\pi\over 4} -{i\over2}\ln{3\over 2}~~~~~~~~\mbox{as $n\rightarrow \infty$}~.
\label{3/2case-scalar}
\eneq
However, $\eta$ is a negative real constant which means that the WKB conditions (\ref{3/2case}) and (\ref{3/2case-scalar}) will result in QNMs with positive imaginary frequency.  Since we chose the perturbations to depend on time as $e^{-i\omega t}$, positive imaginary frequency indicates instability.  This is rather surprising since it has been shown that these black holes are stable against metric perturbations \cite{Ishibashi1, Ishibashi2}.  The appearance of these frequencies could be a consequence of the assumption, made throughout the analytic calculations of the highly real QNMs, that the modes are purely real with zero damping. This makes the analytic calculation insensitive to the sign of the imaginary part of the frequency.  Since we know these black holes are stable, we discard those frequencies with positive imaginary part.  We believe this should not occur in numerical calculations where the imaginary term is not ignored.
%However, $\eta$ is a negative real constant which means that the WKB conditions (\ref{3/2case}) and (\ref{3/2case-scalar}) will result in QNMs with positive damping rates.   Since we chose the perturbations to depend on time as $e^{-i\omega t}$, the QNMs with positive damping rates cannot be physical and they need to be discarded. 

We now move along the path shown in Fig. \ref{schem-d4-AdS-real} with the choice of $f_2$ dominant on the pair of Stokes lines ending at the pole and $f_1$ dominant on the pair looping around the horizon which results in the WKB condition
\beeq
e^{2i\omega \eta} = -i e^{2i\bar{\beta}}  {e^{i\gamma} \over 1+e^{2i\gamma}}~.
\label{fourdreal}
\eneq
Also taking the path in the lower half of the complex plane through points $b'$ and $c'$ will result in the WKB condition
\beeq
e^{2i\omega \eta} = i e^{2i\bar{\beta}} e^{i\gamma} \left({1 \over 1+e^{2i\gamma}}\right)~.
\eneq
These results translate to
\beeq
\omega \eta = \bar{\beta} - n\pi  \mp {\pi\over 4}+{i\over2}\ln\left[ 2\cos\left({J\over 4}\pi\right)\right]~~~~\mbox{as $n\rightarrow \infty$}
\eneq
in four spacetime dimensions.  After entering the values for $J$ and $\bar{\beta}$ in $D=4$ for tensor, vector, and scalar perturbations using Eqs.\ (\ref{J}), (\ref{Jinfinite}), and (\ref{beta}), it is easy to show that the highly real QNM frequency for tensor and vector perturbations satisfies the condition
\beeq
\omega \eta = -n\pi \mp {\pi\over 4} +{i\over2}\ln{2}~
\label{2case}
\eneq
and for scalar perturbations satisfies
\beeq
\omega \eta = -n\pi -{\pi\over2}\mp {\pi\over 4} +{i\over2}\ln{2}~~~~~~~~\mbox{as $n\rightarrow \infty$}~.
\label{2case-scalar}
\eneq
These results with the choice of $+\pi/4$ are identical to the results that we found in Eqs.\ (\ref{wkb-complex}) and (\ref{wkb-complex-d=4}).  The conditions (\ref{2case}) and (\ref{2case-scalar}) with a negative real $\eta$ will result in a negative damping rate which physically makes sense.  Therefore, for large black holes in four spacetime dimensions we can use condition (\ref{2case}) to evaluate the highly real QNM frequency of tensor and vector perturbations, where
\beeq
\omega =4T_H \sin({\pi \over 3})\left[ n\pi \pm {\pi\over 4} -{i\over2}\ln{2}\right]~.
\label{highly-real-WKB}
\eneq 
For scalar perturbations we use condition (\ref{2case-scalar}) to get
\beeq
\omega =4T_H \sin({\pi \over 3})\left[ n\pi +{\pi \over 2}\pm {\pi\over 4} -{i\over2}\ln{2}\right]~.
\label{highly-real-WKB-scalar}
\eneq 
It is important to note that the damping rate in this region for all perturbations is
\beeq
|\omega_I| =2T_H \sin({\pi \over 3})\ln{2}=1.2 T_H~.
\label{damping-rate}
\eneq  
The significance of this result lies in the fact that the damping rate in this asymptotic region of the QNM frequency spectrum is less than the damping rate calculated by Horowitz and Hubeny \cite{Horowitz} for the lowest QNM frequency where they found $|\omega_I|=11.16T_H$.  This means that in four spacetime dimensions the highly real QNMs decay at a slower rate than any other regions of the QNM spectrum and therefore they are the most relevant in the context of AdS/CFT correspondence.  In other words, we need to use Eq.\ (\ref{damping-rate}) to calculate the thermalization timescale in the CFT.

In the case of small black holes with a horizon radius much less than the AdS radius, we can evaluate $\eta$ analytically which gives
\beeq
\eta \sim -{\pi \over 2\sqrt{|\lambda|}}~.
\label{small-eta2}
\eneq  
To make $\eta$ single valued in the small black hole limit, we introduce the same branch cut that resulted in a negative real $\eta$ for large and intermediate black holes.  Using (\ref{small-eta2}), we find the QNM frequency of tensor and vector perturbations for small black holes to be
\beeq
\omega =2\sqrt{|\lambda|}\left[ n \pm {1\over 4} -{i\over2\pi}\ln{2}\right]~,
\label{highly-real-WKB-small}
\eneq  
and for scalar perturbations
\beeq
\omega =2\sqrt{|\lambda|}\left[ n +{1\over2}\pm {1\over 4} -{i\over2\pi}\ln{2}\right]~.
\label{highly-real-WKB-small-scalar}
\eneq  
Note that the highly real QNM frequency (\ref{highly-real-WKB-small}) for small black holes behaves as $2\sqrt{|\lambda|}n$, which is the large $n$ normal mode frequency of a pure AdS space with no black holes \cite{Cardoso-K-L, Natario-S, Konoplya-0}.

As mentioned earlier, the topology of the Stokes/anti-Stokes lines shown in Fig. \ref{schem-d4-AdS-real} is not generic; it changes in higher spacetime dimensions.  For example, in six dimensions shown in Fig. \ref{numerical3} we need to cross eight Stokes lines versus four Stokes lines in four dimensions.  Fortunately a pattern emerges as we derive the WKB condition for higher spacetime dimensions.  
In general, for even spacetime dimensions $D$, the WKB condition can be written as
\beeq
e^{2i\omega \eta} = \mp i e^{2i\bar{\beta}} e^{i\gamma} {1+e^{2i\gamma}+e^{4i\gamma}+ \cdots + e^{2i(D-4)\gamma} \over 1+e^{2i\gamma}+e^{4i\gamma}+ \cdots + e^{2i(D-3)\gamma}}~.
\eneq
These results translate to
\beeq
\omega \eta = \bar{\beta} - n\pi  \mp {\pi\over 4}-{i\over2}\ln\left[e^{i\gamma} {1+e^{2i\gamma}+e^{4i\gamma}+ \cdots + e^{2i(D-4)\gamma} \over 1+e^{2i\gamma}+e^{4i\gamma}+ \cdots + e^{2i(D-3)\gamma}}\right]~~~~\mbox{as $n\rightarrow \infty$}~.
\eneq
After entering the values for $\bar{\beta}$ and $\gamma$ for tensor, vector, and scalar perturbations using Eqs.\ (\ref{J}), (\ref{Jinfinite}), (\ref{beta-bar}), and (\ref{gamma}), it is easy to show that all types of perturbations in any dimension $D>3$ (except scalar perturbations in $D=4$ and $5$) follow the WKB condition
\beeq
\omega \eta = -n\pi - {\pi\over 4}(D\pm 1) +{i\over2}\ln{D-2 \over D-3}~.
\label{WKB-real-all-D}
\eneq
For large black holes we use Eq.\ (\ref{WKB-real-all-D}) to derive the general equation for the highly real QNM frequency in $D$ dimensions where
\beeq
\omega =4T_H \sin({\pi \over D-1})\left[ n\pi + {\pi\over 4}(D\pm 1) -{i\over2}\ln{D-2 \over D-3}\right]~.
\label{real-WKB-large-D}
\eneq  
Using Eqs.\ (\ref{small-eta2}) and (\ref{WKB-real-all-D}), we can also find the highly real QNM frequency for small black holes in $D$ dimensions:
\beeq
\omega =2 \sqrt{|\lambda|}\left[ n + {1\over 4}(D\pm 1) -{i\over2\pi}\ln{D-2 \over D-3}\right]~.
\label{real-WKB-small-D}
\eneq

\begin{figure}[tb]
\begin{center}
\includegraphics[height=3.5cm]{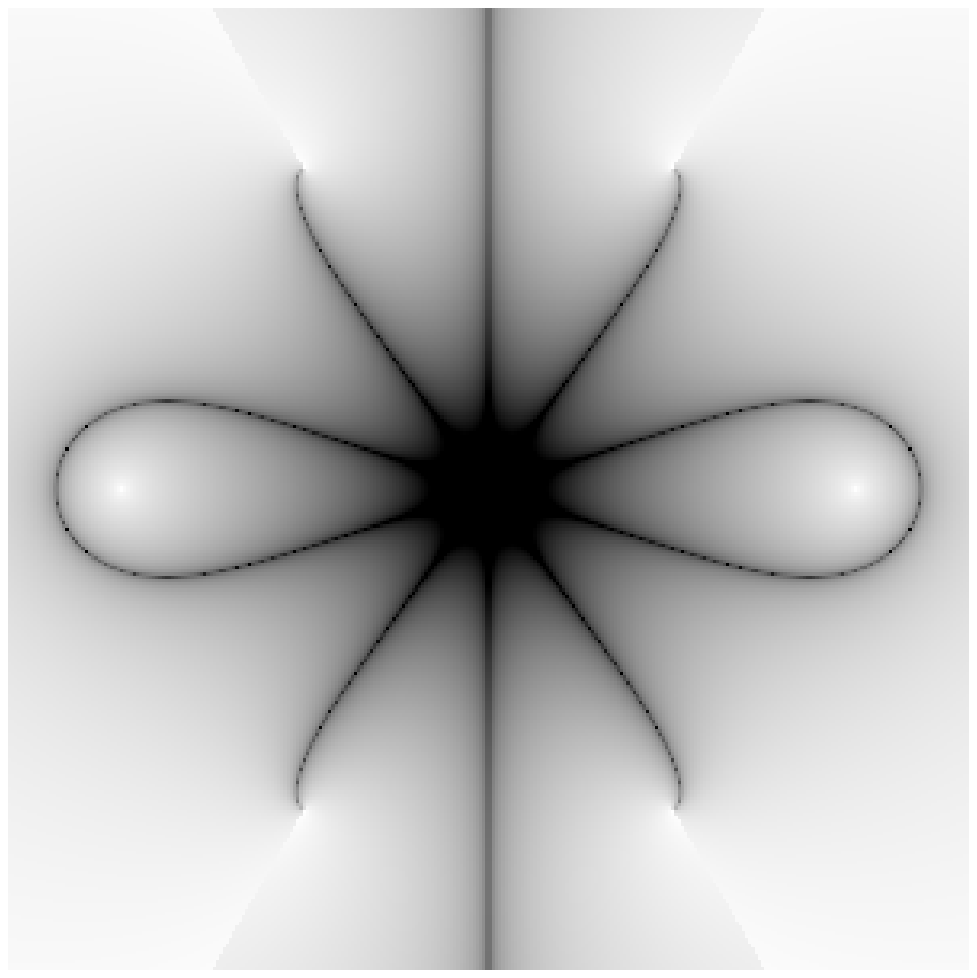}
\includegraphics[height=3.5cm]{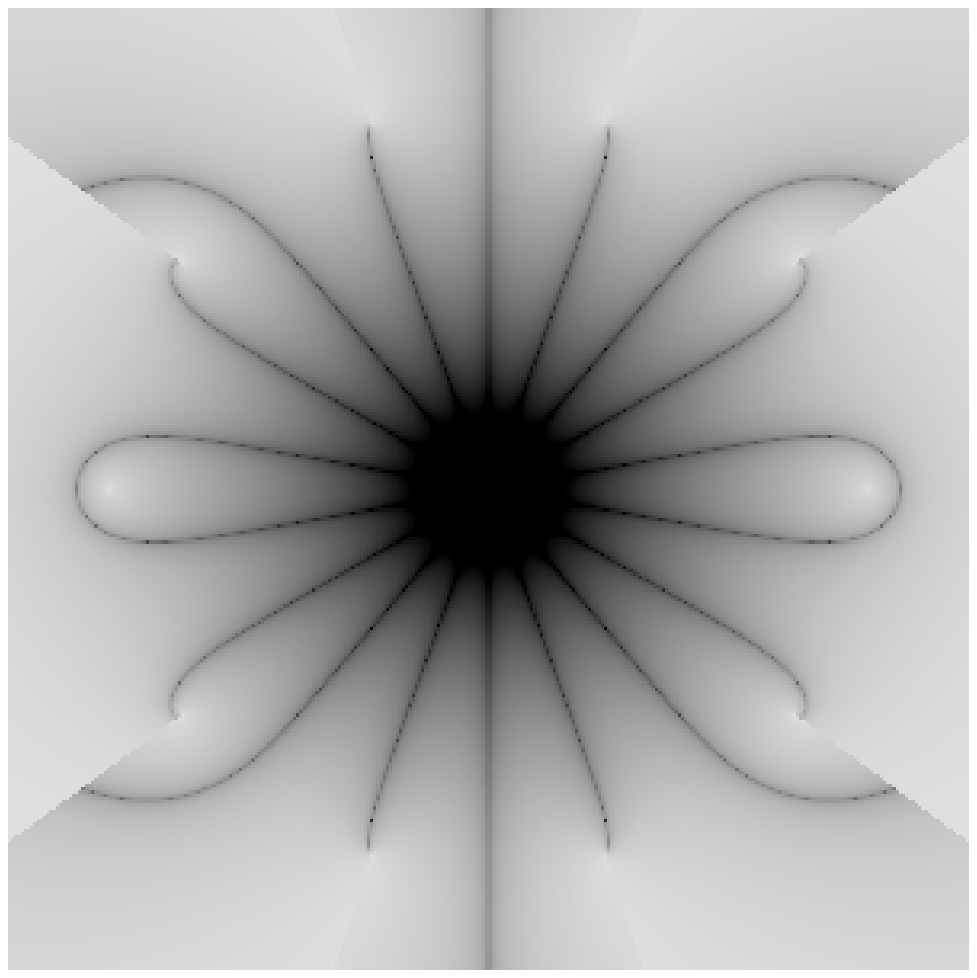}
\includegraphics[height=3.5cm]{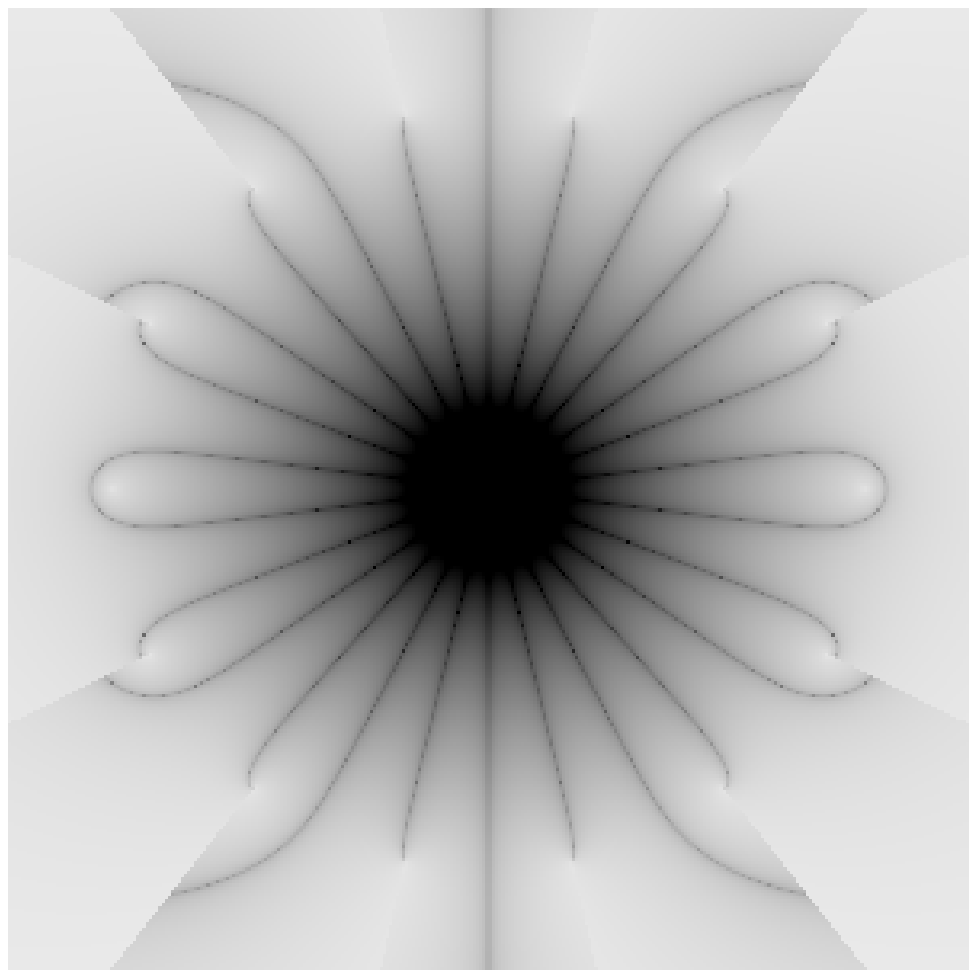}\\
\includegraphics[height=3.5cm]{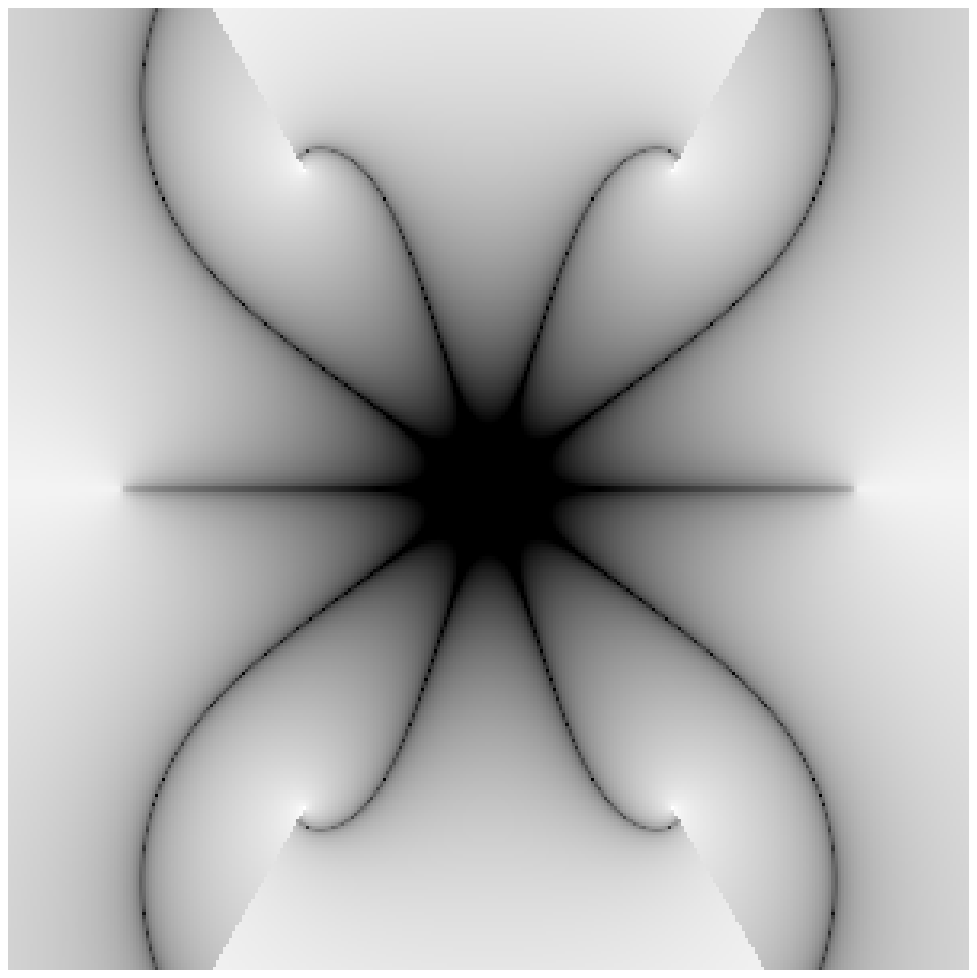}
\includegraphics[height=3.5cm]{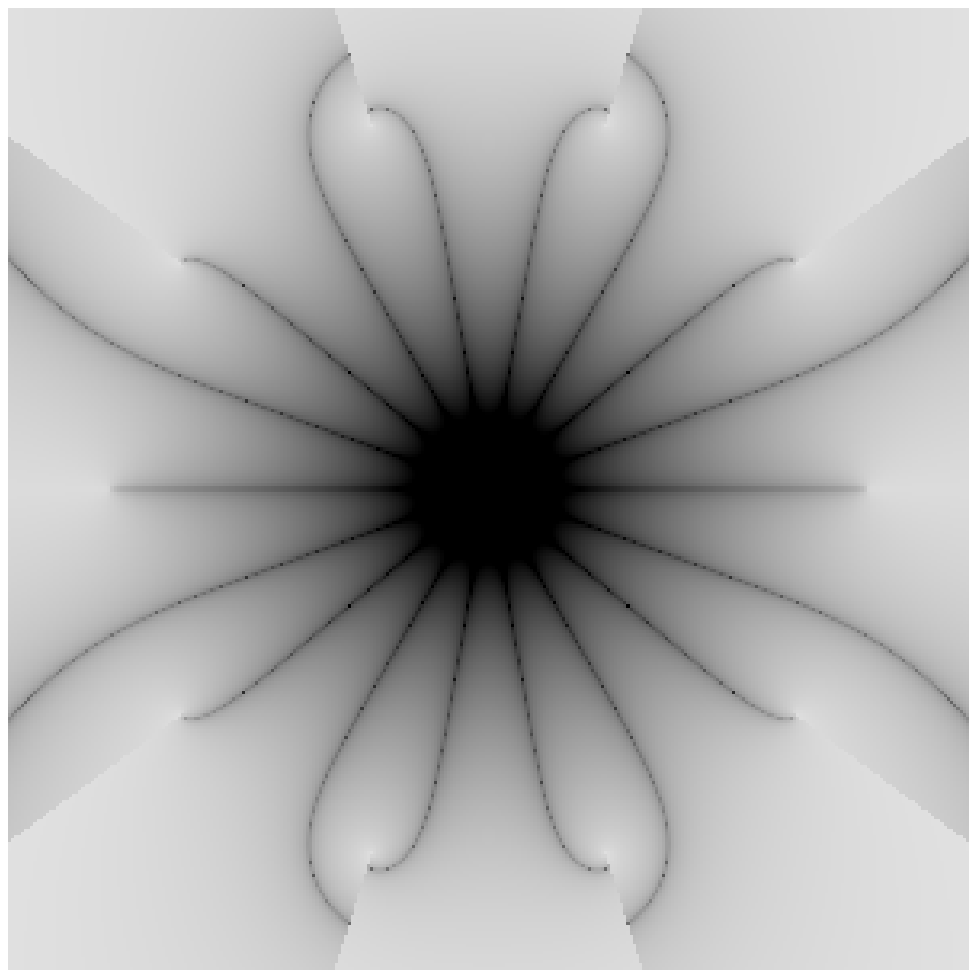}
\includegraphics[height=3.5cm]{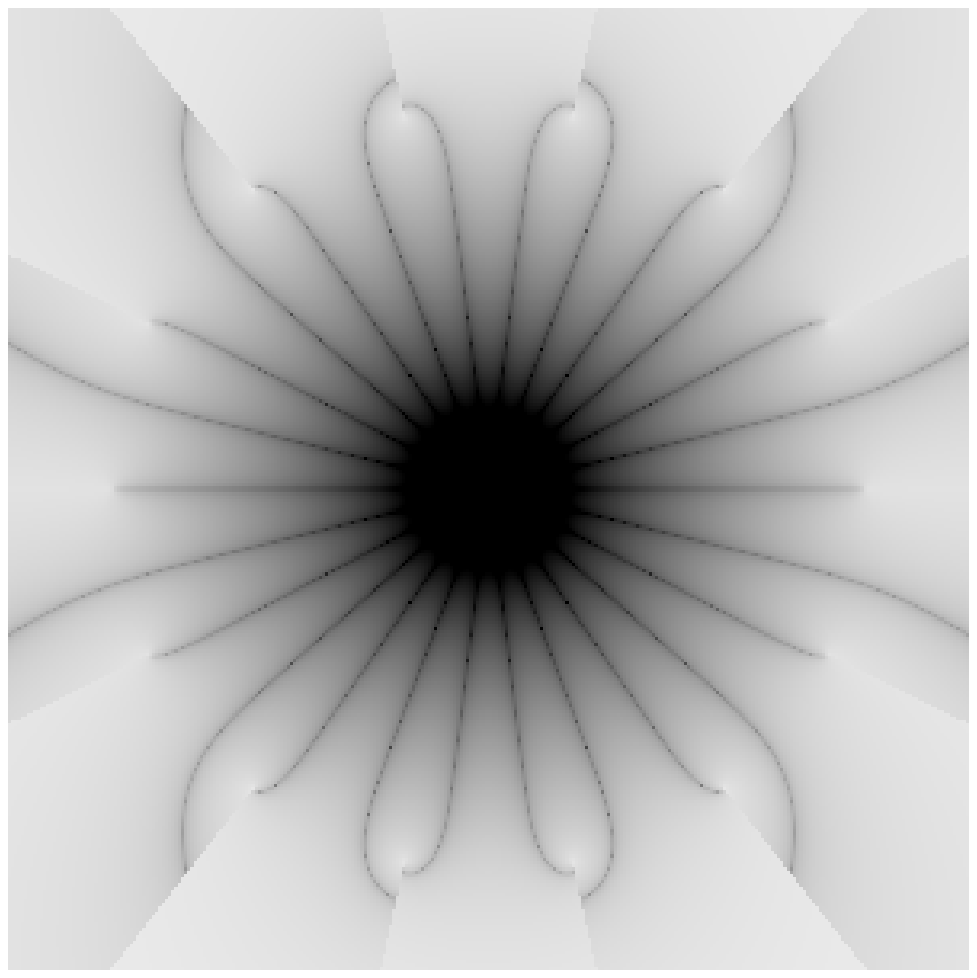}
\end{center}
\caption{Numerically generated pictures of the Stokes and anti-Stokes lines for the highly damped QNMs of Schwarzschild-AdS black holes.  The top three pictures from left to right show the anti-Stokes lines in the complex $r$-plane in $D=7$, $D=11$, and $D=15$ respectively.  The bottom three pictures show the corresponding Stokes lines.}
\label{numeric-high-damp}
\end{figure}

\sxn{Highly Damped Quasinormal Modes}
\vskip 0.3cm

The highly damped QNMs which we define as a region of the spectrum where the damping rate goes to infinity while the real frequency approaches a finite value occurs in spacetime dimensions $D=7, 11, 15, 19, \dots$.  The topology of the Stokes/anti-Stokes lines for spacetime dimensions $7$, $11$, and $15$ is shown in Fig. \ref{numeric-high-damp}.  In order to determine the WKB condition on the QNM frequency we follow a path along anti-Stokes lines similar to the one which was followed by Andersson and Howls \cite{Andersson} for Schwarzschild black holes.  We start on the unbounded anti-Stokes line that extends to infinity along the positive imaginary axis.  Note that the zero of the function $Q$ on the positive imaginary axis is of the same kind as $t_1$ in the highly real case explained in the previous section.  Also, ${\omega \over |\lambda| r} \ll -1$ on the unbounded anti-Stokes line that extends to infinity along the positive imaginary axis.  Therefore, the solution on this anti-Stokes line is of the type (\ref{Psi-a-short}) where $c_1$ and $c_2$ are (\ref{c1-real}) and (\ref{c2-real}) respectively.  We follow along anti-Stokes lines toward the bounded anti-Stokes line which encircles the event horizon.  After crossing a number of Stokes lines in the vicinity of the zeros of the function $Q$ we loop around the pole at the event horizon in the clockwise direction.  Then we return to the same unbounded anti-Stokes line on which we initiated our path.  The initial solution and the final solution are related to each other according to
\beeq
\Psi_{final}=\chi \Psi_{initial}~,
\eneq
where $\chi$ is the monodromy of this contour. Since this contour circles only the pole at the horizon, the  monodromy of this contour must be the same as the monodromy of a contour in the vicinity of the pole at the event horizon. This latter monodromy is determined by the boundary condition of ingoing waves at the horizon, which gives
\beeq
\chi = e^{-i \Gamma}~,
\eneq
where $\Gamma$ is the integral of the function $Q$ along the contour encircling the pole at the event horizon in the negative direction.  Using the residue theorem to evaluate $\Gamma$ yields
\beeq
\Gamma=\oint Q dy= -2\pi i \mathop{Res}_{r=|R_H|} Q= -2\pi i {\omega \over f'(R_H)}= -i{\omega \over 2 T_H} ~.
\label{Gamma}
\eneq
Applying the above method to Schwarzschild-AdS black holes in seven spacetime dimensions leads to the WKB condition of the form
\beeq
e^{-2i\bar{\beta}} e^{2i\omega \eta} e^{-i\gamma}= {3+e^{2i\Gamma} \over 2i\left(2+e^{2i\Gamma}\right)}~.
\eneq
After entering the values for $\bar{\beta}$ and $\gamma$ for tensor, vector, and scalar perturbations using Eqs.\ (\ref{J}), (\ref{Jinfinite}), (\ref{beta-bar}), and (\ref{gamma}), it is easy to show that for large black holes in $D=7$, where $\eta={i\over 2T_H}$, all types of perturbations follow the WKB condition of the simple form
\beeq
e^{2i\Gamma}={-1\pm \sqrt{17} \over 2}~,
\eneq
and it immediately follows that
\beeq
{\omega \over T_H} =\ln \left({-1+ \sqrt{17} \over 2}\right) -i(2n\pi) ~,
\eneq
or
\beeq
{\omega \over T_H} = \ln \left({1+ \sqrt{17} \over 2}\right)-i(2n+1)\pi ~~~~\mbox{as $n\rightarrow \infty$}~.
\eneq
It is not clear to us why we get two different answers for the QNM frequency and which one is the correct one.
In the small black hole limit where $\eta \rightarrow -{\pi \over 2|\lambda|}$, we notice that the term $e^{2i\omega \eta} \rightarrow 0$.  This means that for small black holes, the WKB condition is
\beeq
e^{2i\Gamma}=-3~,
\eneq
which results in the same QNM frequency as in a Schwarzschild black hole in flat spacetime:
\beeq
{\omega \over T_H} = \ln(3)-i(2n+1)\pi ~~~~\mbox{as $n\rightarrow \infty$}~.
\eneq
In the highly damped region of the QNM spectrum, it seems natural that small black holes in AdS space should resemble Schwarzschild black holes.  The results may look good so far, but when we go to the next dimension where the highly damped limit appears again things start to become complex.  
In spacetime dimension eleven, the WKB condition on the QNM frequency that we get is
\beeq
e^{-2i\bar{\beta}} e^{2i\omega \eta} e^{-i\gamma}= {5+3e^{2i\Gamma} \over 2i\left(3+2e^{2i\Gamma}\right)}~.
\eneq
For large black holes in $D=11$, where $\eta={i\over 4T_H \sin(\pi/10)}$, we can solve the above equation numerically after entering the values of $\bar{\beta}$ and $\gamma$ for tensor, vector, and scalar perturbations.  The result is
\beeq
e^{2i\Gamma} = 1.1504 ~,        
\eneq
which translates to
\beeq
{\omega \over T_H} = \ln(1.1504)-i(2n\pi) ~~~~\mbox{as $n\rightarrow \infty$}~.
\eneq
For small black holes, where $\eta \rightarrow -{2\pi \over \lambda}$, the term $e^{2i\omega \eta} \rightarrow 0$ and the WKB condition is
\beeq
e^{2i\Gamma}=-{5 \over 3}~.
\eneq
This condition is not the same WKB condition that we found for Schwarzschild black holes.
  
We also evaluated the WKB condition for the highly damped QNMs in higher dimensions where a pattern emerges and the general form of the WKB condition can be written as
\beeq
e^{-2i\bar{\beta}} e^{2i\omega \eta} e^{-i\gamma}= {{D-1 \over 2}+{D-5 \over 2}e^{2i\Gamma} \over 2i\left({D+1 \over 4}+{D-3 \over 4}e^{2i\Gamma}\right)}~ 
\eneq
for dimensions $D=7, 11, 15, \dots$.

%It is not clear to us how to interpret the results in this section.  The fact that in $D>7$ spacetime dimensions the highly damped QNMs of small Shwarzschild-AdS black holes do not match with the Schwarzschild black hole QNMs may be an indication that these modes are not physical.  Further investigation is necessary for this case.

\begin{figure}[tb]
\begin{center}
\includegraphics[height=3.5cm]{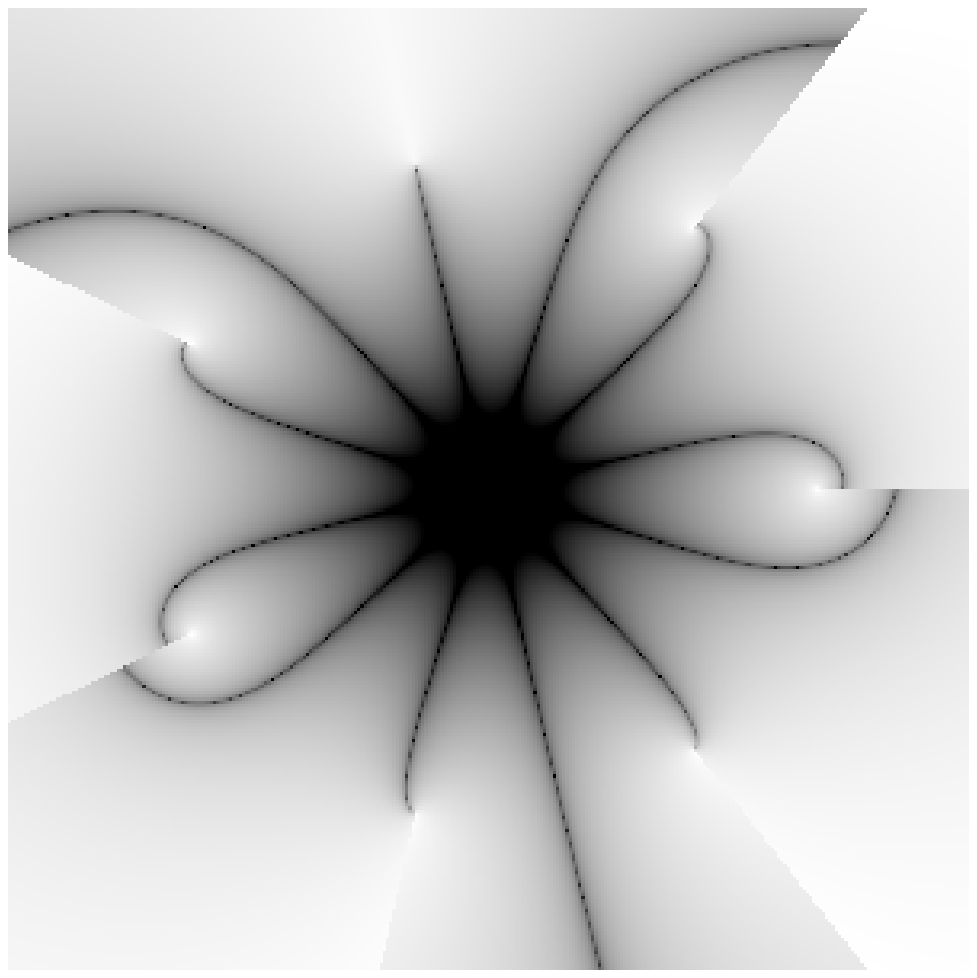}
\includegraphics[height=3.5cm]{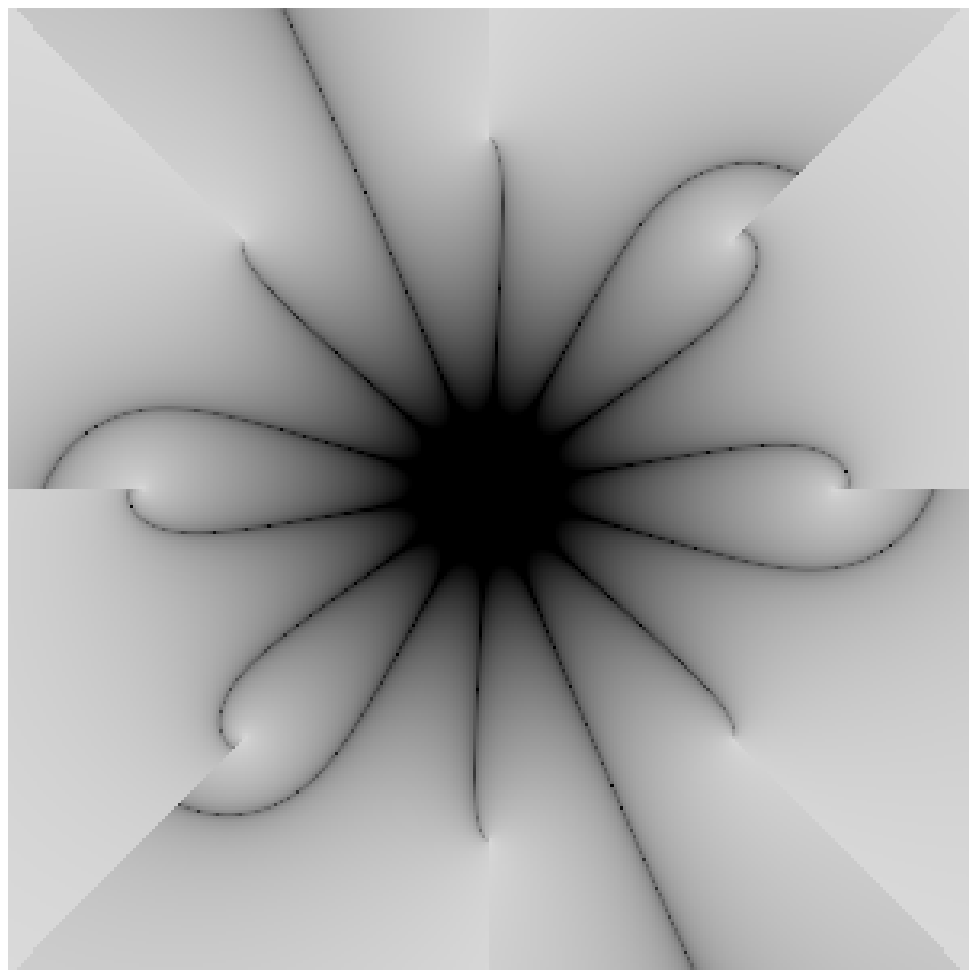}\\
\includegraphics[height=3.5cm]{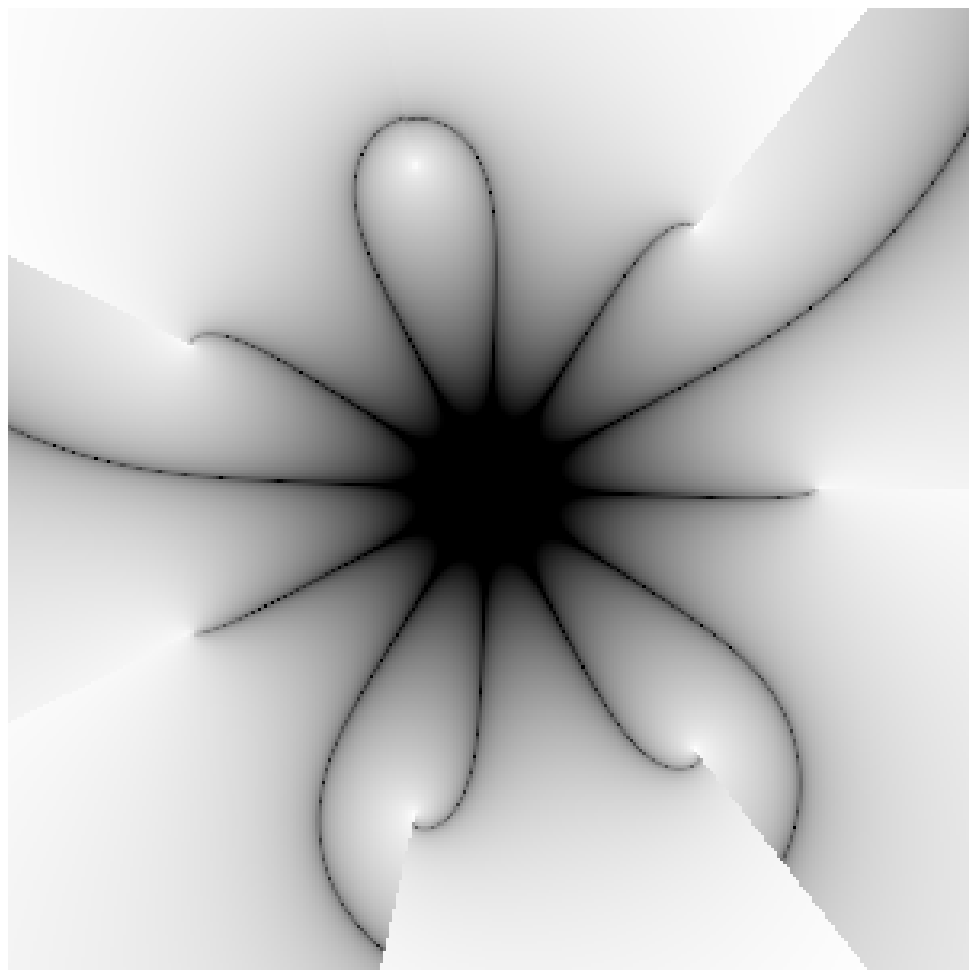}
\includegraphics[height=3.5cm]{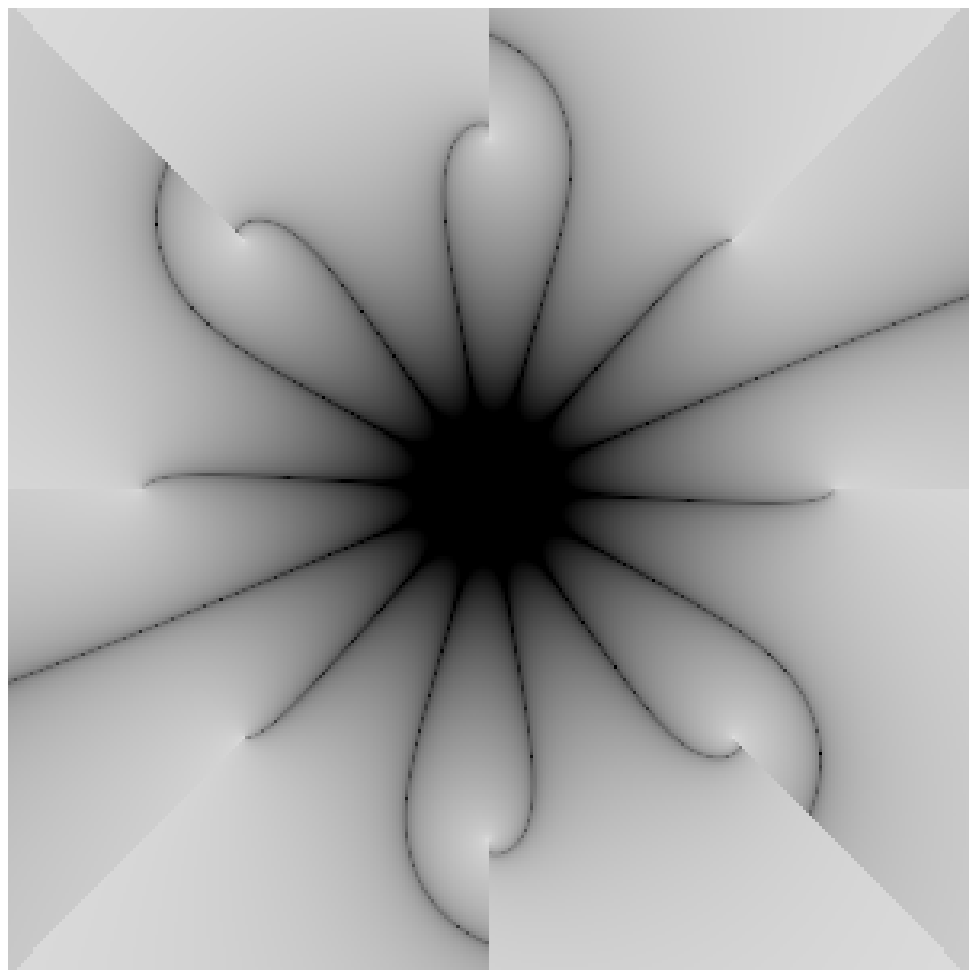}
\end{center}
\caption{Numerically generated pictures of the Stokes and anti-Stokes lines for large Schwarzschild-AdS black holes when $\arg(\omega)=-{3\pi \over D-1}$.  The top two pictures from left to right show the anti-Stokes lines in the complex $r$-plane in $D=8$ and $D=9$ respectively.  The bottom two pictures show the corresponding Stokes lines.}
\label{numerical4}
\end{figure}

\begin{figure}[tb]
\begin{center}
\includegraphics[height=3.5cm]{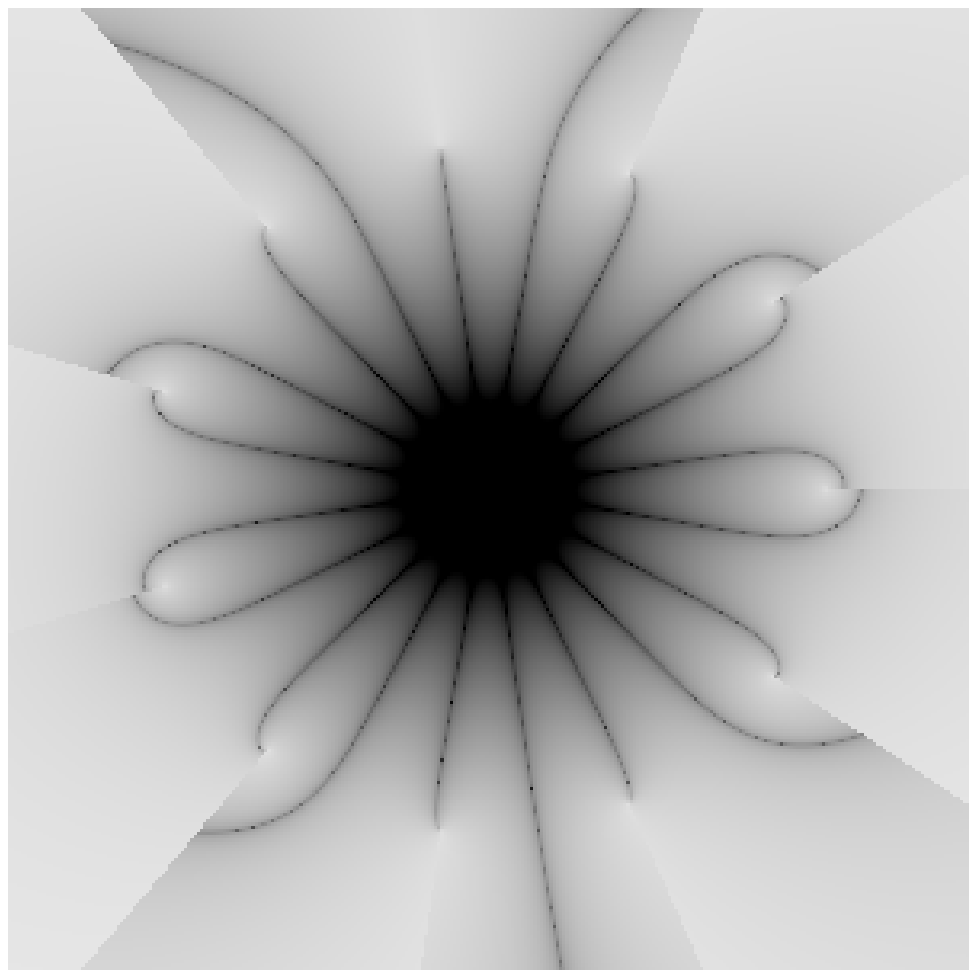}
\includegraphics[height=3.5cm]{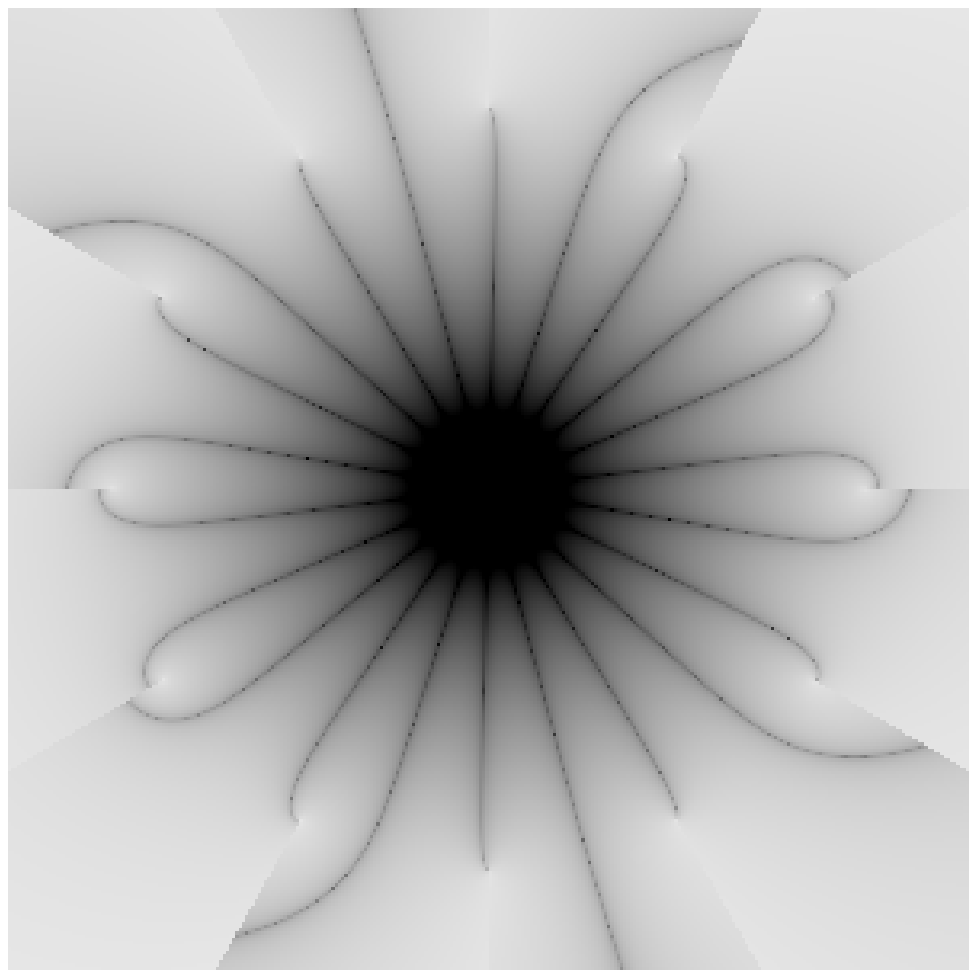}\\
\includegraphics[height=3.5cm]{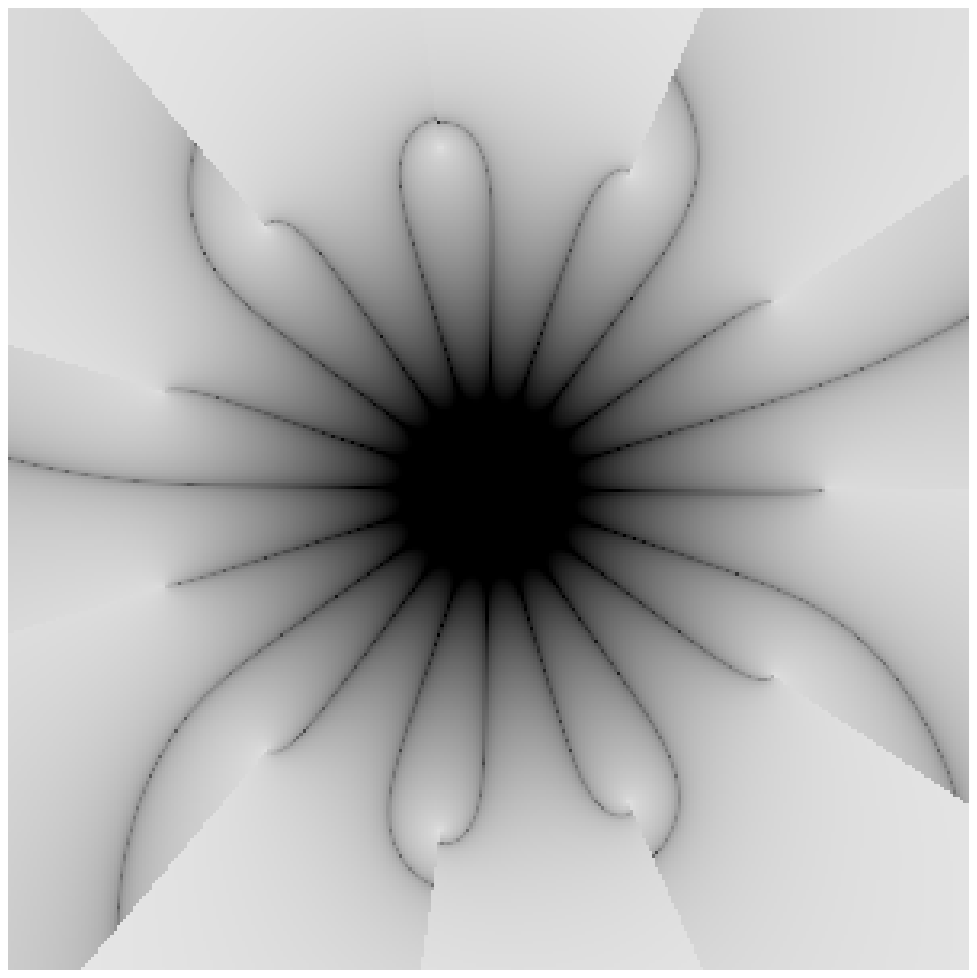}
\includegraphics[height=3.5cm]{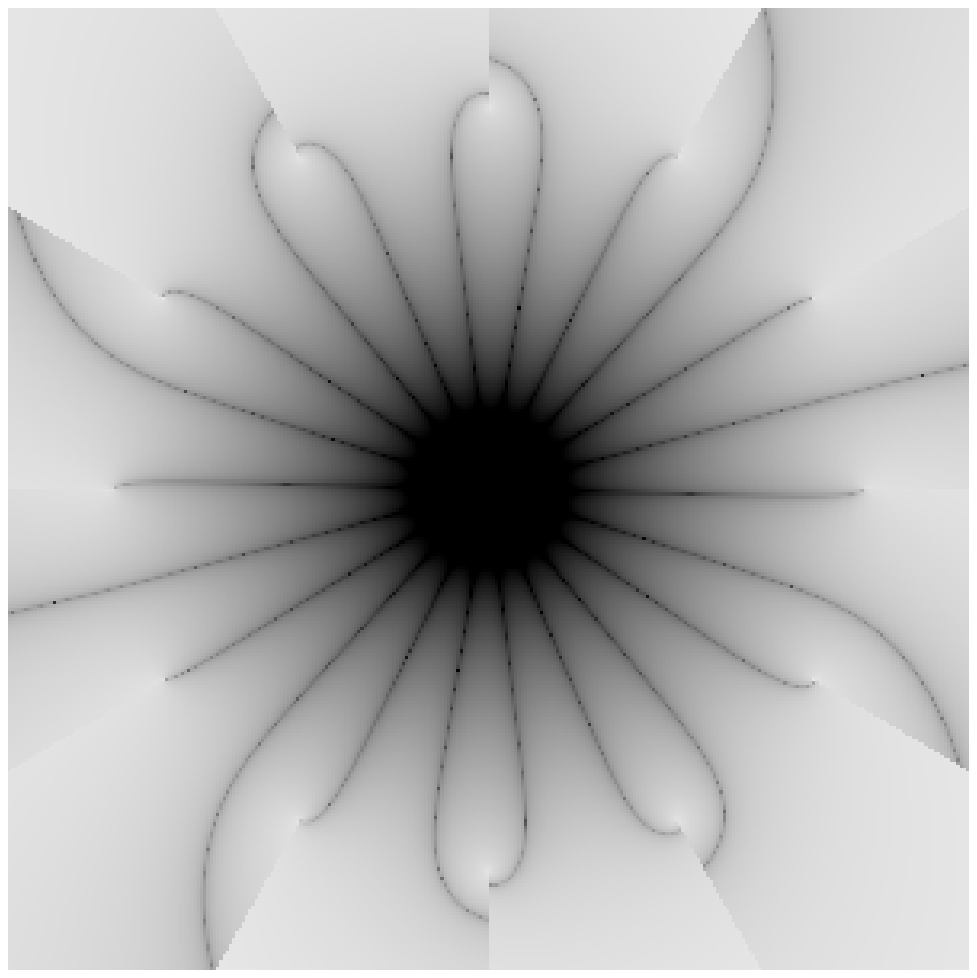}
\end{center}
\caption{Numerically generated pictures of the Stokes and anti-Stokes lines for large Schwarzschild-AdS black holes when $\arg(\omega)=-{5\pi \over D-1}$.  The top two pictures from left to right show the anti-Stokes lines in the complex $r$-plane in $D=12$ and $D=13$ respectively.  The bottom two pictures show the corresponding Stokes lines.}
\label{numerical5}
\end{figure}

\sxn{Other Possible Asymptotic Quasinormal Modes}
\vskip 0.3cm

In higher spacetime dimensions, new possible asymptotic regions of the QNM spectrum emerge in which both the real and imaginary part of the QNM frequency approach infinity similar to the case explained in section 3.  

In spacetime dimensions $D\ge 8$, a new choice of $\eta$ appears.  For large black holes, this $\eta$ has an argument of ${3\pi \over D-1}$.  For this particular $\eta$, the QNM frequency which allows the presence of an anti-Stokes line stretching to infinity takes the form
\beeq
\omega = |\omega| e^{-{3 i\pi \over D-1}}~,
\label{omega-alt1}
\eneq
in the case of large black holes.  The argument of $\omega$ will change for intermediate and small black holes, but the topology of Stokes/anti-Stokes lines will remain similar to the topology shown in Fig. \ref{numerical4}.  This topology is generic for all spacetime dimensions greater than seven.  The path that we take to determine the WKB condition is similar to the one we took in Fig. \ref{schem-d4-AdS}, with the exception that we need to cross two extra pairs of Stokes lines.  On the first pair of Stokes lines $f_2$ is dominant and on the second pair of Stokes lines $f_1$ is dominant.  After crossing these Stokes lines we get to an anti-Stokes line neighboring the anti-Stokes line which ends at the event horizon.  The rest of the calculation stays the same as in section 3 and we arrive at the WKB condition
\beeq
e^{2i\omega \eta} = i e^{2i\beta} e^{i\gamma}{1+e^{2i\gamma}+ e^{4i\gamma} \over 1+e^{2i\gamma}+e^{4i\gamma}+e^{6i\gamma} }~.
\label{D>7}
\eneq
This result translates to
\beeq
\omega \eta = \beta + n\pi  + {\pi\over 4}-{i\over2}\ln\left[e^{i\gamma}{1+e^{2i\gamma}+ e^{4i\gamma} \over 1+e^{2i\gamma}+e^{4i\gamma}+e^{6i\gamma}}\right]~~~~\mbox{as $n \rightarrow \infty$}~.
\eneq
After entering the values for $J$, $\beta$ and  $\gamma$ for tensor, vector, and scalar perturbations using Eqs.\ (\ref{J}), (\ref{Jinfinite}), and (\ref{beta}), it is easy to show the asymptotic QNM frequency for all types of perturbations in all dimensions is
\beeq
\omega \eta = n\pi + {\pi\over 4} (D+1)+{i\over2}\ln{4\over 3}~.
\label{4/3case}
\eneq

In spacetime dimensions $D\ge 12$, another new choice of $\eta$ emerges.  For large black holes this $\eta$ has an argument of ${5\pi \over D-1}$.  For this particular $\eta$, the QNM frequency which allows the presence of an anti-Stokes line stretching to infinity takes the form
\beeq
\omega = |\omega| e^{-{5 i\pi \over D-1}}~,
\eneq
in the case of large black holes.  The argument of $\omega$ will change for intermediate and small black holes, but the topology of the Stokes/anti-Stokes lines will remain similar to the topology shown in Fig. \ref{numerical5}.  Note that this topology is generic for all spacetime dimensions greater than eleven.  The path that we take to determine the WKB condition is similar to the one we took in Fig. \ref{schem-d4-AdS}, with the exception that we need to cross four extra pairs of Stokes lines.  Repeating the calculations will give the WKB condition
\beeq
e^{2i\omega \eta} = i e^{2i\beta} e^{i\gamma}{1+e^{2i\gamma}+ e^{4i\gamma}+e^{6i\gamma}+ e^{8i\gamma} \over 1+e^{2i\gamma}+e^{4i\gamma}+e^{6i\gamma} +e^{8i\gamma}+ e^{10i\gamma}}~.
\label{D>11}
\eneq
After entering the values for $J$, $\beta$ and  $\gamma$ for tensor, vector, and scalar perturbations using Eqs.\ (\ref{J}), (\ref{Jinfinite}), and (\ref{beta}), it is easy to show the asymptotic QNM frequency for all types of perturbations in all dimensions is
\beeq
\omega \eta = n\pi + {\pi\over 4} (D+1)+{i\over2}\ln{6\over 5}~~~~\mbox{as $n\rightarrow  \infty$}~.
\label{6/5case}
\eneq

It is not difficult to see that in spacetime dimensions $D\ge 16$, the new choice of $\eta$ for large black holes has an argument of ${7\pi \over D-1}$.  This $\eta$ results in $\arg(\omega)=-{7\pi \over D-1}$ for large black holes. The good news is that in the WKB conditions (\ref{D>7}) and (\ref{D>11}) a pattern can be recognized.  This pattern makes it easy to predict the WKB condition for the last case to be
\beeq
e^{2i\omega \eta} = i e^{2i\beta} e^{i\gamma}{1+e^{2i\gamma}+ e^{4i\gamma}+e^{6i\gamma}+ e^{8i\gamma} +e^{10i\gamma}+ e^{12i\gamma} \over 1+e^{2i\gamma}+e^{4i\gamma}+e^{6i\gamma} +e^{8i\gamma}+ e^{10i\gamma}+e^{12i\gamma}+ e^{14i\gamma}}~,
\label{D>15}
\eneq
and so on.

\sxn{Conclusions and Discussions}
\vskip 0.3cm

We have calculated explicitly the QNM frequencies of Schwarzschild-AdS black holes in the asymptotic region with infinitely large overtone numbers in arbitrary spacetime dimensions greater than three.  In this limit, we used an analytic phase-integral method based on the WKB approximation to confirm the results obtained by Cardoso {\it et al} \cite{Cardoso-N-S} for four spacetime dimensions and by Natario and Schiappa \cite{Natario-S} for arbitrary spacetime dimensions.  In addition, we have shown that this analytic technique implies the existence of other regions of the asymptotic QNM frequency spectrum which have not appeared earlier in the literature.  Among these is the highly real region of the QNM spectrum which appears only in even spacetime dimensions for black holes of any size.  We define the highly real QNMs as the modes where the real part of the frequency approaches infinity while the damping rate approaches a finite value.  These modes resemble the normal modes in a pure AdS space.  Another asymptotic region found in this paper is the highly damped region of the QNM spectrum which appears in spacetime dimensions $D=7,11,15,\dots$ for black holes of any size.  We define the highly damped QNMs as the modes where the damping rate approaches infinity while the real part of the frequency approaches a finite value.  %We are not sure if these asymptotic regions belong to the same QNM frequency spectrum which is explored earlier in the literature, for example in \cite{Cardoso-K-L}, or they belong to a new class of QNM frequency spectra.  

The big question is, do these newly discovered modes actually exist?  These modes have not been previously discovered using numerical techniques.  That could certainly be an indication that they do not exist.  As far as the analytic techniques are concerned, however, we see no reason to discard these regions of the QNM spectrum. Thus, our work leads us to the conclusion that there is an important shortcoming in these analytic methods.  More specifically, while these analytic methods predict many solutions, there doesn't seem to be a mechanism built into these methods to exclude certain solutions (other than the obviously non-physical solutions).  On the other hand, these analytic methods have been very successful to date in the calculation of the asymptotic QNMs of numerous black holes.  If these new modes do indeed exist, it could be that previous numerical research has not attempted to find QNMs in these regions.  Or, it could be that the numerical techniques have not been sufficient to detect these modes.  In the later case, new numerical methods would have to be developed.

For the sake of discussion, we will present some arguments in favor of the existence of these modes. 

One argument in favor of these modes is that since perturbing a pure AdS space results in normal mode frequencies, it would seem natural to have part of the QNM spectrum of a black hole in AdS space resembling such modes.

Another argument is the fact that other classes of QNM frequency spectra also exist in other black hole spacetimes.  For example, in Schwarzschild black holes, assuming the perturbations depend on time as $e^{-i\omega t}$, we have two different QNM frequency spectra with two different asymptotic regions.  These two spectra consist of modes with

a) positive real frequency and negative damping rate 

b) negative real frequency and negative damping rate

\noindent Of these choices, only (a) is physical.  These choices appear both in numerical and analytical calculations.  The same phenomenon happens in all the other black hole spacetimes in which only one spectrum is physical and all the others are ignored.  In the case of Schwarzschild-AdS black holes, as we showed, more than one of the asymptotic regions of the QNM frequency spectrum turn out to be physical and therefore cannot be ignored.

\begin{figure}[tb]
\begin{center}
\includegraphics[height=6cm]{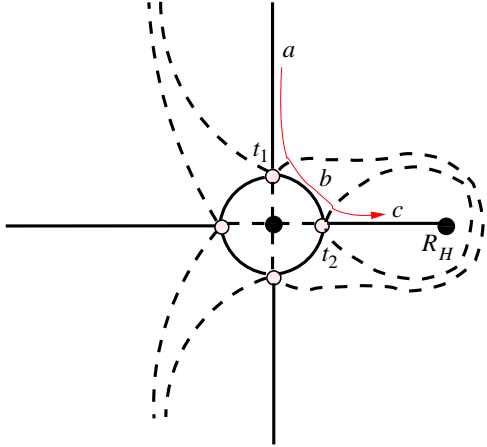}
\end{center}
\caption{Schematic presentation of the Stokes (dashed) and anti-Stokes (solid) lines in the complex $r$-plane for Schwarzschild black holes in four spacetime dimensions in the highly real limit when $|\omega_R| \gg |\omega_I|$.  The open circles are the zeros and the filled circles are the poles of the function $Q$.  $R_H$ is the event horizon.  The thin arrow which starts on the anti-Stokes line that stretches to infinity at point $a$ and ends on the anti-Stokes line that connects to the event horizon at point $c$ is the path we take to determine the WKB condition on QNM frequencies.}
\label{schem-schwarz-4d-real}
\end{figure}

One more argument that can be made in support of the existence of these regions is that the analytic techniques used to calculate the asymptotic QNMs are very restrictive.  For example, if we look for the highly real QNMs in Schwarzschild black holes we get a Stokes/anti-Stokes line structure of the form shown in Fig. \ref{schem-schwarz-4d-real}.  If we follow a path shown by the arrow in this figure it is easy to see that we can never come up with a WKB condition to determine the QNM frequencies.  In fact we can try the same calculation looking for any other region of the asymptotic QNM frequency spectrum and we will fail to derive a sensible WKB condition.  Therefore, if certain physical regions of the QNM spectrum can be calculated via the analytic technique employed in this paper then there is reason to believe those regions should exist.

The last argument is that the highly real modes are supported by the universal bound on the relaxation time $\tau$ of a perturbed thermodynamic system suggested by Hod in \cite{Hod2,Hod3}.  According to this universal bound, which is conjectured from information theory and thermodynamic considerations, the relaxation time of a perturbed black hole satisfies $\tau \ge {1 \over \pi T_{H}}$. 
From this, it follows that the damping rate of at least one of the black hole QNMs satisfies $|\omega_I| \le \pi T_H$.
No previously discovered QNMs conform to this bound, however, the highly real modes with a damping rate of $1.2T_H$, Eq.\ (\ref{damping-rate}), do conform to this bound.

In the remainder of this paper we discuss some consequences of both the QNMs which we already know exist and those newer modes that have been discussed in this paper.

Can we establish a connection between the asymptotic QNMs of Schwarzschild-AdS black holes and the quantum structure of their event horizon?  It is helpful if we consider a toy model such as a string fixed at two ends.  Assume the string is composed of discrete lattice points which are apart from each other with a quantum of length, i.e. the minimum possible distance allowed in nature.  If we perturb such a system, we will get normal modes with wavelengths which are multiples of the lattice spacing.  The frequency of every mode is a multiple of the fundamental frequency which again carries information about the lattice spacing.  The highest possible mode for such a system will have a wavelength comparable to the lattice spacing.  In order to go to the next mode after the highest mode we have to increase the length of the string or add lattice points to the system.  By analogy, we should expect that the perturbations of a black hole with a quantized horizon area should carry information regarding the quantum of the horizon area.  Since the black hole perturbations are damped, in order to minimize the possible modifications of the frequency of the oscillating system due to damping we need to look for a region of the QNM frequency spectrum where the oscillation frequency is much larger than the damping rate.  In other words, the highly real QNMs of Schwarzschild-AdS black holes is the most promising region of the spectrum to establish a connection, if there is any, between the QNMs and quantum behavior of the horizon area.  Suppose we accept that there is a relation between the highly real QNMs and the quantum of area $\Delta A$.  What would this relation look like?  The horizon area of a large Schwarzschild-AdS black hole is given by
\beeq
A=A_{D-2} r_H^{D-2} \approx A_{D-2} \left| \left({2 \mu \over |\lambda|}\right)^{1\over D-1}\right|^{D-2}~.
\eneq
From this we get
\beeq
\Delta A \approx  {2A_{D-2} \over |\lambda|} {D-2 \over D-1}\left| \left({2 \mu \over |\lambda|}\right)^{-{1\over D-1}}\right| \Delta \mu \approx {4 \over  T_H }\Delta M~,
\label{DA=DM}
\eneq
which is valid as long as the black hole is large.  We now apply the new interpretation of the QNM spectrum proposed by Maggiore \cite{Maggiore} who considers black hole perturbations in terms of a collection of damped harmonic oscillators where we need to use the proper frequency $\omega_0$ of the undamped oscillators rather than the frequency of the QNM ($\omega_R$).  The beauty of the highly real region of the QNM spectrum is that we do not need to go through the effort of distinguishing between these two frequencies as Maggiore did, because in this region $\omega_0 \sim \omega_R \sim \omega$.  This can provide an important testing ground for Maggiore's interpretation.  Now, according to this interpretation the absorbed energy $\Delta M$ is equal to the transition energy from the highly real QNM frequency $\omega_n\sim (\omega_R)_n $ to the highly real QNM frequency $\omega_{n-1}\sim (\omega_R)_{n-1}$ which means 
\beeq
\Delta M={\omega_R}_{n}-{\omega_R}_{n-1}=4 \pi T_H \sin(\pi/(D-1))~,
\label{DM-real}
\eneq
once we use Eq.\ (\ref{highly-real-WKB}) or (\ref{highly-real-WKB-scalar}).  This leads to
\beeq
\Delta A = 16 \pi \sin\left({\pi \over D-1}\right) ~,
\label{A-quanta}
\eneq
and we have an equispaced area spectrum of the form
$A = N \Delta A$,
where $N$ is a large integer.
Since entropy is equal to a quarter of the area of the event horizon for Schwarzschild-AdS black holes \cite{Page}, we find the spacing in the entropy spectrum to be
\beeq
\Delta S = 4 \pi \sin\left({\pi \over D-1}\right) ~.
\label{S-quanta}
\eneq
Let us now do the same calculation for the case of asymptotic QNMs where both the real and the imaginary part of the frequency approach infinity.  In these cases we see from Eq.\ (\ref{wkb-complex}) that
\beeq
\omega_0=\sqrt{\omega_R^2+\omega_I^2}\approx {n\pi \over |\eta|}= 4n \pi T_H \sin(\pi/(D-1))~,
\eneq
This result leads to the same spacing in the equispaced area and entropy spectrum found in Eqs.\ (\ref{A-quanta}) and (\ref{S-quanta}).  It is not difficult to show that for all the different asymptotic regions of the QNM frequency spectrum including the highly damped ones and for all types of perturbations the same  spacing in the area and entropy spectrum emerges.  Therefore, the results (\ref{A-quanta}) and (\ref{S-quanta}) are universal which shows that Maggiore's interpretation is extremely successful in the case of Schwarzschild-AdS black holes.

Another important consequence of the highly real QNM frequencies is their relevance to the AdS/CFT correspondence.  We showed that the damping rate of the highly real QNMs in four spacetime dimensions ($1.2T_H$, Eq.\ (\ref{damping-rate})), is actually less than the lowest damping rate found by Horowitz and Hubeny \cite{Horowitz}.  Therefore, in four dimensions the highly real QNMs decay at a slower rate than any other regions of the QNM spectrum and they are the most relevant in calculating the thermalization timescale in the CFT.  
Thus, the highly real QNMs of Schwarzschild-AdS black holes are not only useful in the context of quantum gravity, as is the case for other asymptotic QNMs, but they also have a physical application in studying many systems with two spatial dimensions in condensed matter physics (see for example Herzog {\it et al} \cite{Herzog}).  One puzzling issue, however, is the fact that the highly real QNMs only appear in even spacetime dimensions.  This leads to a fundamentally different thermalization behavior in the dual CFT in even and odd dimensions.  We do not have any intuition for why this would be the case.

Let us now return to Hod's interpretation \cite{Hod} of the highly damped QNMs.  According to Hod's proposal, when the damping rate approaches infinity the real part of the QNM frequency approaches a finite value which carries information about the black hole quantum area spectrum.  If we restrict ourselves to such a region of the QNM frequency spectrum and only consider metric perturbations of black holes in spacetime dimensions $D\ge4$, Hod's proposal seems to be very successful since $\ln(3)$ appears for all metric perturbations in every spacetime dimension (Kerr black holes seem to be an exception to our argument but we believe further investigation of the Kerr case is necessary).  It is only when we try to generalize Hod's proposal to include perturbations of fields coupled to the background of a black hole or when we demand that the natural logarithm of an integer should appear in the QNM spectrum of black holes in which the highly damped region does not exist things go wrong.  In this paper we showed that, beside Schwarzschild black holes in $D\ge 4$, such a region can also appear in Schwarzschild-AdS black holes in dimensions $D=7, 11, 15, 19, \dots$.  Assuming that this region of the QNM frequency spectrum is the most relevant to Hod's interpretation in \cite{Hod}, we can equate the absorbed energy $\Delta M$ to the real part of the highly damped QNM frequency of Schwarzschild-AdS black holes in $D=7$, i.e., 
$\Delta M={\omega_R}=\ln\left({\pm 1+ \sqrt{17} \over 2}\right)T_H$.
This together with Eq.\ (\ref{DA=DM}) leads to the equispaced area spectrum with spacing
\beeq
\Delta A = 4\ln\left({\pm 1+ \sqrt{17} \over 2}\right)  ~.
\eneq  
The corresponding spacing in the entropy spectrum is 
\beeq
\Delta S = \ln\left({\pm 1+ \sqrt{17} \over 2}\right)~.
\eneq  
Unfortunately, we do not get the natural logarithm of an integer which allows for an elegant statistical interpretation \cite{Hod, Bekenstein-1} of the Bekenstein-Hawking entropy spectrum; but, based on Maggiore's arguments \cite{Maggiore}, the appearance of the logarithm of an integer may not be necessary after all.  
%Our arguments for the highly damped QNMs of Schwarzschild-AdS black holes are too speculative at this stage.  We do not have much confidence in the existence of this region of the QNM frequency spectrum due to the reasons explained in section 5.  

One last observation is that the term $\ln(2)$ appears in many asymptotic regions of the QNM frequency spectra of Schwarzschild-AdS black holes.  As mentioned earlier, Hod in \cite{Hod} argues that the appearance of the natural logarithm of an integer in the asymptotic QNM frequency of Schwarzschild black holes allows for an elegant statistical interpretation of the resulting entropy spectrum.  Is it possible to claim the same thing for the term $\ln(2)$ which appears in Schwarzschild-AdS black holes?  Unfortunately, in the case of Schwarzschild-AdS black holes, $\ln(2)$ does not appear exclusively in the real part of the asymptotic QNM frequency.  It appears both in the real and imaginary parts of the frequency and in the case of highly real QNMs, it appears in the imaginary part of the frequency only.  This makes it difficult to establish a connection between the term $\ln(2)$ and quantum gravity.  It seems that if the appearance of $\ln(2)$ in Schwarzschild-AdS black hole QNMs is just a sheer coincidence , then the results of this paper may have a negative impact on Hod's arguments which try to associate a deeper meaning to the appearance of the natural logarithm of an integer in asymptotic QNMs.  This issue needs to be studied closely.

As a final remark, we would like to stress that considering the interesting consequences of the existence of the highly real QNMs, further numerical and analytical investigations are warranted to prove the existence or nonexistence of these modes.

\vskip .5cm

\leftline{\bf Acknowledgments}
We are grateful to Gabor Kunstatter for sharing his invaluable insights on the topic of black hole quasinormal modes during numerous useful discussions. 
We also thank Shahar Hod for very useful comments on this paper.

%%%%%%%%%%%%%%%%%%%%%%%%%%%%%%

% A useful Journal macro
\def\jnl#1#2#3#4{{#1}{\bf #2} (#4) #3}

\def\Zphys{{\em Z.\ Phys.} }
\def\jssc{{\em J.\ Solid State Chem.} }
\def\jpsJ{{\em J.\ Phys.\ Soc.\ Japan} }
\def\ptps{{\em Prog.\ Theoret.\ Phys.\ Suppl.\ } }
\def\PTP{{\em Prog.\ Theoret.\ Phys.\  }}
\def\LNC{{\em Lett.\ Nuovo.\ Cim.\  }}
\def\LRR{{\em Living \ Rev.\ Relative.} }
\def\JMP{{\em J. Math.\ Phys.} }
\def\NPB{{\em Nucl.\ Phys.} B}
\def\NP{{\em Nucl.\ Phys.} }
\def\PLB{{\em Phys.\ Lett.} B}
\def\PL{{\em Phys.\ Lett.} }
\def\PRL{\em Phys.\ Rev.\ Lett. }
\def\PRB{{\em Phys.\ Rev.} B}
\def\PRD{{\em Phys.\ Rev.} D}
\def\PR{{\em Phys.\ Rev.} }
\def\PRe{{\em Phys.\ Rep.} }
\def\AP{{\em Ann.\ Phys.\ (N.Y.)} }
\def\RMP{{\em Rev.\ Mod.\ Phys.} }
\def\ZPC{{\em Z.\ Phys.} C}
\def\SCI{\em Science}
\def\CMP{\em Comm.\ Math.\ Phys. }
\def\MPLA{{\em Mod.\ Phys.\ Lett.} A}
\def\IJMPB{{\em Int.\ J.\ Mod.\ Phys.} B}
\def\cmp{{\em Com.\ Math.\ Phys.}}
\def\JPA{{\em J.\  Phys.} A}
\def\CQG{\em Class.\ Quant.\ Grav.~}
\def\ATMP{\em Adv.\ Theoret.\ Math.\ Phys.~}
\def\PRSA{{\em Proc.\ Roy.\ Soc.\ Lond.} A }
\def\IJTP{\em Int.\ J.\ Theor.\ Phys.~}
\def\ibid{{\em ibid.} }
\vskip 1cm

\leftline{\bf References}

\renewenvironment{thebibliography}[1]
        {\begin{list}{[$\,$\arabic{enumi}$\,$]}  
% {\arabic{enumi}.}
        {\usecounter{enumi}\setlength{\parsep}{0pt}
         \setlength{\itemsep}{0pt}  \renewcommand{\baselinestretch}{1.2}
         \settowidth
        {\labelwidth}{#1 ~ ~}\sloppy}}{\end{list}}

%%%%%%%%%%%%%%%%%%

\end{document}